\documentclass[11pt]{article}

\usepackage[margin=1in]{geometry}
\usepackage{amsmath,amssymb,mathtools,amsthm,mathrsfs}
\usepackage{booktabs,tabularx,array}
\usepackage{enumitem}
\usepackage{microtype}
\usepackage{graphicx}
\usepackage{needspace}
\usepackage{setspace}
\usepackage[authoryear,round]{natbib}
\usepackage{tikz}
\usetikzlibrary{arrows.meta}

\setlist[itemize]{leftmargin=1.5em,itemsep=0.2em,topsep=0.25em}
\setlist[enumerate]{leftmargin=1.7em,itemsep=0.2em,topsep=0.25em}

\theoremstyle{plain}
\newtheorem{theorem}{Theorem}
\newtheorem{proposition}{Proposition}

\newtheorem{corollary}{Corollary}

\theoremstyle{definition}
\newtheorem{definition}{Definition}
\newtheorem{assumption}{Assumption}

\theoremstyle{remark}
\newtheorem{remark}{Remark}

\newcommand{\E}{\mathbb{E}}
\makeatletter
\newcommand{\px}{\mathpalette\pxaux{}}
\newcommand{\pxaux}[2]{\mathord{\vbox{\offinterlineskip\ialign{\hfil##\hfil\cr\hbox{\kern.12ex\scalebox{\ifx#1\scriptstyle .88\else\ifx#1\scriptscriptstyle .73\else 1.10\fi\fi}{$\m@th\scriptscriptstyle\circ$}}\cr\noalign{\kern.08ex}\hbox{$\m@th#1x$}\cr}}}}
\makeatother
\newcommand{\Prob}{\mathbb{P}}

\newcommand{\T}{\mathcal{T}}
\newcommand{\Sg}{S}

\newcommand{\R}{\mathbb{R}}

\begin{document}

\title{A Theory of Dynamic Mechanism Design with Endogenous Information and Hidden Actions}
\author{Guoqiang Tian\thanks{Department of Economics, Texas A\&M University, USA. Email: gtian@tamu.edu.}}
\date{First draft: March 2026; Revised December 1, 2026}
\maketitle

\begin{abstract}
This paper studies how the dynamic mechanism design problem changes when contracting, information-production technology, and disclosure are jointly optimized in the presence of hidden actions. Our framework allows information acquisition, learning, disclosure, and private choices to interact through dynamic incentives. We accommodate correlated private information, interdependent preferences, and potentially infinite-dimensional allocations. We introduce joint perfect Bayesian equilibrium, establish the Dynamic Revelation Principle, and recursively characterize incentive compatibility. Under private values with informational interdependence, we characterize joint incentives through choice obedience, reporting-envelope first-order conditions, and global integral monotonicity. We derive an exact dynamic information-rent identity, dynamic revenue equivalence, and dynamic virtual surplus. Comparing first-best, two second-best, and third-best contracts distinguishes information rents, hidden-action implementation costs, and direct information-production expenditures, revealing nonadditive interactions of either sign and branched outcomes. Periodwise disclosure optimality need not imply joint optimality. Our results reveal how and why established results change and under what conditions they remain valid. Applications concern auctions and financial contracting.

\medskip\noindent
\textbf{JEL Classification:} D82, D83, D86, C73.

\noindent
\textbf{Keywords:} dynamic mechanism design, joint institutional design, endogenous information, hidden actions, joint incentive compatibility, optimal contracting, information disclosure, implementation costs.
\end{abstract}

\pagebreak
\setcounter{tocdepth}{2}
\tableofcontents

\pagebreak

\section{Introduction}

Economic theory has achieved fundamental advances by isolating particular economic forces. Screening analyzes allocations under private information; moral hazard studies incentives for hidden actions; dynamic mechanism design characterizes intertemporal incentives and rents; and information acquisition, experimentation, and information design examine how information is produced and used. These approaches are scientifically indispensable and complementary: by holding other economic margins fixed, they isolate individual causal and incentive relationships. Yet institutional design must also account for interactions among these margins when they are jointly determined in the economic system. Actions produce information; disclosure changes the return to acquiring it; information affects future incentives and choices. \emph{Endogenous information} encompasses the acquisition, production, learning, use, and disclosure of information through economic and institutional choices. Their joint determination poses a problem that cannot generally be resolved by optimizing each margin separately.

This problem reflects the fundamental principle of incentive-based mechanism design, rooted in Hurwicz's analysis of structural invariance: relations observed under one institutional arrangement may change when the rules governing economic behavior change \citep{Hurwicz1962}. The Lucas critique further revealed the significance of this problem for macroeconomic policy evaluation: new policies alter expectations and decision rules, so behavioral relations estimated under an earlier policy need not predict the effects of the new one \citep{Lucas1976}. This methodological tradition is reaffirmed by \citet{Sargent2025Lucas}, who emphasizes equilibrium consistency in the analysis of policy interventions. The same principle applies to institutional design. Allocation, incentive, and information rules shape the strategic responses on which institutional performance depends.

Recent results demonstrate why the interaction of endogenous margins matters for institutional design. \citet{Tian2026} establishes three impossibility results: endogenous resource constraints can make an informational frontier unattainable; sequential incentive and disclosure design need not achieve the joint optimum; and Blackwell-more-informative technologies need not increase maximal implementable value without implementation replicability. In auctions, the distribution-independent reveal-low/pool-high optimum of \citet{BHMSW2022} need not survive endogenous costly verification, even with uniform or binary valuations \citep{TianAuction2026}. The auction application in Section~\ref{subsec:auction-application} provides a distinct reversal: endogenous acquisition yields first-price revenue above second-price revenue under the specified equilibria and common primitives, in contrast with the fixed-information ranking of \citet{MilgromWeber1982} under its hypotheses. 

The foregoing considerations motivate three questions of institutional design: how do established results change when previously restricted institutional margins are jointly considered, what economic forces account for these changes, and under what conditions do the established results survive? The generality of our framework serves to address these questions rather than being an end in itself. Our analysis proceeds at three levels.

First, at the level of economic perspective, we study joint optimization of contracting, information technology, and disclosure in general economic environments with \emph{joint allocation constraints}. Reporting and action incentives constrain these choices. We examine how dynamic private information and privately implemented hidden actions change institutional design when institutional rules, actual implementation, and endogenous information jointly determine equilibrium outcomes.

Second, we extend the general allocation-space approach of \citet{PavanSegalToikka2014} (PST) to dynamic economies with endogenous information and privately implemented hidden actions. General, potentially infinite-dimensional allocation spaces accommodate consumption, production, private and public goods, and other economic activities under joint allocation constraints. It allows multiple agents, interdependent preferences, correlated information, and history-dependent transitions without independence or Markovianity. Information technology may be chosen by the principal or by one or more agents, depending on the economic environment. Mechanism rules and privately implemented actions jointly determine realized outcomes, information production, and future economic opportunities.

Third, we establish dynamic implementation and joint incentive characterizations, and derive their implications for information rents, optimal contracting, and disclosure. We introduce \emph{JPBE} in the indirect-mechanism environment. For \emph{direct mechanisms}, we define \emph{joint perfect Bayesian incentive compatibility} (JPBIC) and establish the \emph{Dynamic Revelation Principle} under JPBE and the theorem's causal representation conditions. Under these conditions, indirect equilibrium outcomes admit truthful history-reporting representations with contingent private-action implementation. Reporting current types differs from reporting complete private histories; nonimplementability under the former restriction requires separate verification when actions are privately implemented. A necessary and sufficient recursive characterization of JPBIC then connects reporting optimality, post-report choice obedience, consistent beliefs, and continuation payoffs. Crucially, reporting incentives must account for joint deviations in reports and subsequent private choices. Under the stated causal representation conditions, these implementation results allow correlated and history-dependent type transitions.

Under a private-values payoff specification, direct payoff dependence on other agents' types is restricted, while informational interdependence, endogenous information production, disclosure, and hidden actions remain. Theorem~\ref{thm:integral-monotonicity} characterizes JPBIC through \emph{choice obedience} (CO), \emph{reporting-envelope first-order condition} (RE-FOC), and \emph{global integral monotonicity} (GIM), using the recursive characterization of Theorem~\ref{thm:recursive-ic}. CO is equivalent to CO-FOC under its stated concavity and differentiability conditions.

Theorem~\ref{thm:dynamic-virtual-surplus} establishes an \emph{exact dynamic information-rent identity}. The principal's expected payoff equals economic surplus less dynamic information rents and lowest-type continuation rents, even when initial private types and signals are correlated. The identity also implies \emph{dynamic revenue equivalence} for mechanisms sharing marginal continuation-rent functions. Focusing next on independent controlled private-type processes, we derive explicit impulse-response and \emph{dynamic virtual-surplus} representations. Hidden information generates dynamic rent adjustments, whereas hidden action constrains the economic choices that can be induced; its implementation cost is not simply an additional virtual-type term.

We characterize four contracting regimes: \emph{first best} (1B), \emph{second best with hidden information} ($2B^{\mathrm{HI}}$), \emph{second best with hidden action} ($2B^{\mathrm{HA}}$), and \emph{third best with both frictions} (3B). Their implementable outcome sets form a branched inclusion hierarchy, with no general value ranking between the two second-best branches. In the Coase--Williamson tradition, the comparison distinguishes informational rents, hidden-action implementation costs, and direct information-production expenditures. We derive two decompositions of the first-best-to-third-best value loss, depending on the order in which the incentive frictions are introduced. Their interaction is not generally additive, and its sign is unrestricted. A finite-horizon quadratic specification then identifies dynamic virtual-type and obedience distortions and characterizes how their interaction changes optimal economic choices.

Theorems~\ref{thm:joint-design-existence}--\ref{thm:information-design-characterization} address joint optimization of contracting, information technology, and disclosure: attainment and continuity, \emph{institutional dominance}, and conditional and joint optimality conditions. Unlike static disclosure design (\citealp{TianOID2026}), periodwise linearity under finite-model and fixed-assessment restrictions need not extend to joint intertemporal optimization. The principal determines an information technology $\mathcal Y\in Y$ and commits to it ex ante. Agents choose their actions, which jointly with the selected kernel determine the conditional distribution of produced information and private signals. Information production and disclosure trade off the principal's gains from improved information against production, disclosure, and implementation costs. Information produced by economic activity bounds the disclosure policies available to the institution: public signals must arise through garblings of the underlying experiment, not merely satisfy an abstract Bayes-plausibility condition. Disclosure changes continuation rewards and information-producing actions: current-payoff replication need not preserve continuation incentives, and optimal incentive and disclosure design are generally nonseparable. We identify conditions for separation. Correlated private signals also provide peer-information incentives: under finite-type \emph{predictive separation}, appropriate scoring rules can discipline reporting deviations, whereas a fixed finite quadratic score need not deter arbitrarily small deviations when types are continuous. Optimal disclosure is characterized through its effects on economic surplus, dynamic information rents, and lowest-type continuation rents. The analysis provides conditions for interior and boundary policies and \emph{bang--bang} rules. In the quadratic example, full disclosure strictly raises the principal's maximized payoff over partial or no disclosure in all four contracting regimes under the stated conditions.

The applications establish concrete consequences for institutional design. In interdependent-value auctions with endogenous information acquisition, auction formats change acquisition incentives, equilibrium bids, and informational rents. We derive an interdependent-value envelope incorporating posterior-response effects and a seller-revenue decomposition separating allocated value, real acquisition costs, and bidders' rents. Endogenous information acquisition can reverse the classical revenue ranking of \citet{MilgromWeber1982}; under Gaussian signals, no efficient first-price equilibrium exists within the specified equilibrium class. In financial contracting, we construct a fee menu that jointly implements truthful reports and hidden evidence production and derive an explicit optimal disclosure policy that may entail no disclosure, interior disclosure, or full disclosure according to the economic value and cost of verifiable evidence.

Taken together, these results reveal how the joint determination of contracting, information production, disclosure, and private actions changes not only the conditions for established economic results but also the properties of optimal institutions. They identify the economic forces underlying these changes and the conditions under which established results remain valid.

\paragraph{Related literature.} First, consider the informational and incentive approach to economic institutions. The structural-invariance problem of \citet{Hurwicz1962} and the mechanism-design program of \citet{Hurwicz1973} provide a foundation for studying how economic rules induce strategic responses. General implementation and revelation theory examines which social outcomes can be supported through strategic communication and incentives \citep{HurwiczShapiro1978,Myerson1981,Myerson1982,GreenLaffont1986,Forges1986,EpsteinPeters1999,BergemannMorris2005}. The distinction between equilibrium outcomes of indirect mechanisms and their direct representations is particularly important here because private action choices and information generated after reports can affect continuation behavior. Dynamic implementation further clarifies the role of equilibrium restrictions \citep{Penta2015}. \citet{SugayaWolitzky2021} study communication revelation in multistage games. Our analysis identifies conditions for preserving conditional economic-path laws across mechanisms when private actions affect information production and future incentives.

The second strand is dynamic screening, mechanism design, and recursive contracting. Continuing relationships with private information are studied by \citet{BaronBesanko1984}, \citet{CourtyLi2000}, and \citet{Battaglini2005}; efficient dynamic mechanisms and dynamic pivot rules are developed by \citet{BergemannValimaki2010} and \citet{AtheySegal2013}. \citet{PavanSegalToikka2014} provide the principal benchmark for the general economic allocation-space approach, dynamic envelopes, controlled impulse responses, integral monotonicity, and virtual surplus. Further developments study dynamic contracting, variational methods, and the limits of first-order approaches \citep{KakadeLobelNazerzadeh2013,BergemannPavan2015,GarrettPavan2015,BergemannStrack2015,BattagliniLamba2019,BergemannValimaki2019,MakrisPavan2021}. \citet{Pavan2017} examines robustness and endogenous types in dynamic mechanism design. The recent survey by \citet{Pavan2026} synthesizes dynamic contracting, including environments with learning, experimentation, and evolving private information. Recovering PST's virtual-surplus representation need not recover their optimal contracting problem when choice obedience remains binding.

A third strand concerns adverse selection, moral hazard, and their interaction. Foundational contracting models analyze private information together with hidden economic choices \citep{LaffontTirole1986,LaffontTirole1988,Picard1987,GuesneriePicardRey1989,CaillaudGuesnerieRey1992,MengTian2013}. Recursive and continuous-time agency models characterize hidden-action incentives through continuation promises and contractible performance information \citep{DeMarzoSannikov2006,Sannikov2008,Williams2011}. Dynamic interactions and interdependent economic incentives are also studied by \citet{MengTian2021} and \citet{MengSunTian2022}. Whereas \citet{DoepkeTownsend2006} and \citet{EsoSzentes2017} analyze dynamic contracting with hidden information and hidden actions, we additionally endogenize information production and disclosure and characterize their joint effects on incentives and optimal institutions. The benchmark comparisons further connect to second-best analysis \citep{LipseyLancaster1956}: information rents and the incentive costs of hidden-action implementation interact in the set of attainable contracts rather than forming independent additive deductions.

A fourth strand develops information comparison, persuasion, and institutional information design. The comparison of experiments originates with \citet{Blackwell1953}; Bayesian persuasion and Bayes-correlated equilibrium make information structures and obedience central design objects \citep{KamenicaGentzkow2011,BergemannMorris2016,BergemannMorris2019}. Screening with persuasion provides a related information-policy benchmark \citep{BergemannHeumannMorrisJPE2026}; dynamic information design examines sequential signals, feedback, and disclosed evidence \citep{Ely2017,DovalEly2020,ElySzydlowski2020}. Complementing \citet{ElyGeorgiadisRayo2025}, we examine how endogenous information production and disclosure jointly affect screening rents and private-action implementation. Disclosure is constrained by evidence generated through privately implemented economic activity. This distinction is central to both technological restrictions on information and the nonseparability of contracting and disclosure.

The fifth strand concerns experimentation, learning, and endogenous information acquisition. Costly information acquisition has been studied as an incentive and mechanism-design problem \citep{BergemannValimaki2002,Persico2000,BergemannShiValimaki2009}; dynamic experimentation and learning connect economic choices to subsequent evidence and opportunities \citep{HornerSkrzypacz2017,ElyGeorgiadisRayo2025,Manso2011}. Together with the preceding strand on information design and disclosure, these theories address distinct aspects of endogenous information. They inform the distinction between an information technology, the private choice of information-generating activity, and an institutional disclosure rule. The present environment allows the same privately implemented choice to affect current allocations, future information, and subsequent economic states, while its implementation remains subject to JPBIC.

A sixth strand studies correlated information, evidence, and strategic communication. Mechanisms with verifiable or partially verifiable information restrict permitted claims and disclosures \citep{GreenLaffont1986,BenPorathDekelLipman2019,BesterStrausz2001,BesterStrausz2007}. One-sided restrictions on reports also arise in \citet{Tian1989}, where agents can understate but not overstate endowments, and in the explicit withholding mechanism of \citet{Tian1993}. Agency communication has also been analyzed as an incentive instrument \citep{MelumadReichelstein1989}. These restrictions on reports or claims differ from the present dynamic implementation problem, which also incorporates privately chosen information-producing actions and their continuation effects. Correlated reports can support informational inference and surplus extraction under additional conditions \citep{CremerMcLean1988}; peer-information and monitoring methods provide related incentive instruments \citep{MillerResnickZeckhauser2005,QiuCarrollAllen2026}. Our analysis distinguishes the finite-type predictive-separation conditions under which peer scores reinforce reporting incentives from the continuous-type limitations of a fixed finite quadratic penalty, while separately retaining the choice-obedience problem.

Finally, the applications connect the general theory to auctions and financial contracting. Classic affiliated-value auction analysis establishes influential revenue comparisons \citep{MilgromWeber1982}; information acquisition, disclosure, and bidding are jointly relevant to their scope \citep{BergemannShiValimaki2009,EsoSzentes2007,BHMSW2022,TianAuction2026}. The latter two papers also show the importance of distinguishing distribution-independent optimal-disclosure results from their performance under endogenous costly verification. In financial contracting, costly verification and evidence-dependent claims relate to \citet{Townsend1979}, \citet{Innes1990}, and \citet{DeMarzoSannikov2006}. Across these applications, the central economic question is how institutional rules change the informational and behavioral responses that determine implementable outcomes and optimal policies.

The remainder of the paper is organized as follows. Section 2 specifies the economic environment and information technology. Section 3 establishes equilibrium, revelation, and recursive JPBIC. Section 4 characterizes joint screening incentives; Section 5 studies optimal contracting and transaction costs; and Section 6 analyzes joint information-technology and disclosure design. Section 7 develops the applications, and Section 8 concludes.

\section{Economic Environment, Information Production, and Disclosure}
\label{sec:environment}

We study dynamic screening with hidden actions, endogenous information production, and disclosure.

\subsection{Economic primitives and preferences}
\label{subsec:primitives}

Time is discrete, $t=1,2,\ldots,T$, where $T$ may be finite or infinite. There is one principal and a finite set $N=\{1,\ldots,n\}$ of agents. All parties discount future payoffs by $\beta\in(0,1]$, with $\beta<1$ if $T=\infty$. The principal can commit to a dynamic mechanism at date $0$.

Let $X_t\subseteq\overline X_t:=\prod_{i\in N}\overline X_{it}$ be the \emph{feasible allocation space}, with individual components $x_{it}\in \overline X_{it}$. A positive $p_{it}$ is a payment from agent $i$ to the principal. An \emph{allocation} specifies consumption bundles and production plans for private or public goods and services. The spaces are general measurable spaces, possibly multidimensional or infinite-dimensional; $X_t$ may impose resource, capacity, technological, or inter-agent constraints.

At each date $t$, let $\theta_t\in\Theta_t$ denote a potentially unobserved \emph{payoff-relevant state}. Agent $i$ has a privately observed \emph{screening type} $\tau_{it}\in\T_{it}=[\underline\tau_{it},\overline\tau_{it}]\subset\R$ and an additional \emph{private signal} $s_{it}\in\Sg_{it}$. Let $\tau_t=(\tau_{it})_{i\in N}\in\T_t:=\prod_{i\in N}\T_{it}$ and $s_t=(s_{it})_{i\in N}\in\Sg_t:=\prod_{i\in N}\Sg_{it}$. After observing their private types and signals, agents send messages to the mechanism that determines a \emph{recommended allocation} $\px_{t}\in X_t$ and a vector of monetary transfers $p_t=(p_{it})_{i\in N}\in\R^n$. Let $x_t$ denote the \emph{realized allocation}, which may differ from $\px_{t}$ because of agents' hidden actions.

The type $\tau_{it}$ may represent a valuation, marginal cost, or productivity, while the signal $s_{it}$ may convey information about $\theta_t$ or other agents that is not summarized by the current type. Agent $i$ observes both $\tau_{it}$ and $s_{it}$ before date-$t$ communication; the signal is generated in the preceding period, or drawn initially at $t=1$. He need not observe $\theta_t$ or the other agents' type and signal profiles, $\tau_{-i,t}$ and $s_{-i,t}$, respectively. The principal does not directly observe the agents' private types or signals. The current type need not summarize all previously acquired private information; additional information-sufficiency conditions are stated where needed for subsequent results.

For time-indexed objects and spaces, write $b_t^s:=(b_r)_{r=t}^{s}$ and $B_t^s:=\prod_{r=t}^{s}B_r$ for $1\le t\le s\le T$. Let $b^s:=b_1^s$ and $B^s:=B_1^s$ for $s\ge 1$, with agent indices retained. At $s=0$, set $b^0:=\varnothing$ and $B^0:=\{\varnothing\}$. Hidden-action deviations need not be public, though subsequent signals or disclosures may reveal their consequences. For private-history reports, hats distinguish reported from actual values. A tilde distinguishes random variables from realizations, e.g., $\tilde\theta_t$ from $\theta_t$, with the same convention for types, signals, allocations, transfers, and disclosed information.

All primitive, history, message, allocation, and outcome spaces are standard Borel spaces. All mappings are measurable with respect to their relevant $\sigma$-algebras, unless otherwise stated. Here $\Delta(B)$ denotes the set of probability measures on a measurable space $B$. We write $\mathbb P(B\mid c;\eta)$ for the probability of a measurable event $B$, where $c$ denotes conditioning information and $\eta$ denotes parameters. The notation also applies to $\mu$. Primitive uncertainty is defined on a probability space $(\Omega,\mathcal A,\Prob)$ with an initial joint distribution
$\mathbb P_1\in\Delta(\Theta_1\times\T_1\times\Sg_1)$
for $(\tilde\theta_1,\tilde\tau_1,\tilde s_1)$. The initial distribution permits correlation among agents' types and signals and with the underlying state. Initial signals may be degenerate. The probability space supports subsequent primitive shocks and mechanism randomization.

We maintain a \emph{common-prior Bayesian environment}: primitive laws, preferences, allocation spaces, and information technologies are common knowledge, while distinct observations induce history-dependent posterior beliefs wherever Bayes' rule applies. Off-path beliefs are specified in the equilibrium assessment (Section~3). The principal's committed mechanism and disclosure rules are common knowledge; induced probability laws depend on these rules and agents' strategies.

Let
\[
u_{it}:\overline X_t\times\T_t\times\Theta_t\longrightarrow\R,
\qquad
u_t^{P}:\overline X_t\times\T_t\times\Theta_t\longrightarrow\R
\]
denote agent $i$'s and the principal's \emph{gross period payoff functions}, respectively. Dependence on the full type profile accommodates \emph{interdependent values}, while dependence on the joint allocation accommodates strategic and technological interactions. The principal's payoff may depend on privately observed types without requiring him to observe their realizations. The \emph{private-values restriction} imposes $u_{it}(x_t,\tau_t,\theta_t)=\bar u_{it}(x_t,\tau_{it},\theta_t)$ for all $i,t$; this restriction underlies the analyses in Sections~4--6, subject to their respective additional assumptions.

For a realized outcome path, intertemporal payoffs are
\begin{equation}\label{eq:lifetime-payoffs}
\begin{aligned}
U_i&=\sum_{t=1}^T\beta^{t-1}[u_{it}(x_t,\tau_t,\theta_t)-p_{it}],\qquad i\in N,\\
U^P&=\sum_{t=1}^T\beta^{t-1}\left[u_t^{P}(x_t,\tau_t,\theta_t)+\sum_{i\in N}p_{it}\right].
\end{aligned}
\end{equation}
The principal's expected payoff is $\Pi:=\E[U^P]$. Expected payoffs are evaluated under the probability laws induced by the primitive laws, the committed mechanism, and agents' strategies. Conditions ensuring finite expected continuation payoffs, particularly for an infinite horizon, are imposed before the corresponding incentive results.

\subsection{Information dynamics and disclosure}
\label{subsec:technology}

Economic states and information evolve over time. Their evolution is described by five related stochastic kernels: the \emph{state-transition kernel} $\mathcal Q_{t+1}$, the \emph{information-production technology} $\mathcal Y_{t+1}$, the \emph{disclosure kernel} $\mathcal D_{t+1}$, the \emph{type-transition kernel} $\mathcal F_{t+1}^{\tau}$, and the \emph{induced joint-transition kernel} $\mathcal K_{t+1}$. Let $d_t\in\overline{D}_t$ denote the \emph{publicly disclosed information}, where $\overline{D}_t$ is its disclosed-information space, $d_1=\varnothing$, and $d^t=(d_1,\ldots,d_t)$ is its history in $\overline{D}^t$. The disclosed information is generated by $\mathcal D_t$, a kernel taking values in probability distributions over $\overline{D}_t$. The state- and type-transition kernels $\mathcal Q_{t+1}$ and $\mathcal F_{t+1}^{\tau}$ are primitives. The principal chooses the information technology $\mathcal Y=(\mathcal Y_{t+1})_{t<T}\in Y$ and disclosure rule $\mathcal D=(\mathcal D_{t+1})_{t<T}\in D$, where $Y$ and $D$ are the respective classes of kernel sequences with the domains and timing specified below. Further restrictions are stated where needed. Their composition with the primitive transition kernels induces $\mathcal K_{t+1}$.

Probability kernels use measure notation, e.g., $\mathcal Q_{t+1}(d\theta_{t+1}\mid\cdot)$ for the conditional distribution of the next-period state.

For $t<T$, the underlying state evolves according to
\begin{equation}\label{eq:state-kernel}
\mathcal Q_{t+1}:\Theta^t\times\overline X^t\times\overline{D}^t\longrightarrow\Delta(\Theta_{t+1}),
\qquad
\tilde\theta_{t+1}\sim \mathcal Q_{t+1}(\cdot\mid\theta^t,x^t,d^t).
\end{equation}
Throughout the transition and information-production kernels, the history $x^t$ denotes \emph{realized} allocations, including deviations. Dependence on $x^t$ and $d^t$ accommodates investment effects, learning-by-doing, and previously disclosed information.

Let $y_{t+1}\in\overline{Y}_{t+1}$ denote the \emph{produced information} generated by the information technology and made available to the disclosure rule. The technology jointly determines this information and the next private-signal profile $s_{t+1}\in\Sg_{t+1}$ through the kernel
\begin{equation}\label{eq:raw-signal-kernel}
\mathcal Y_{t+1}:\Theta^{t+1}\times\T^t\times\overline X^t\times\overline{D}^t
\longrightarrow\Delta(\overline{Y}_{t+1}\times\Sg_{t+1}),
\end{equation}
with
$(\tilde y_{t+1},\tilde s_{t+1})\sim \mathcal Y_{t+1}(\cdot\mid\theta^{t+1},\tau^t,x^t,d^t).$
The kernel permits correlation between produced information and private signals and among agents' signals. Produced information and private signals may be degenerate. The signal $s_{i,t+1}$ becomes privately observable to agent $i$ before date-$(t+1)$ communication, while produced information is processed through the disclosure rule.

The committed mechanism transforms produced information into \emph{disclosed information} $d_{t+1}\in\overline{D}_{t+1}$ according to the principal's disclosure rule. Let $m^t$ be the history of messages received by the mechanism, with message spaces specified in Section~3. The \emph{disclosure kernel} is
\begin{equation}\label{eq:disclosure-kernel}
\mathcal D_{t+1}(\cdot\mid \px^t,p^t,d^t,m^t,y_{t+1})\in\Delta(\overline{D}_{t+1}),
\qquad
\tilde d_{t+1}\sim \mathcal D_{t+1}(\cdot\mid \px^t,p^t,d^t,m^t,y_{t+1}).
\end{equation}
The principal commits to $\mathcal D_{t+1}$ at date $0$. The disclosure rule may condition on produced information and the mechanism's observable history, including its recommendations, but not directly on uncommunicated private types, signals, or hidden-action deviations.

The next private-type profile follows
\begin{equation}\label{eq:type-kernel}
\begin{aligned}
\mathcal F_{t+1}^{\tau}&:\Theta^{t+1}\times\T^t\times\Sg^{t+1}\times\overline X^t\times\overline{D}^{t+1}\longrightarrow\Delta(\T_{t+1}),\\
\tilde\tau_{t+1}&\sim \mathcal F_{t+1}^{\tau}(\cdot\mid\theta^{t+1},\tau^t,s^{t+1},x^t,d^{t+1}),\qquad \tilde\tau_{t+1}\in\T_{t+1}.
\end{aligned}
\end{equation}
The kernel $\mathcal F_{t+1}^{\tau}$ specifies the conditional joint distribution of future private types. It allows type evolution to depend on the underlying state, previous types, allocations, produced signals, and disclosed information. The general formulation permits dependence across agents. Restrictions on this dependence are imposed only for results that require them.

Together, $\mathcal Y_{t+1}$ and $\mathcal F_{t+1}^{\tau}$ allow allocations to influence future private information and types. Disclosed information enters subsequent information histories and may also influence future private types without directly altering the current state-transition kernel.

The induced joint-transition kernel composes the four kernels in the following order:
\begin{equation}\label{eq:induced-kernel-composition}
\begin{aligned}
&\mathcal K_{t+1}(d\theta_{t+1},dy_{t+1},ds_{t+1},dd_{t+1},d\tau_{t+1}\mid\theta^t,\tau^t,s^t,x^t,\px^t,p^t,d^t,m^t\,;\,\mathcal Y_{t+1},\mathcal D_{t+1})\\
&\quad=\mathcal Q_{t+1}(d\theta_{t+1}\mid\theta^t,x^t,d^t)\,
\mathcal Y_{t+1}(dy_{t+1},ds_{t+1}\mid\theta^{t+1},\tau^t,x^t,d^t)\\
&\qquad\quad\cdot \mathcal D_{t+1}(dd_{t+1}\mid \px^t,p^t,d^t,m^t,y_{t+1})\,
\mathcal F_{t+1}^{\tau}(d\tau_{t+1}\mid\theta^{t+1},\tau^t,s^{t+1},x^t,d^{t+1}).
\end{aligned}
\end{equation}
Equation~\eqref{eq:induced-kernel-composition} factors the one-step joint law given the current history by the chain rule of probability. State transitions, information production, and type transitions use $x^t$, whereas disclosure uses $\px^t$. The initial law, mechanism, and strategies determine the full path distribution.\footnote{The composed kernel generates a controlled path law satisfying the law of iterated conditional expectations, including under deviations. This identity follows from conditionalization within a single path law; consistency with separately specified next-period continuation beliefs is required in Theorem~\ref{thm:recursive-ic}.}

\textbf{Information production and disclosure} are distinct but interacting margins. Through $\mathcal Y_{t+1}$, realized allocations affect the raw-information experiment, while disclosure determines its Blackwell garbling. Thus, future information and incentives are jointly shaped by institutional design and agents' behavior \citep{BergemannMorris2019}.

\begin{remark}[Special cases and scope]
The framework accommodates four special cases, subject to their respective timing and incentive restrictions: \emph{dynamic screening} (exogenous $\mathcal Y,\mathcal D$ and appropriate continuation sufficiency, but possibly allocation-dependent types); \emph{endogenous acquisition} ($x^t$ affects information production, with restricted $\mathcal D$); \emph{information design} (fixed raw experiment, chosen $\mathcal D$); and \emph{moral hazard/feedback} (hidden $x^t$ changes subsequent information and requires obedience).

These reductions provide benchmarks for identifying how opening institutional margins changes established results; exact equivalence requires the corresponding timing, commitment, and incentive assumptions.

\end{remark}

\section{Mechanisms, Dynamic Incentives, and Revelation}
\label{sec:mechanisms}

This section examines how hidden actions and endogenous information change the conditions for dynamic revelation and recursive incentive compatibility. Throughout the paper, we use the superscript $\mathrm{I}$ to distinguish objects associated with an indirect mechanism from their counterparts in the general environment, while retaining the original notation without this superscript for the corresponding objects in a direct mechanism.

\subsection{Indirect mechanisms and joint perfect Bayesian equilibrium}\label{subsec:jpcbe}
\label{subsec:histories}

An \emph{indirect mechanism} is $\Gamma^I=(M;g,\mathcal Y,\mathcal D^I)$, where $M=(M_t)_{t=1}^T$, $g=(g_t)_{t=1}^T$, and $\mathcal D^I=(\mathcal D^I_{t+1})_{t<T}$ are message spaces, allocation--transfer rules, and disclosure rules, respectively; $\mathcal Y\in Y$ is the information technology, and $\mathcal D^I$ satisfies the specified disclosure restrictions. At date $0$ the principal commits to all rules, including off-path rules. At date $t$, $m_{it}\in M_{it}$, $m_t=(m_{it})_{i\in N}\in M_t$, and $m^t=(m_1,\ldots,m_t)$. 

The \emph{allocation--transfer rules} are stochastic kernels
\begin{equation}\label{eq:indirect-allocation}
g_t(\cdot\mid \px^{t-1},p^{t-1},d^t,m^t)\in\Delta(X_t\times\R^n),\qquad t=1,\ldots,T,
\end{equation}
The kernel draws $(\tilde{\px}_{t},\tilde p_t)$ and announces $(\px_{t},p_t)$, while $\mathcal D^I_{t+1}$ conditions on $(\px^t,p^t,d^t,m^t)$ and produced information as in \eqref{eq:disclosure-kernel}; neither rule observes hidden-action deviations.

Under an indirect mechanism, the within-period interaction follows five stages.

\begin{enumerate}
    \item \textbf{State, types, and private signals.} The current state $\theta_t$ and types $\tau_t$ are realized; the private signals $s_t$ have been generated before date-$t$ messages are submitted. Agent $i$ observes his private information, including $(\tau_i^t,s_i^t)$, previous messages $m_i^{t-1}$, and his own remembered actions, together with the public history $(\px^{t-1},p^{t-1},d^t)$. The principal observes this public history and previously received messages.

    \item \textbf{Messages and mechanism decisions.} Each agent submits $m_{it}\in M_{it}$ to the mechanism, which then draws $(\px_{t},p_t)$ according to \eqref{eq:indirect-allocation} and publicly announces the recommended allocation and transfer vector to all agents.

    \item \textbf{Actions and current payoffs.} After observing $(\px_{t},p_t)$, agents privately choose actions, potentially resulting in $x_t\ne \px_{t}$. Current payoffs follow the period-payoff convention in Section~\ref{subsec:primitives}.

    \item \textbf{Information production and disclosure.} For $t<T$, the realized allocation $x_t$ enters both the state transition \eqref{eq:state-kernel} and the production of produced information and private signals \eqref{eq:raw-signal-kernel}. The disclosure rule $\mathcal D^I_{t+1}$ processes the produced information using observable history, after which next-period types are drawn according to \eqref{eq:type-kernel}.

    \item \textbf{Feedback and history updating.} The disclosed information $d_{t+1}$ becomes available, and agent $i$ privately observes his next type $\tau_{i,t+1}$ and signal $s_{i,t+1}$ before the next message. Histories update accordingly; hidden actions do not become public merely because they affect the primitives.
\end{enumerate}

At $T$, messages, recommendations, transfers, and actions remain, but no information transition follows. For $t<T$, the \emph{transition roadmap} is
\[
\begin{aligned}
&(\theta_t,\tau_t,s_t,\px^{t-1},p^{t-1},d^t)\longrightarrow m_t
\longrightarrow(\px_{t},p_t)\longrightarrow x_t\\
&\qquad\longrightarrow(\theta_{t+1},\tilde y_{t+1},s_{t+1},d_{t+1},\tau_{t+1})
\longrightarrow m_{t+1}\longrightarrow\cdots.
\end{aligned}
\]
Primitive transitions depend on realized $x_t$, while disclosure also conditions on recommended $\px_{t}$ as in \eqref{eq:induced-kernel-composition}. The corresponding decision-stage histories and spaces, including agent $i$'s pre-report and post-report histories $h_{it}^{-}$ and $h_{it}$, are:
{\footnotesize
\[
\begin{aligned}
h_t&=(\px^{t-1},p^{t-1},d^t)\in H_t:=\left(\prod_{k=1}^{t-1}(X_k\times\R^n)\right)\times\prod_{k=2}^{t}\overline{D}_k,\\[3pt]
h_{Pt}&=(h_t,m^t)\in H_{Pt}(M):=H_t\times\prod_{k=1}^{t}M_k,\\[3pt]
h_{it}^{-}&=(h_t,\tau_i^t,s_i^t,x_i^{t-1},m_i^{t-1})\in H_{it}^{-}(M):=H_t\times\prod_{k=1}^{t-1}(\T_{ik}\times\Sg_{ik}\times\overline X_{ik}\times M_{ik})\times\T_{it}\times\Sg_{it},\\[3pt]
h_{it}&=(h_{it}^{-},m_{it},\px_{t},p_t)\in H_{it}(M):=H_{it}^{-}(M)\times M_{it}\times X_t\times\R^n.
\end{aligned}
\]
}
$H_1=\{\varnothing\}$.

A pure strategy profile $(m(\cdot),x(\cdot))$ consists of
\begin{equation}\label{eq:action-strategy}
m_{it}(\cdot):H_{it}^{-}(M)\longrightarrow M_{it},\qquad
x_{it}(\cdot):H_{it}(M)\longrightarrow \overline X_{it},\qquad i\in N,\ t=1,\ldots,T.
\end{equation}
In an implementing equilibrium, the private-choice strategy obeys $x_{it}(h_{it})=\px_{it}$ for every $i\in N$, $t=1,\ldots,T$, and realized post-report history $h_{it}$.

Let $W$ denote the space of \emph{complete paths}:
\[
W
=
\left(\prod_{t=1}^{T}
\bigl(\Theta_t\times\T_t\times\Sg_t\times X_t\times\overline X_t\times\R^n\times\overline{D}_t\bigr)\right)
\times\left(\prod_{t=2}^{T}\overline{Y}_t\right),
\]
with generic coordinates $(\theta^T,\tau^T,s^T,\px^T,x^T,p^T,d^T,y_2^T)$ and the corresponding product $\sigma$-algebra. Both recommended and realized allocations enter $W$; $d_1$ is fixed and message labels are excluded. The initial distribution and ordered kernels under $\Gamma^I$ and $(m(\cdot),x(\cdot))$ generate $\mathbb P_W(\cdot\,;\,\Gamma^I,m(\cdot),x(\cdot))\in\Delta(W)$.

At either decision stage, let $h$ denote agent $i$'s private history in $H_{it}^{-}(M)\sqcup H_{it}(M)$, where $\sqcup$ denotes disjoint union.

A \emph{belief system} is $\nu=(\nu_i)_{i\in N}$, with $\nu_i=(\nu_{it})_{t=1}^{T}$, where $\nu_{it}:H_{it}^{-}(M)\sqcup H_{it}(M)\to\Delta(W)$ assigns beliefs respecting observations and primitive restrictions. Starting from the marginal of $\nu_{it}(\cdot\mid h)$ over past and current economic uncertainty, the mechanism, subsequent strategies, and primitive kernels generate $\mathbb P_{it}(\cdot\mid h;\Gamma^I,m(\cdot),x(\cdot),\nu_i)$.
By \eqref{eq:induced-kernel-composition}, deviations change future distributions, not information already observed. On-path beliefs conditional on private message histories come from the induced complete-history law, marginalized to economic paths in $W$.\footnote{Since $W$ excludes message labels, its marginal distribution alone does not determine conditionals given message histories. Off-path beliefs remain subject to observational and primitive-consistency requirements.}

For any continuation strategy pair $(m_i(\cdot),x_i(\cdot))$ consistent with decisions already taken, the distinct \emph{indirect continuation payoffs} at the pre-report and post-report stages are defined by the same conditional payoff integral:
{\small
\begin{align}
&\begin{cases}
V_{it}^{I-}\bigl((h_t,\tau_i^t,s_i^t,x_i^{t-1},m_i^{t-1}),
(m_i(\cdot),x_i(\cdot))\bigr)&\text{(pre-report)},\\[2pt]
V_{it}^{I}\bigl((h_t,\tau_i^t,s_i^t,x_i^{t-1},m_i^t,\px_{t},p_t),
(m_i(\cdot),x_i(\cdot))\bigr)&\text{(post-report)}
\end{cases}\notag\\
&\quad=\int\sum_{s=t}^{T}\beta^{s-t}[u_{is}(\tilde x_s,\tilde\tau_s,\tilde\theta_s)-\tilde p_{is}]\,
 d\mathbb P_{it}\bigl(\cdot\mid h;\Gamma^I,(m_i(\cdot),m_{-i}(\cdot)),(x_i(\cdot),x_{-i}(\cdot)),\nu_i\bigr),\notag\\
&\qquad h=h_{it}^{-},h_{it}\ \text{in the respective rows}.\label{eq:continuation-plan}
\end{align}
}
For finite $T$, set $V_{i,T+1}^{-}=V_{i,T+1}^{**}=0$.

\begin{definition}[Joint perfect Bayesian equilibrium]
\label{def:multiagent-pbe}
An assessment $(m^*(\cdot),x^*(\cdot),\nu^*)$ is a \emph{joint perfect Bayesian equilibrium} (JPBE) of $\Gamma^I$ if it satisfies the following conditions.
\begin{enumerate}
\item \textbf{Joint sequential rationality.} For every $i,t$ and every continuation deviation $(\hat m_i(\cdot),\hat x_i(\cdot))$ consistent with the decisions already taken at the relevant stage,
\begin{equation}\label{eq:indirect-equilibrium}
\begin{aligned}
V_{it}^{I-}(h_{it}^{-},(m_i^*(\cdot),x_i^*(\cdot)))
&\geq V_{it}^{I-}(h_{it}^{-},(\hat m_i(\cdot),\hat x_i(\cdot)))\quad\text{(pre-report)},\\
V_{it}^{I}(h_{it},(m_i^*(\cdot),x_i^*(\cdot)))
&\geq V_{it}^{I}(h_{it},(\hat m_i(\cdot),\hat x_i(\cdot)))\quad\text{(post-report)}.
\end{aligned}
\makeatletter\protected@write\@auxout{}{\string\newlabel{eq:indirect-post-equilibrium}{{\theequation}{\thepage}{}{}{}}}\makeatother
\end{equation}

    \item \textbf{Belief consistency.}
    \begin{equation}\label{eq:belief-consistency}
    \nu_{it}^*(\cdot\mid h)=\mathbb P_W(\cdot\mid h;\Gamma^I,m^*(\cdot),x^*(\cdot)),\quad
    \text{a.e. }h\in H_{it}^{-}(M)\sqcup H_{it}(M),\ \forall i,t.
    \end{equation}
\end{enumerate}
Off path, $\nu^*$ assigns probability one to histories consistent with observations, primitive domains, and kernel supports.
Let $\operatorname{JPBE}(\Gamma^I)$ denote the set of such assessments.
\end{definition}

% Obedience terminology (formerly a numbered remark).
Recommended continuation actions $\px(\cdot)$ are \emph{obedient} if \eqref{eq:indirect-equilibrium}, evaluated at $x^*(\cdot)=\px(\cdot)$, holds for every $i,t$ and post-report history $h_{it}$, following the recommendation-obedience idea of \citet{BergemannMorris2016}.

\begin{definition}[Implementation in JPBE]\label{def:jpbe-implementation}
An indirect mechanism $\Gamma^I$ \emph{implements} $\px(\cdot)$ in JPBE if there exist a message strategy profile $m^*(\cdot)$ and a belief system $\nu^*$ such that $(m^*(\cdot),\px(\cdot),\nu^*)\in\operatorname{JPBE}(\Gamma^I)$.
\end{definition}

\begin{assumption}[Regularity of the dynamic environment]
\label{ass:regularity}
The following conditions hold for every mechanism, equilibrium assessment, and unilateral continuation deviation under consideration.\footnote{Attainment and uniform-tail requirements are maintained restrictions on the relevant continuation problems; integrability alone does not imply either. Verification depends on the specified action domains and controlled continuation laws.}

\begin{enumerate}
    \item \textbf{Payoff integrability.} For every $i\in N$, $\mathbb E[|U_i|+|U^P|]<\infty$, and the continuation payoffs in \eqref{eq:continuation-plan} are finite.

    \item \textbf{Attainment.} At every $h_{it}$, $V_{it}^{I}$ attains its maximum over $x_{it}(\cdot)$ for each $x_{i,t+1}^{T}(\cdot)$.

    \item \textbf{Continuity at infinity.} If $T=\infty$, for every $i$,
    \[
    \lim_{k\to\infty}
    \sup_{h\in H_{it}^{-}(M)\sqcup H_{it}(M),\,(\hat{m}_i(\cdot),\hat{x}_i(\cdot))}
    \mathbb E_{h}^{\hat{m}_i(\cdot),\hat{x}_i(\cdot)}
    \left[
    \left|
    \sum_{s=t+k}^{\infty}\beta^{s-t}
    \bigl(u_{is}(\tilde x_s,\tilde\tau_s,\tilde\theta_s)-\tilde p_{is}\bigr)
    \right|
    \right]=0.
    \]
    The analogous uniform tail condition holds for the principal.
\end{enumerate}
\end{assumption}

\subsection{Direct mechanisms, JPBIC, and dynamic revelation}\label{subsec:direct-revelation}
\label{subsec:mechanisms-direct}

Let $\vartheta_i^t:=(\tau_i^t,s_i^t,x_i^{t-1})\in\mathcal V_{it}:=\T_i^t\times\Sg_i^t\times\overline X_i^{t-1}$ denote agent $i$'s \emph{private economic history} at date $t$. The direct \emph{pre-report history} is $(h_t,\vartheta_i^t)$; the \emph{post-report history}, after the mechanism announces $(\px_{t},p_t)$ and before private action $x_{it}$, is $(h_t,\vartheta_i^t,\widehat\vartheta_{it},\px_{t},p_t)$. Let $\mathcal V^{T}:=\prod_{i\in N}\mathcal V_{iT}$.

\begin{definition}[Direct mechanism]\label{def:direct-mechanism}
A \emph{history-based direct mechanism} (hereafter, \emph{direct mechanism}) is $\Gamma=(\mathcal V^{T};\allowbreak \mathcal X,\allowbreak \mathcal P,\allowbreak \mathcal Y,\allowbreak \mathcal D)$, with joint allocation--transfer kernels $(\mathcal X_t,\mathcal P_t)(\cdot\mid h_t,\widehat\vartheta^t)\in\Delta(X_t\times\R^n)$, an information technology $\mathcal Y\in Y$, and a committed disclosure kernel $\mathcal D\in D$ satisfying the information and timing restrictions in \eqref{eq:disclosure-kernel}.
\end{definition}

Use $\vartheta$ for truthful reporting: $\widehat\vartheta_{it}=\vartheta_i^t$. Define the \emph{report-conditional and jointly optimized continuation values} by
\begin{equation}\label{eq:report-deviation-direct}
\begin{gathered}
V_{it}^{*}(h_t,\vartheta_i^t,\widehat\vartheta_{i,t}^{T}(\cdot))
:=\sup_{x_{i,t}^{T}(\cdot)}V_{it}^{-}\bigl(h_t,\vartheta_i^t,(\widehat\vartheta_{i,t}^{T}(\cdot),x_{i,t}^{T}(\cdot))\bigr),\\
V_{it}^{**}(h_t,\vartheta_i^t)
:=\sup_{\widehat\vartheta_{i,t}^{T}(\cdot)}V_{it}^{*}(h_t,\vartheta_i^t,\widehat\vartheta_{i,t}^{T}(\cdot))
=\sup_{(\widehat\vartheta_{i,t}^{T}(\cdot),x_{i,t}^{T}(\cdot))}V_{it}^{-}\bigl(h_t,\vartheta_i^t,(\widehat\vartheta_{i,t}^{T}(\cdot),x_{i,t}^{T}(\cdot))\bigr).
\end{gathered}
\makeatletter\protected@write\@auxout{}{\string\newlabel{eq:report-deviation-value}{{\theequation}{\thepage}{}{}{}}}\makeatother
\end{equation}

\begin{definition}[JPBIC and truthful implementability]
\label{def:joint-ic}\label{def:dynamic-implementability}
A direct mechanism $\Gamma$ satisfies \emph{joint perfect Bayesian incentive compatibility} (JPBIC) if $\px(\cdot)$ is \emph{truthfully implementable} in JPBE, i.e., $(\vartheta,\px(\cdot),\nu)\in\operatorname{JPBE}(\Gamma)$ for some belief system $\nu$.\footnote{JPBIC requires only that a truthful-reporting JPBE implement $\px$; other JPBE may not implement $\px$.}
\end{definition}

\label{subsec:revelation}
\begin{assumption}[Conditional probability consistency]\label{ass:controlled-history}
\leavevmode\par\nobreak
\begin{enumerate}
\item \textbf{Cross-mechanism correspondence.}
For every implementing equilibrium $(m^*(\cdot),\allowbreak \px(\cdot),\allowbreak\nu^*)\allowbreak\in\operatorname{JPBE}(\Gamma^I)$, there exists a belief system $\nu$ in a corresponding direct mechanism $\Gamma$ such that
\begin{equation}\label{eq:revelation-probability-consistency}
\begin{aligned}
\mathbb P_{it}\!(A\mid h_t,\vartheta_i^t;\Gamma,\widehat\vartheta_i(\cdot),\widehat x_i(\cdot),\nu_i)
&=\mathbb P_{it}\!(A\mid h_{it}^{-};\Gamma^I,\widehat m_i(\cdot),\widehat x_i(\cdot),\nu_i^*),\\
\mathbb P_{it}\!(A\mid h_t,\vartheta_i^t,\widehat\vartheta_{it},\px_{t},p_t;\Gamma,\widehat\vartheta_i(\cdot),\widehat x_i(\cdot),\nu_i)
&=\mathbb P_{it}\!(A\mid h_{it};\Gamma^I,\widehat m_i(\cdot),\widehat x_i(\cdot),\nu_i^*),\;\forall i,t,A.
\end{aligned}
\end{equation}
\item \textbf{Continuation conditioning.} For all $i,t<T,A$,\allowbreak $(h_t,\vartheta_i^t,\widehat\vartheta_{it},\px_{t},p_t)$ and $(\widehat\vartheta_{i,t}^{T}(\cdot),\allowbreak x_{i,t}^{T}(\cdot))$, the following holds:
{\footnotesize
\begin{equation}\label{eq:recursive-controlled-conditioning}
\begin{aligned}
&\mathbb P_{it}\bigl(A\mid h_t,\vartheta_i^t,\widehat\vartheta_{it},\px_{t},p_t,
(\tilde h_{t+1},\tilde\vartheta_i^{t+1});(\widehat\vartheta_{i,t}^{T}(\cdot),x_{i,t}^{T}(\cdot))\bigr)\\[-2pt]
&\quad=\mathbb P_{i,t+1}\bigl(A\mid (\tilde h_{t+1},\tilde\vartheta_i^{t+1});
(\widehat\vartheta_{i,t+1}^{T}(\cdot),x_{i,t+1}^{T}(\cdot))\bigr),\quad\mathbb P_{it}\text{-a.s.}
\end{aligned}
\end{equation}
}
\end{enumerate}
\end{assumption}
\begin{theorem}[Dynamic Revelation Principle]\label{thm:dynamic-revelation}
Under Assumptions~\ref{ass:regularity} and \ref{ass:controlled-history}.1, an indirect mechanism $\Gamma^I$ implements $\px(\cdot)$ in JPBE if and only if there exists a direct mechanism $\Gamma$ truthfully implementing $\px(\cdot)$ in JPBE.
\footnote{In finite multistage games, \citet{SugayaWolitzky2021} show that the communication revelation principle can fail under sequential equilibrium but holds under conditional probability perfect Bayesian equilibrium. Assumption~\ref{ass:controlled-history}.1 imposes the conditional-law correspondence used here.}
\end{theorem}
\begin{proof}
For the ``only if'', by Definition~\ref{def:jpbe-implementation}, take $(m^*(\cdot),\allowbreak \px(\cdot),\allowbreak\nu^*)\allowbreak\in\operatorname{JPBE}(\Gamma^I)$. For this assessment, recursively define the composite message history by
\[
m_{it}(\widehat\vartheta_i^t):=m_{it}^*\bigl(h_t,\widehat\vartheta_{it},m_i^{t-1}(\widehat\vartheta_i^{t-1})\bigr),
\]
where the public-history arguments are implicit in the recursion. The associated direct mechanism $\Gamma=(\mathcal V^T;\allowbreak \mathcal X,\allowbreak \mathcal P,\allowbreak \mathcal Y,\allowbreak \mathcal D)$ retains the same information technology $\mathcal Y$ and is given by
\begin{equation}\label{eq:direct-composite-rules}
\begin{aligned}
(\mathcal X_t,\mathcal P_t)(\cdot\mid h_t,\widehat\vartheta^t)
&:=g_t\bigl(\cdot\mid h_t,m^t(\widehat\vartheta^t)\bigr),\\
\mathcal D_{t+1}(\cdot\mid \px^t,p^t,d^t,\widehat\vartheta^t,y_{t+1})
&:=\mathcal D_{t+1}^I\bigl(\cdot\mid \px^t,p^t,d^t,m^t(\widehat\vartheta^t),y_{t+1}\bigr).
\end{aligned}
\end{equation}
Substitution defines an indirect reporting deviation $\widehat m_i(\cdot)$, with the same actual actions. Assumption~\ref{ass:controlled-history}.1 and source sequential rationality give
\begin{gather*}
V_{it}^{-}\bigl(h_t,\vartheta_i^t,(\vartheta_{i,t}^{T}(\cdot),\px_{i,t}^{T}(\cdot))\bigr)=V_{it}^{I-}\bigl(h_{it}^{-},(m_i^*(\cdot),\px_{i}(\cdot))\bigr)\\
\ge V_{it}^{I-}\bigl(h_{it}^{-},(\widehat m_i(\cdot),\widehat x_i(\cdot))\bigr)=V_{it}^{-}\bigl(h_t,\vartheta_i^t,(\widehat\vartheta_{i,t}^{T}(\cdot),\widehat x_{i,t}^{T}(\cdot))\bigr),\quad\forall(\widehat\vartheta_{i,t}^{T}(\cdot),\widehat x_{i,t}^{T}(\cdot)).
\end{gather*}
At each post-report history, the same argument gives
\begin{gather*}
V_{it}\bigl(h_t,\vartheta_i^t,\widehat\vartheta_{it},\px_{t},p_t,(\vartheta_{i,t+1}^{T}(\cdot),\px_{i,t}^{T}(\cdot))\bigr)=V_{it}^{I}\bigl(h_{it},(m_i^*(\cdot),\px_{i}(\cdot))\bigr)\\
\ge V_{it}^{I}\bigl(h_{it},(\widehat m_i(\cdot),\widehat x_i(\cdot))\bigr)=V_{it}\bigl(h_t,\vartheta_i^t,\widehat\vartheta_{it},\px_{t},p_t,(\widehat\vartheta_{i,t+1}^{T}(\cdot),\widehat x_{i,t}^{T}(\cdot))\bigr),\quad\forall(\widehat\vartheta_{i,t+1}^{T}(\cdot),\widehat x_{i,t}^{T}(\cdot)).
\end{gather*}
These comparisons establish joint sequential rationality. Assumption~\ref{ass:controlled-history}.1 transfers Bayesian updating and observation consistency, yielding $(\vartheta,\px(\cdot),\nu)\in\operatorname{JPBE}(\Gamma)$, so $\Gamma$ truthfully implements $\px(\cdot)$. The converse follows since every direct mechanism is an indirect mechanism.
\end{proof}

Assumption~\ref{ass:controlled-history}.1 bundles the requirements for cross-mechanism consistency of conditional economic-path distributions. The following proposition separates these requirements into verifiable sufficient conditions on controlled Markov dynamics, conditional beliefs, causal local correspondence, and continuation compatibility. With hidden actions and endogenous information, matching recommended allocations need not preserve realized actions and subsequent information. The controlled Markov dynamics and current-type restrictions of \citet[Section~3.3]{PavanSegalToikka2014} alone are therefore insufficient to guarantee the cross-mechanism consistency required for dynamic revelation.

\begin{proposition}[Sufficient Conditions for Cross-Mechanism Correspondence]\label{prop:cross-mechanism-sufficient}
\leavevmode\par\nobreak
For every $(m^*(\cdot),\px(\cdot),\nu^*)\in\operatorname{JPBE}(\Gamma^I)$, suppose (i)--(iv) hold for the corresponding direct mechanism $\Gamma$, at all paired histories (including off path) and unilateral deviations.
\begin{enumerate}[label=\textup{(\roman*)},leftmargin=*,itemsep=0.15em,topsep=0.15em]
\item \textbf{Controlled Markov dynamics (M).}\footnote{The benchmark of Pavan, Segal, and Toikka (2014) uses allocation-history-dependent Markov type transitions. Agents may condition reports on their full private histories, although the mechanism elicits only current types. Under the corresponding restrictions, its type-transition kernel is a special case of (M). Our framework permits full-history reports, endogenous information production and disclosure, and hidden actions.} For $r<T$, the kernels satisfy
{\small
\begin{equation*}
\begin{aligned}
\mathcal Q_{r+1}(\cdot\mid\theta^r,x^r,d^r)
&=\mathcal Q_{r+1}(\cdot\mid\theta_r,x^r,d^r),\\
\mathcal Y_{r+1}(\cdot\mid\theta^{r+1},\tau^r,x^r,d^r)
&=\mathcal Y_{r+1}(\cdot\mid\theta_r,\theta_{r+1},\tau_r,x^r,d^r),\\
\mathcal F_{r+1}^{\tau}(\cdot\mid\theta^{r+1},\tau^r,s^{r+1},x^r,d^{r+1})
&=\mathcal F_{r+1}^{\tau}(\cdot\mid\theta_r,\theta_{r+1},\tau_r,s_r,s_{r+1},x^r,d^{r+1}).
\end{aligned}
\end{equation*}
}

\item \textbf{Conditional belief correspondence (C1).} There is one direct belief system $\nu$ such that, at every pair of corresponding histories,
\begin{equation*}
\begin{aligned}
\nu_{ir}(B\mid h_r,\vartheta_i^r)
&=\nu^*_{ir}(B\mid h_{ir}^{-}),&&B\in\mathscr B_r^{-},\\
\nu_{ir}(B\mid h_r,\vartheta_i^r,\widehat\vartheta_{ir},\px_{r},p_r)
&=\nu^*_{ir}(B\mid h_{ir}),&&B\in\mathscr B_r,
\end{aligned}
\end{equation*}
where
\[
\begin{aligned}
\mathscr B_r^{-}&:=\sigma(\tilde\theta^r,\tilde\tau^r,\tilde s^r,\tilde{\px}^{r-1},\tilde x^{r-1},\tilde p^{r-1},\tilde d^r,\tilde y_2^r),\\
\mathscr B_r&:=\mathscr B_r^{-}\vee\sigma(\tilde{\px}_{r},\tilde p_r).
\end{aligned}
\]
The beliefs respect observations and primitive supports and satisfy Bayes' rule at reached histories.

\item \textbf{Causal local correspondence (C2).} Every unilateral continuation $(\widehat\vartheta_i(\cdot),\widehat x_i(\cdot))$ has a single causal indirect message continuation $\widehat m_i(\cdot)$, with the other agents following their corresponding equilibrium continuations. For every $r\ge t$ and corresponding histories,
{\small
\begin{equation*}
\begin{aligned}
(\mathcal X_r,\mathcal P_r)(\cdot\mid h_r,\widehat\vartheta^r)
&=g_r(\cdot\mid h_r,\widehat m^r),\\
\mathcal D_{r+1}(\cdot\mid \px^r,p^r,d^r,\widehat\vartheta^r,y_{r+1})
&=\mathcal D^I_{r+1}(\cdot\mid \px^r,p^r,d^r,\widehat m^r,y_{r+1}),\quad r<T.
\end{aligned}
\end{equation*}
}
Moreover, for all $B\in\mathcal B(\overline X_r)$, the conditional laws of the realized allocation agree given corresponding post-report histories and the realized economic prefix:
{\small
\begin{equation*}
\begin{aligned}
&\mathbb P_{ir}(\tilde x_r\in B\mid\mathscr B_r,h_r,\vartheta_i^r,\widehat\vartheta_{ir},\px_{r},p_r;\Gamma,\widehat\vartheta_i(\cdot),\widehat x_i(\cdot),\nu_i)\\
&\quad=\mathbb P_{ir}(\tilde x_r\in B\mid\mathscr B_r,h_{ir};\Gamma^I,\widehat m_i(\cdot),\widehat x_i(\cdot),\nu_i^*).
\end{aligned}
\end{equation*}
}

\item \textbf{Recursive continuation compatibility (C3).} Let $\sim_r^{-}$ and $\sim_r^{A}$ denote the pre- and post-report history correspondences induced by C2, preserving economic coordinates. For $r\ge t$,
{\small
\begin{equation*}
\begin{aligned}
(h_r,\vartheta_i^r)\sim_r^{-} h_{ir}^{-}
\quad\text{implies}\quad
(h_r,\vartheta_i^r,\widehat\vartheta_{ir},\px_{r},p_r)
\sim_r^{A}(h_{ir}^{-},\widehat m_{ir},\px_{r},p_r).
\end{aligned}
\end{equation*}
}
For $r<T$, every supported paired transition, including deviations, satisfies
{\small
\begin{equation*}
\begin{aligned}
&(h_r,\vartheta_i^r,\widehat\vartheta_{ir},\px_{r},p_r)\sim_r^{A}h_{ir}\\
&\quad\text{implies }(h_{r+1},\vartheta_i^{r+1})\sim_{r+1}^{-}
(h_{r+1},\tau_i^{r+1},s_i^{r+1},x_i^r,\widehat m_i^r).
\end{aligned}
\end{equation*}
}
\end{enumerate}
Then Assumption~\ref{ass:controlled-history}.1 holds.
\end{proposition}

\begin{proof}[Proof of Proposition~\ref{prop:cross-mechanism-sufficient}]
Fix $i,t$, paired pre-report histories $(h_t,\vartheta_i^t)\sim_t^-h_{it}^-$, and an arbitrary unilateral continuation $(\widehat\vartheta_i(\cdot),\widehat x_i(\cdot))$. By C2, choose one causal indirect continuation $\widehat m_i(\cdot)$, with other agents following their paired equilibrium continuations. C1 gives $\nu_{it}(\cdot\mid h_t,\vartheta_i^t)|_{\mathscr B_t^-}=\nu_{it}^*(\cdot\mid h_{it}^-)|_{\mathscr B_t^-}$.

For $r\ge t$, suppose that $(h_r,\vartheta_i^r)\sim_r^-h_{ir}^-$ and that the corresponding conditional economic-prefix laws agree on $\mathscr B_r^-$. By C2, $(\mathcal X_r,\mathcal P_r)(\cdot\mid h_r,\widehat\vartheta^r)=g_r(\cdot\mid h_r,\widehat m^r)$; its conditional realized-allocation equality also holds at paired post-report histories, including histories after unilateral deviations. The conditional chain rule therefore equates the joint laws of $(\tilde{\px}_{r},\tilde p_r,\tilde x_r)$ given the common economic prefix. The last equality concerns actual $x_r$, not merely recommended $\px_{r}$.

For $r<T$, the ordered kernel $\mathcal K_{r+1}=\mathcal Q_{r+1}\mathcal Y_{r+1}\mathcal D_{r+1}\mathcal F_{r+1}^{\tau}$ is defined by \eqref{eq:induced-kernel-composition}. The two mechanisms share $\mathcal Q_{r+1}$, $\mathcal Y_{r+1}$, and $\mathcal F_{r+1}^{\tau}$; C2 equates $\mathcal D_{r+1}$ at paired message histories. Thus, conditional on the same realized $(\theta^r,\tau^r,s^r,x^r,\px^r,p^r,d^r)$, the composed one-step laws agree. Combining this with the preceding decision--action correspondence gives equal conditional distributions of $(\tilde{\px}_{r},\tilde p_r,\tilde x_r,\tilde\theta_{r+1},\tilde y_{r+1},\tilde s_{r+1},\tilde d_{r+1},\tilde\tau_{r+1})$. C3 preserves the correspondence at $(h_{r+1},\vartheta_i^{r+1})\sim_{r+1}^-h_{i,r+1}^-$, so induction applies to subsequent deviations as well.

At paired post-report histories, start instead from C1's equal conditional laws on $\mathscr B_t$ and apply the same realized-action and transition correspondence. Both inductions yield equality of finite-dimensional economic-path distributions and hence, because their cylinders generate $\mathcal B(W)$, both equalities in \eqref{eq:revelation-probability-consistency} for all $A\in\mathcal B(W)$, including $T=\infty$. This is Assumption~\ref{ass:controlled-history}.1.
\end{proof}

\subsection{Recursive characterization of JPBIC}
\label{subsec:sufficiency}

The preceding definitions impose joint optimality of reporting and private-action strategies. The following theorem characterizes JPBIC through choice obedience and \emph{history-reporting incentive compatibility}, under JPBE-consistent beliefs.

\begin{theorem}[Recursive Characterization of JPBIC]\label{thm:recursive-ic}
Under Assumptions~\ref{ass:regularity} and \ref{ass:controlled-history}.2, $\Gamma$ is JPBIC if and only if there exists a belief system $\nu$ satisfying the belief-consistency requirements of JPBE such that:

\noindent\textbf{1. Choice obedience.} For all $i,t,(h_t,\vartheta_i^t,\widehat\vartheta_{it},\px_{t},p_t)$,
\begin{equation}\label{eq:recursive-action}
\begin{gathered}
\px_{it}\in\arg\max_{x_i\in\overline X_{it}}
V_{it}\bigl(h_t,\vartheta_i^t,\widehat\vartheta_{it},\px_{t},p_t,
(\vartheta_{i,t+1}^{T}(\cdot),(x_i,\px_{i,t+1}^{T}(\cdot)))\bigr).
\end{gathered}
\end{equation}

\noindent\textbf{2. History-reporting incentive compatibility.}
\begin{equation}\label{eq:recursive-reporting}
\begin{gathered}
V_{it}^{-}\bigl(h_t,\vartheta_i^t,(\vartheta_{i,t}^{T}(\cdot),\px_{i,t}^{T}(\cdot))\bigr)
=V_{it}^{*}\bigl(h_t,\vartheta_i^t,\vartheta_{i,t}^{T}(\cdot)\bigr)\\
=V_{it}^{**}(h_t,\vartheta_i^t)
\ge V_{it}^{*}\bigl(h_t,\vartheta_i^t,\widehat\vartheta_{i,t}^{T}(\cdot)\bigr),\quad\forall i,t,(h_t,\vartheta_i^t),\widehat\vartheta_{i,t}^{T}(\cdot).
\end{gathered}
\end{equation}
\end{theorem}

\begin{proof}
\textbf{Necessity.} Suppose $(\vartheta,\px(\cdot),\nu)\in\operatorname{JPBE}(\Gamma)$. Sequential rationality gives condition 1. The definitions in \eqref{eq:report-deviation-direct} and reporting optimality give
\begin{equation*}
\begin{gathered}
V_{it}^{**}(h_t,\vartheta_i^t)
\ge V_{it}^{*}\bigl(h_t,\vartheta_i^t,\vartheta_{i,t}^{T}(\cdot)\bigr)\\
\ge V_{it}^{-}\bigl(h_t,\vartheta_i^t,(\vartheta_{i,t}^{T}(\cdot),\px_{i,t}^{T}(\cdot))\bigr)
\ge V_{it}^{**}(h_t,\vartheta_i^t),\quad\forall i,t,(h_t,\vartheta_i^t).
\end{gathered}
\end{equation*}
All inequalities are equalities, yielding condition 2.

\textbf{Sufficiency.} Let $\nu$ satisfy JPBE belief consistency and both conditions. For all $i,t,(h_t,\vartheta_i^t)$ and $(\widehat\vartheta_{i,t}^{T}(\cdot),\widehat x_{i,t}^{T}(\cdot))$, condition 2 and the definition of $V_{it}^{**}$ give
\begin{equation*}
\begin{gathered}
V_{it}^{-}\bigl(h_t,\vartheta_i^t,(\vartheta_{i,t}^{T}(\cdot),\px_{i,t}^{T}(\cdot))\bigr)
=V_{it}^{**}(h_t,\vartheta_i^t)\\
\ge V_{it}^{-}\bigl(h_t,\vartheta_i^t,(\widehat\vartheta_{i,t}^{T}(\cdot),\widehat x_{i,t}^{T}(\cdot))\bigr).
\end{gathered}
\end{equation*}
Thus truthful reporting and recommended actions are jointly optimal at every pre-report history. For all $i,t<T,(h_t,\vartheta_i^t,\widehat\vartheta_{it},\px_{t},p_t)$ and $(\widehat\vartheta_{i,t+1}^{T}(\cdot),\widehat x_{i,t}^{T}(\cdot))$, substitute the $t+1$ comparison using Assumption~\ref{ass:controlled-history}.2, then apply condition 1. Suppressing the fixed history arguments,
\begin{equation*}
\begin{gathered}
V_{it}\bigl((\widehat\vartheta_{i,t+1}^{T}(\cdot),\widehat x_{i,t}^{T}(\cdot))\bigr)
\le V_{it}\bigl((\vartheta_{i,t+1}^{T}(\cdot),(\widehat x_{it},\px_{i,t+1}^{T}(\cdot)))\bigr)\\
\le V_{it}\bigl((\vartheta_{i,t+1}^{T}(\cdot),\px_{i,t}^{T}(\cdot))\bigr).
\end{gathered}
\end{equation*}
The first inequality follows from condition 2 and Assumption~\ref{ass:controlled-history}.2; the second follows from condition 1. At $t=T$, condition 1 gives the comparison directly. Together with belief consistency, these inequalities yield $(\vartheta,\px(\cdot),\nu)\in\operatorname{JPBE}(\Gamma)$.
\end{proof}

Assumption~\ref{ass:controlled-history}.2 bundles cross-date consistency requirements for conditional economic-path distributions. The following proposition separates them into verifiable sufficient conditions on controlled Markov dynamics, current-type rules, recorded reports and actual actions, common Bayesian filtering, and causal continuation. Because joint deviations affect subsequent information and private histories, the controlled Markov dynamics and current-type restrictions of \citet[Section~3.3]{PavanSegalToikka2014} alone do not ensure the continuation consistency required for recursive incentive compatibility.

\begin{proposition}[Sufficient Conditions for Continuation Conditioning]\label{prop:continuation-sufficient}
For $\Gamma,\nu$ and every $i,r$, suppose the following hold at $(h_r,\vartheta_i^r)$ and $(h_r,\vartheta_i^r,\widehat\vartheta_{ir},\px_{r},p_r)$ under every $(\widehat\vartheta_{i,r}^{T}(\cdot),x_{i,r}^{T}(\cdot))$:
\begin{enumerate}[label=\textup{(\roman*)},leftmargin=*,itemsep=0.12em,topsep=0.12em]
\item \textbf{Controlled Markov dynamics.} Condition M of Proposition~\ref{prop:cross-mechanism-sufficient} holds.

\item \textbf{Current-type rules and recorded decisions.} For $\widehat\tau^r=\widehat\tau'{}^r$,
{\small
\[
\begin{aligned}
(\mathcal X_r,\mathcal P_r)(\cdot\mid h_r,\widehat\vartheta^r)
&=(\mathcal X_r,\mathcal P_r)(\cdot\mid h_r,\widehat\vartheta'{}^r),\\
\mathcal D_{r+1}(\cdot\mid \px^r,p^r,d^r,\widehat\vartheta^r,y_{r+1})
&=\mathcal D_{r+1}(\cdot\mid \px^r,p^r,d^r,\widehat\vartheta'{}^r,y_{r+1}),\quad r<T.
\end{aligned}
\]
}
At $r<T$, $d_{r+1}$ records $\widehat\tau^r$ through a fixed rule, and $\vartheta_i^{r+1}=(\tau_i^{r+1},s_i^{r+1},x_i^r)$ records actual private actions.

\item \textbf{Common controlled Bayesian filtering.} Fix $(\widehat\vartheta_{-i}(\cdot),x_{-i}(\cdot))$, $\mathbb P_1$, $\Gamma$, and $(\mathcal K_{k+1})_{k<r}$, suppressed in $\mathbb P$. For $B\in\mathscr B_r^{-}$ and $B\in\mathscr B_r$, respectively,
{\footnotesize
\begin{equation*}
\begin{aligned}
\nu_{ir}(B\mid h_r,\vartheta_i^r)&=\mathbb P(B\mid h_r,\vartheta_i^r;\widehat\tau_i^{r-1},x_i^{r-1}),\\
\nu_{ir}(B\mid h_r,\vartheta_i^r,\widehat\vartheta_{ir},\px_{r},p_r)&=\mathbb P(B\mid h_r,\vartheta_i^r,\widehat\vartheta_{ir},\px_{r},p_r;\widehat\tau_i^r,x_i^{r-1}).
\end{aligned}
\end{equation*}
}
Both are regular conditionals of the same controlled joint-kernel law, with $\widehat\tau_i$ and $x_i$ fixed as observed controls; the selected versions define $\nu_{ir}$ at the stated histories.

\item \textbf{Causal continuation.} At every resulting $(h_{r+1},\vartheta_i^{r+1})$,
\[
\left.\bigl(\widehat\vartheta_{i,r}^{T}(\cdot),x_{i,r}^{T}(\cdot)\bigr)\right|_{(h_{r+1},\vartheta_i^{r+1})}
=\bigl(\widehat\vartheta_{i,r+1}^{T}(\cdot),x_{i,r+1}^{T}(\cdot)\bigr),\quad r<T,
\]
with the same $\Gamma$ and other agents' strategies.
\end{enumerate}
Then Assumption~\ref{ass:controlled-history}.2 holds.
\end{proposition}

\begin{proof}
Fix $i,t<T$, $(h_t,\vartheta_i^t,\widehat\vartheta_{it},\px_{t},p_t)$, and an arbitrary continuation $(\widehat\vartheta_{i,t}^{T}(\cdot),x_{i,t}^{T}(\cdot))$. Denote the next pre-report history by $(\tilde h_{t+1},\tilde\vartheta_i^{t+1})$. By (ii) and the definitions of the histories,
\[
\begin{gathered}
\widehat\tau_i^t\text{ is recorded in }\tilde d_{t+1},\qquad
\tilde\vartheta_i^{t+1}=(\tilde\tau_i^{t+1},\tilde s_i^{t+1},\tilde x_i^t),\\
\tilde h_{t+1}=(\tilde{\px}^t,\tilde p^t,\tilde d^{t+1}).
\end{gathered}
\]
Thus the next history retains the current conditioning history, the realized report coordinates that affect mechanism rules, and the actual $x_{it}$. The current report and announcement are already fixed at the initial post-report history; they are not resampled from the continuation strategy.

Before the next pre-report history, the only factor from agent $i$'s strategy is
\[
\mathbf1\!\left\{x_{it}=x_{it}(h_t,\vartheta_i^t,\widehat\vartheta_{it},\px_{t},p_t)\right\}=1
\]
for $\mathbb P_{it}$-almost every realized successor, since $x_{it}$ is contained in $\tilde\vartheta_i^{t+1}$. Thus the factor cancels from the conditional Bayes numerator and denominator. At later dates $r>t$, the pure-strategy factors for $\widehat\tau_{ir}$ and $x_{ir}$ likewise equal $1$ given the corresponding complete private histories and recorded controls. By (ii), the unreported coordinates of $\widehat\vartheta_{ir}$ do not affect $(\mathcal X_r,\mathcal P_r,\mathcal Y_{r+1},\mathcal D_{r+1})$. The remaining joint-law factors are $\mathbb P_1$, $(\mathcal X_r,\mathcal P_r)$, the other agents' fixed strategies, and $\mathcal K_{r+1}$, evaluated at actual $x^r$.

Part (iii) selects the corresponding controlled regular conditional version at each resulting pre-report history. For $B\in\mathscr B_{t+1}^{-}$, this gives
{\footnotesize
\[
\begin{aligned}
&\mathbb P_{it}\bigl(B\mid h_t,\vartheta_i^t,\widehat\vartheta_{it},\px_{t},p_t,
(\tilde h_{t+1},\tilde\vartheta_i^{t+1});
(\widehat\vartheta_{i,t}^{T}(\cdot),x_{i,t}^{T}(\cdot))\bigr)\\
&\qquad=\nu_{i,t+1}(B\mid\tilde h_{t+1},\tilde\vartheta_i^{t+1}),
\qquad\mathbb P_{it}\text{-a.s.}
\end{aligned}
\]
}
The equality holds for all $B$ using a countable determining class for the standard Borel economic prefix and extension to $\mathscr B_{t+1}^{-}$. It is asserted $\mathbb P_{it}$-almost surely under each continuation, not pointwise at every zero-probability successor. At off-path initial histories the common conditional versions are those specified in (iii).

By (iv), the plan following $(\tilde h_{t+1},\tilde\vartheta_i^{t+1})$ is the restriction of the original plan. The two conditional constructions consequently start with the same law on $\mathscr B_{t+1}^{-}$ and use identical subsequent rules $(\mathcal X_r,\mathcal P_r,\mathcal Y_{r+1},\mathcal D_{r+1})$, actions, other-agent strategies and primitive kernels. Part (i) preserves the stated controlled dependence, with the same realized histories under both constructions. Iterating the common one-period kernels yields equality on all finite-dimensional economic cylinders; these generate $\mathcal B(W)$. Hence, for every $A\in\mathcal B(W)$,
{\footnotesize
\[
\begin{aligned}
&\mathbb P_{it}\bigl(A\mid h_t,\vartheta_i^t,\widehat\vartheta_{it},\px_{t},p_t,
(\tilde h_{t+1},\tilde\vartheta_i^{t+1});
(\widehat\vartheta_{i,t}^{T}(\cdot),x_{i,t}^{T}(\cdot))\bigr)\\
&\quad=\mathbb P_{i,t+1}\bigl(A\mid(\tilde h_{t+1},\tilde\vartheta_i^{t+1});
(\widehat\vartheta_{i,t+1}^{T}(\cdot),x_{i,t+1}^{T}(\cdot))\bigr),
\quad\mathbb P_{it}\text{-a.s.}
\end{aligned}
\]
}
This is Assumption~\ref{ass:controlled-history}.2.
\end{proof}

\section{Dynamic Screening with Endogenous Information Production}
\label{sec:incentives}

We now characterize joint incentives under the private-values payoff specification $u_{it}(x_t,\tau_t,\theta_t)=\bar u_{it}(x_t,\tau_{it},\theta_t)$, while allowing correlated private information, endogenous information production, learning, disclosure, and hidden actions. We examine how these additional margins alter dynamic screening and when established characterizations remain valid. The following conditions concern the mechanism's induced continuation problems.

\subsection{Choice Obedience and First-Order Conditions}
\label{subsec:production-value}

To characterize truthful implementability of a direct mechanism $\Gamma$, consider $(\vartheta,\px(\cdot),\nu)$, consisting of truthful reporting, recommended actions, and consistent beliefs. At each post-report history $(h_t,\vartheta_i^t,\widehat\vartheta_{it},\px_{t},p_t)$, fix truthful future reporting $\tau_{i,t+1}^{T}$ and $x_{i,t+1}^{T}(\cdot)=\px_{i,t+1}^{T}(\cdot)$. The post-report choice payoff is
\begin{equation}\label{eq:action-deviation-value}
{\small
\begin{aligned}
&V_{it}(\tau_{it},(\tau_i,(x_i,x_{i,t+1}^{T}(\cdot))))=\int\bigl[u_{it}(\tilde x_t,\tau_{it},\tilde\theta_t)-p_{it}\\
&\qquad+\beta V_{i,t+1}^{-}(\tilde h_{t+1},\tilde\vartheta_i^{t+1},(\tau_{i,t+1}^{T},x_{i,t+1}^{T}(\cdot)))\bigr]\\
&\qquad d\mathbb P_{it}(\cdot\mid h_t,\vartheta_i^t,\widehat\vartheta_{it},\px_{t},p_t;\Gamma,(\tau_{i,t+1}^{T},x_{i,t+1}^{T}(\cdot)),\nu_i,x_i).
\end{aligned}
}
\end{equation}
Choice obedience requires that $\px_{it}$ be a best response $x_{it}^{\mathrm{br}}$, where
\begin{equation}\label{eq:action-best-response}
V_{it}(\tau_{it},(\tau_i,(x_{it}^{\mathrm{br}},x_{i,t+1}^{T}(\cdot))))
=\max_{x_i\in \overline X_{it}}V_{it}(\tau_{it},(\tau_i,(x_i,x_{i,t+1}^{T}(\cdot)))).
\end{equation}
At the pre-report history $(h_t,\vartheta_i^t)$, where $s_{it}$ is observed, the report-deviation value $V_{it}^{*}$ is given by \eqref{eq:report-deviation-direct}, which optimizes the entire post-report action plan following a report, whereas \eqref{eq:action-best-response} optimizes $x_i$ for a fixed continuation plan. Since $x_i$ affects current payoffs and future information, we characterize choice obedience by a directional first-order condition (CO-FOC).

For the following first-order analysis, assume that $\overline X_{it}$ is a nonempty convex subset of a real Banach space. For $x_i,x_i'\in\overline X_{it}$, define the \emph{right directional derivative} $D_{x_i'-x_i}g(x_i)=\lim_{\epsilon\downarrow0}[g(x_i+\epsilon(x_i'-x_i))-g(x_i)]/\epsilon$. Let $\mu_{it}(d\omega\mid h_t,\vartheta_i^t,\widehat\vartheta_{it},\px_{t},p_t)$ be a measure independent of $x_i$, write $\varphi_{it}(\cdot\mid h_t,\vartheta_i^t,\widehat\vartheta_{it},\px_{t},p_t;x_i)=d\mathbb P_{it}(\cdot\mid h_t,\vartheta_i^t,\widehat\vartheta_{it},\px_{t},p_t;x_i)/d\mu_{it}(\cdot\mid h_t,\vartheta_i^t,\widehat\vartheta_{it},\px_{t},p_t)$, with the current conditioning $\sigma((h_t,\vartheta_i^t,\widehat\vartheta_{it},\px_{t},p_t),\tilde\theta_t,\tilde x_{-i,t})$.

\begin{assumption}[Differentiability and choice obedience]
\label{ass:controlled-differentiability}
For each $(h_t,\vartheta_i^t,\widehat\vartheta_{it},\px_{t},p_t)$:
\begin{enumerate}
\item \textbf{Differentiability and conditional probability regularity.} For every $x_i,x_i'\in\overline X_{it}$, $\varphi_{it}(\cdot\mid h_t,\vartheta_i^t,\widehat\vartheta_{it},\px_{t},p_t;x_i)>0$, and $D_{x_i'-x_i}\varphi_{it}$, $D_{x_i'-x_i}u_{it}$, and $D_{x_i'-x_i}V_{i,t+1}^{-}$ exist $\mu_{it}$-a.e., with arguments as defined above. The derivative of $[u_{it}(\tilde x_t,\tau_{it},\tilde\theta_t)-p_{it}+\beta V_{i,t+1}^{-}(\tilde h_{t+1},\tilde\vartheta_i^{t+1},(\tau_{i,t+1}^{T},x_{i,t+1}^{T}(\cdot)))]\varphi_{it}(\omega\mid h_t,\vartheta_i^t,\widehat\vartheta_{it},\px_{t},p_t;x_i)$ admits an integrable dominating bound permitting differentiation under the conditional integral defining $V_{it}$.
For all $x_i,x_i'\in\overline X_{it}$,
\[
\begin{aligned}
&\mathbb P_{it}(d\tilde\theta_t,d\tilde x_{-i,t}\mid h_t,\vartheta_i^t,\widehat\vartheta_{it},\px_{t},p_t;x_i)\\
&\quad=\mathbb P_{it}(d\tilde\theta_t,d\tilde x_{-i,t}\mid h_t,\vartheta_i^t,\widehat\vartheta_{it},\px_{t},p_t;x_i').
\end{aligned}
\]
Moreover,
\[
\mathbb E\!\left[D_{x_i'-x_i}\log\varphi_{it}(\omega\mid h_t,\vartheta_i^t,\widehat\vartheta_{it},\px_{t},p_t;x_i)
\,\middle|\,\sigma((h_t,\vartheta_i^t,\widehat\vartheta_{it},\px_{t},p_t),\tilde\theta_t,\tilde x_{-i,t})\right]=0.
\]
\item $V_{it}(\tau_{it},\allowbreak(\tau_i,\allowbreak(\cdot,\allowbreak x_{i,t+1}^{T}(\cdot))))$ is concave on $\overline X_{it}$ and right directionally differentiable along every segment in $\overline X_{it}$.
\end{enumerate}
\end{assumption}

Differentiability concerns local private choices, not the general allocation spaces of Section~2. At $(h_t,\vartheta_i^t,\widehat\vartheta_{it},\px_{t},p_t)$, define the \emph{directional score} $\ell_{it}^{x_i,x_i'-x_i}(\omega\mid h_t,\vartheta_i^t,\widehat\vartheta_{it},\px_{t},p_t)=D_{x_i'-x_i}\log \varphi_{it}(\omega\mid h_t,\vartheta_i^t,\widehat\vartheta_{it},\px_{t},p_t;x_i)$. The derivative follows the $x_i$-controlled path; the conditional mean-zero score eliminates current-payoff score terms.

\begin{proposition}[Choice-obedience first-order characterization]
\label{prop:choice-obedience}
Under Assumption~\ref{ass:controlled-differentiability}, at every post-report history $(h_t,\vartheta_i^t,\widehat\vartheta_{it},\px_{t},p_t)$:
\begin{enumerate}
\item \textbf{General CO-FOC.}
Choice obedience ($x_{it}=\px_{it}$) holds if and only if
\[
D_{x_i-\px_{it}}V_{it}(\tau_{it},(\tau_i,(\px_{it},x_{i,t+1}^{T}(\cdot))))\leq0,
\qquad \forall x_i\in\overline X_{it}.
\]
If $\px_{it}\in\operatorname{int}\overline X_{it}$ and the directional derivatives
are linear in $x_i-\px_{it}$, the preceding inequality is equivalent to
\begin{equation}\label{eq:production-score-general}
\begin{aligned}
\int\Bigl[&D_{x_i-\px_{it}}u_{it}(\tilde x_t,\tau_{it},\tilde\theta_t)+\beta V_{i,t+1}^{-}(\tilde h_{t+1},\tilde\vartheta_i^{t+1},\allowbreak(\tau_{i,t+1}^{T},x_{i,t+1}^{T}(\cdot)))\ell_{it}^{\px_{it},x_i-\px_{it}}(\omega\mid h_t,\vartheta_i^t,\widehat\vartheta_{it},\px_{t},p_t)\\
&+\beta D_{x_i-\px_{it}}V_{i,t+1}^{-}(\tilde h_{t+1},\tilde\vartheta_i^{t+1},\allowbreak(\tau_{i,t+1}^{T},x_{i,t+1}^{T}(\cdot)))\Bigr]\\
&\qquad d\mathbb P_{it}(\omega\mid h_t,\vartheta_i^t,\widehat\vartheta_{it},\px_{t},p_t\,;\,\Gamma,\allowbreak(\tau_{i,t+1}^{T},x_{i,t+1}^{T}(\cdot)),\nu_i,\px_{it}) =0.
\end{aligned}
\end{equation}
\item \textbf{Reduced CO-FOC.}
If the future payoff $V_{i,t+1}^{-}(\tilde h_{t+1},\tilde\vartheta_i^{t+1},\allowbreak (\tau_{i,t+1}^{T},x_{i,t+1}^{T}(\cdot)))$ is pointwise independent of $x_i$, the interior CO-FOC reduces to
\begin{equation}\label{eq:production-first-order}
\begin{aligned}
-\int D_{x_i-\px_{it}}u_{it}(\tilde x_t,\tau_{it},\tilde\theta_t)\,d\mathbb P_{it}
={}&\beta\int V_{i,t+1}^{-}(\tilde h_{t+1},\tilde\vartheta_i^{t+1},\allowbreak(\tau_{i,t+1}^{T},x_{i,t+1}^{T}(\cdot)))\\
&\qquad\times\ell_{it}^{\px_{it},x_i-\px_{it}}(\omega\mid h_t,\vartheta_i^t,\widehat\vartheta_{it},\px_{t},p_t)\,d\mathbb P_{it},
\end{aligned}
\end{equation}
Both integrals use $d\mathbb P_{it}$ from \eqref{eq:production-score-general}. At a boundary choice, the equality is replaced by $\geq$ for all $x_i\in\overline X_{it}$.
\end{enumerate}
At $t=T$, terminal obedience requires
\[
\px_{iT}\in\arg\max_{x_i\in\overline X_{iT}}
V_{iT}(h_T,\vartheta_i^{T-1},\tau_{iT},s_{iT},\widehat\vartheta_{iT},\px_{T},p_T,\allowbreak
(\tau_i,\allowbreak (x_i,\allowbreak x_{i,T+1}^{T}(\cdot)))).
\]
\end{proposition}

\begin{proof}
\textbf{Part (1).}
By Assumption~\ref{ass:controlled-differentiability}(ii), CO holds if and only if the directional inequality in part (1) holds.
For $x_i\in\overline X_{it}$, differentiating \eqref{eq:action-deviation-value}
under Assumption~\ref{ass:controlled-differentiability}(i) yields
\[
\begin{aligned}
&D_{x_i-\px_{it}}V_{it}(\tau_{it},(\tau_i,(\px_{it},x_{i,t+1}^{T}(\cdot))))
=\int\bigl[D_{x_i-\px_{it}}u_{it}+\beta D_{x_i-\px_{it}}V_{i,t+1}^{-}\\
&\qquad +(u_{it}-p_{it}+\beta V_{i,t+1}^{-})\,
\ell_{it}^{\px_{it},x_i-\px_{it}}(\omega\mid h_t,\vartheta_i^t,\widehat\vartheta_{it},\px_{t},p_t)\bigr]\,d\mathbb P_{it}.
\end{aligned}
\]
By Assumption~\ref{ass:controlled-differentiability}(i),
$\mathbb E[\ell_{it}^{\px_{it},x_i-\px_{it}}(\omega\mid h_t,\vartheta_i^t,\widehat\vartheta_{it},\px_{t},p_t)\mid\sigma((h_t,\vartheta_i^t,\widehat\vartheta_{it},\px_{t},p_t),\tilde\theta_t,\tilde x_{-i,t})]=0$
and, by the payoff specification with fixed $x_i$, $u_{it}-p_{it}$ is $\sigma((h_t,\vartheta_i^t,\widehat\vartheta_{it},\px_{t},p_t),\tilde\theta_t,\tilde x_{-i,t})$-measurable. Hence $\int(u_{it}-p_{it})\ell_{it}^{\px_{it},x_i-\px_{it}}(\omega\mid h_t,\vartheta_i^t,\widehat\vartheta_{it},\px_{t},p_t)\,d\mathbb P_{it}=0$. At an interior choice, linearity in $x_i-\px_{it}$ turns the preceding inequality
into equality, yielding \eqref{eq:production-score-general}.

\textbf{Part (2).}
By part (2), $D_{x_i-\px_{it}}V_{i,t+1}^{-}=0$;
thus \eqref{eq:production-score-general} reduces to \eqref{eq:production-first-order}.
At a boundary choice, substitution in the directional inequality yields the stated inequality.

At $t=T$, $V_{i,T+1}^{-}=0$; conditional choice optimality gives terminal obedience.
\end{proof}

The CO-FOC extends score-based dynamic moral-hazard conditions to information-producing private actions; see \citet{ElyGeorgiadisRayo2025}. Since reports can alter both actions and information, action obedience and reporting incentive compatibility must hold jointly.

\subsection{Dynamic envelope and controlled impulse responses}
\label{subsec:dynamic-envelope}

The dynamic envelope differentiates the report- and action-optimized continuation payoff in $\tau_{it}$. Controlled impulse responses then separate type propagation from allocation, state, and transfer effects.

Fix $i,t$ and $(h_t,(\tau_i^{t-1},s_i^{t-1},x_i^{t-1}),s_{it})$. Let $\widehat\tau_i:=\widehat\vartheta_{i,t}^{T}(\cdot)$ and $x_i(\cdot):=x_{i,t}^{T}(\cdot)$, and denote the fixed-plan payoff by $V_{it}^{-}(\tau,(\widehat\tau_i,x_i(\cdot)))$. Throughout this subsection, optimization is over the same continuation strategy domain, independent of $\tau$, with past decisions fixed.

\begin{assumption}[Regularity for dynamic envelopes]
\label{ass:envelope-regularity}
Conditional on the fixed history:
\begin{enumerate}
\item \textbf{Uniform Lipschitz continuity.}\footnote{These are maintained regularity conditions on continuation payoffs, not consequences of smooth primitive payoffs alone. With unrestricted contingent plans, fixed-plan expected payoffs may be discontinuous even under regular primitives (see \citealp[footnote~23]{PavanSegalToikka2014}). Primitive sufficient conditions require verification for the particular controlled processes and plan class. For example, Assumption 4 holds in finite-horizon environments with finite, type-independent continuation-plan sets and uniformly bounded differentiable primitive payoffs and controlled transition probabilities with uniformly bounded type derivatives.}
\[
\begin{aligned}
&\left|V_{it}^{-}(\tau,(\widehat\tau_i,x_i(\cdot)))
-V_{it}^{-}(\tau',(\widehat\tau_i,x_i(\cdot)))\right|\\
&\quad\le L_{it}|\tau-\tau'|,
\quad L_{it}<\infty,\quad
\forall\tau,\tau'\in\T_{it},\quad
\forall(\widehat\tau_i,x_i(\cdot)).
\end{aligned}
\]
\item \textbf{Almost-everywhere differentiability and attainment.}
For a.e. $\tau\in\operatorname{int}\T_{it}$,
$\partial_\tau V_{it}^{-}(\tau,(\widehat\tau_i,x_i(\cdot)))$
exists for every $(\widehat\tau_i,x_i(\cdot))$, with
\[
\operatorname*{arg\,max}_{(\widehat\tau_i,x_i(\cdot))}
V_{it}^{-}(\tau,(\widehat\tau_i,x_i(\cdot)))
\ne\varnothing,\qquad \forall\tau\in\T_{it}.
\]
\end{enumerate}
\end{assumption}

Section~3's continuation payoff is
\begin{equation}\label{eq:envelope-primitive}
\begin{aligned}
V_{it}^{-}(\tau_{it},(\widehat\tau_i,x_i(\cdot)))
={}&\int\sum_{s=t}^{T}\beta^{s-t}
\left[u_{is}(\tilde x_s,\tilde\tau_{is},\tilde\theta_s)-\tilde p_{is}\right]\\
&\quad d\mathbb P_{it}(\cdot\mid (h_t,(\tau_i^{t-1},s_i^{t-1},x_i^{t-1})),\tau_{it},s_{it}\,;\,\Gamma,\allowbreak(\widehat\tau_i,x_i(\cdot)),x_{-i}(\cdot),\nu_i).
\end{aligned}
\end{equation}

By Section~\ref{subsec:mechanisms-direct},
\begin{equation}\label{eq:envelope-value}
\begin{aligned}
V_{it}^{**}(\tau_{it})
&=\sup_{\widehat\tau_i}V_{it}^{*}(\tau_{it},\widehat\tau_i)
=\sup_{(\widehat\tau_i,x_i(\cdot))}V_{it}^{-}(\tau_{it},(\widehat\tau_i,x_i(\cdot))).
\end{aligned}
\end{equation}
Here $V_{it}^{*}(\tau_{it},\widehat\tau_i)=\sup_{\hat{x}_i(\cdot)}V_{it}^{-}(\tau_{it},(\widehat\tau_i,\hat{x}_i(\cdot)))$. JPBIC requires truthful reporting to attain $V_{it}^{**}$.

\begin{proposition}[Dynamic Envelope under Joint Reporting and Private-Action Optimization]
\label{prop:envelope}
Under Assumptions~\ref{ass:regularity} and \ref{ass:envelope-regularity}, for every $i,t$, $(h_t,(\tau_i^{t-1},s_i^{t-1},x_i^{t-1}),s_{it})$, and $\tau_{it},\tau_{it}'\in\T_{it}$, the following hold:

\begin{enumerate}
    \item \textbf{Reporting-envelope first-order condition (RE-FOC).} For almost every $\tau_{it}\in\T_{it}$ and any $(\widehat\tau_i,x_i(\cdot))\in\operatorname*{arg\,max}_{(\widehat\tau_i',x_i'(\cdot))} V_{it}^{-}(\tau_{it},(\widehat\tau_i',x_i'(\cdot)))$,
    \[
    \frac{dV_{it}^{**}(\tau_{it})}{d\tau_{it}}
    =
    \frac{\partial V_{it}^{-}(\tau_{it},(\widehat\tau_i,x_i(\cdot)))}{\partial\tau_{it}}.
    \]

    \item \textbf{Integrated envelope identity.}
    \begin{equation}\label{eq:envelope-integrated}
    V_{it}^{**}(\tau_{it})-V_{it}^{**}(\tau_{it}')
    =
    \int_{\tau_{it}'}^{\tau_{it}}\frac{dV_{it}^{**}(\tau)}{d\tau}\,d\tau.
    \end{equation}
\end{enumerate}
\end{proposition}

\begin{proof}
Uniform Lipschitz continuity of $V_{it}^{-}(\cdot,(\widehat\tau_i,x_i(\cdot)))$ implies the same bound for $V_{it}^{*}(\cdot,\widehat\tau_i)$ and $|V_{it}^{**}(\tau)-V_{it}^{**}(\tau')|\leq L_{it}|\tau-\tau'|$ after taking suprema. Hence $V_{it}^{**}$ is absolutely continuous and differentiable a.e.

At an interior point $\tau$ where $V_{it}^{**}$ and all fixed-strategy payoffs are differentiable, take any maximizing pair $(\widehat\tau_i,x_i(\cdot))$. Since $V_{it}^{-}(\tau',(\widehat\tau_i,x_i(\cdot)))\leq V_{it}^{**}(\tau')$ for all $\tau'$, with equality at $\tau$, the difference has an interior minimum, yielding $dV_{it}^{**}(\tau)/d\tau=\partial_{\tau}V_{it}^{-}(\tau,(\widehat\tau_i,x_i(\cdot)))$. Thus (1) holds a.e.; integrating by absolute continuity gives \eqref{eq:envelope-integrated}.
\end{proof}

The envelope requires neither cross-agent independence nor differentiation of the optimizer.

For the controlled-type specialization, initial types and exogenous type shocks are independent across agents, and type propagation is own-agent only; private signals and disclosure may remain correlated. For $(\widehat\tau_i,x_i(\cdot))$, use i.i.d. $\mathrm{Unif}[0,1]$ primitive shocks.

Assume the controlled type process admits the common-shock representation
\begin{equation}\label{eq:shock-representation}
\tilde\tau_{is}=Z_{it}^{is}(\tau_{it},\tilde\varepsilon_t^s),
\qquad s\geq t,\qquad
Z_{it}^{it}(\tau_{it},\tilde\varepsilon_t)=\tau_{it}.
\end{equation}
The dependence on $(h_t,(\tau_i^{t-1},s_i^{t-1},x_i^{t-1}),s_{it},\Gamma,(\widehat\tau_i,x_i(\cdot)))$ is suppressed; there is no cross-agent type propagation.

\begin{assumption}[Regularity for controlled impulse responses]
\label{ass:controlled-impulse-regularity}
For every $i,t$, $(h_t,(\tau_i^{t-1},s_i^{t-1},x_i^{t-1}),s_{it})$, $(\widehat\tau_i,x_i(\cdot))$, and compared $\tau,\tau'\in\T_{it}$, the following hold locally where derivatives are taken:
\begin{enumerate}
\item \textbf{Common-shock consistency.} For every $s\ge t$,
\[
\tilde\varepsilon_t^s\perp\!\!\!\perp\tau_{it}\mid
\bigl((h_t,(\tau_i^{t-1},s_i^{t-1},x_i^{t-1})),s_{it}\bigr).
\]
The conditional joint law of $\tilde\varepsilon_t^s$ is identical for $\tau,\tau'$. Under $(\widehat\tau_i,x_i(\cdot))$ and common primitive-shock realizations, the construction reproduces the complete continuation law for each compared type induced by $\Gamma$, $x_{-i}(\cdot)$, and $(\mathcal Q,\mathcal Y,\mathcal D,\mathcal F^\tau)$.
\item \textbf{Pathwise differentiability.} Almost surely, for every $s\ge t$, $\partial_{\tau_{it}}Z_{it}^{is}$, $\dot\theta_s =\partial_{\tau_{it}}\tilde\theta_s$, $\dot x_s=\partial_{\tau_{it}}\tilde x_s,$ and $\dot p_{is}=\partial_{\tau_{it}}\tilde p_{is}$
exist. The marginal-payoff derivatives in \eqref{eq:controlled-marginal-utility} admit the requisite continuous linear pairings, including on infinite-dimensional local domains, and satisfy the pathwise chain rule.
\item \textbf{Integrability and domination.} For every $s\ge t$, $\mathbb E\!\left[\left|\partial_{\tau_{it}}Z_{it}^{is}\right|\mid h_t,\vartheta_i^t\right]<\infty$.

The products of own-type marginal-payoff coefficients and type-response derivatives are integrable. Integrable domination permits differentiation under conditional expectations and discounted sums, including the full series for $T=\infty$.
\end{enumerate}
\end{assumption}

The common-shock consistency in Assumption~\ref{ass:controlled-impulse-regularity}.1 requires invariance of the conditional joint primitive-shock law across compared types; unconditional independence alone is insufficient.\footnote{The common-shock coupling, controlled derivatives, and integrable bounds are restrictions on the induced path laws. They are not guaranteed by measurability of the primitive kernels; verification in a specified model must account for the continuation policies held fixed during type variation.}

Define the \emph{controlled impulse response} by
\begin{equation}\label{eq:controlled-impulse}
I_{it}^{is}
=
\mathbb E\left[
\left.
\frac{\partial Z_{it}^{is}(\tau_{it},\tilde\varepsilon_t^s)}{\partial\tau_{it}}
\right|\sigma\left((h_t,(\tau_i^{t-1},s_i^{t-1},x_i^{t-1})),s_{it},\tau_{it},\frac{\partial u_{is}}{\partial\tau_{is}}\right)
\right],
\qquad I_{it}^{it}=1,
\end{equation}
where the conditioning information includes the fixed initial information and the date-$s$ marginal payoff coefficient. Holding $(\widehat\tau_i,x_i(\cdot))$ and shocks fixed, $\tau_{i,s+1}=\varrho_i\tau_{is}+\varepsilon_{i,s+1}$ with type-independent shocks gives $I_{it}^{is}=\varrho_i^{s-t}$. The controlled-type restriction leaves state, allocation, transfer, and continuation responses unrestricted; $\dot\theta_s,\dot x_s,\dot p_{is}$ are the directional path derivatives in Assumption~\ref{ass:controlled-impulse-regularity}.

\begin{proposition}[Dynamic Marginal Utility under Controlled Type Evolution]
\label{prop:controlled-marginal-utility}
Under Assumptions~\ref{ass:regularity}, \ref{ass:envelope-regularity}, and \ref{ass:controlled-impulse-regularity}, and the preceding controlled type-process restrictions, for a.e. $\tau_{it}\in\T_{it}$,
\begin{equation}\label{eq:controlled-marginal-utility}
\begin{aligned}
\frac{dV_{it}^{**}(\tau_{it})}{d\tau_{it}}
=&
\int\sum_{s=t}^{T}\beta^{s-t}\Biggl[
\underbrace{\frac{\partial u_{is}}{\partial\tau_{is}}I_{it}^{is}}_{\text{Own-type propagation}}
+\underbrace{\frac{\partial u_{is}}{\partial x_s}\cdot\dot x_s
+\frac{\partial u_{is}}{\partial\theta_s}\cdot\dot\theta_s}_{\text{Allocation and state effects}}
-\underbrace{\dot p_{is}}_{\text{Transfer effect}}
\Biggr]\\
&\qquad d\mathbb P_{it}(\cdot\mid (h_t,(\tau_i^{t-1},s_i^{t-1},x_i^{t-1})),\tau_{it},s_{it}\,;\,\Gamma,(\widehat\tau_i,x_i(\cdot)),x_{-i}(\cdot),\nu_i).
\end{aligned}
\end{equation}
All derivatives and impulse responses are evaluated under an optimizing contingent plan $(\widehat\tau_i,x_i(\cdot))$.
\end{proposition}

\begin{proof}
Fix an optimizing contingent plan $(\widehat\tau_i,x_i(\cdot))$. Under Assumption~\ref{ass:controlled-impulse-regularity}, private values and the chain rule give
\[
\frac{\partial}{\partial\tau_{it}}
\left[u_{is}(x_s,\tau_{is},\theta_s)-p_{is}\right]
=
\frac{\partial u_{is}}{\partial\tau_{is}}
\frac{\partial\tau_{is}}{\partial\tau_{it}}
+\frac{\partial u_{is}}{\partial x_s}\cdot\dot x_s
+\frac{\partial u_{is}}{\partial\theta_s}\cdot\dot\theta_s
-\dot p_{is}.
\]

By Assumption~\ref{ass:controlled-impulse-regularity}, the common primitive-shock representation preserves the continuation law for each compared type. Pathwise differentiability and integrable domination justify differentiation under the conditional integral and discounted sum. By \eqref{eq:controlled-impulse} and conditional iteration,
\[
\int\frac{\partial u_{is}}{\partial\tau_{is}}
\frac{\partial Z_{it}^{is}}{\partial\tau_{it}}\,d\mathbb P_{it}
=
\int\frac{\partial u_{is}}{\partial\tau_{is}}I_{it}^{is}\,d\mathbb P_{it},
\]
where both integrals use the distribution in \eqref{eq:controlled-marginal-utility}. Applying Proposition~\ref{prop:envelope} to the resulting derivative of \eqref{eq:envelope-primitive} establishes \eqref{eq:controlled-marginal-utility}.
\end{proof}

The marginal-utility formula separates own-type propagation from allocation, state, and transfer responses; production is included in the allocation response.

A simpler \emph{PST-type marginal-utility representation} obtains when the aggregate expected discounted non-type marginal effects vanish:
\[
\begin{aligned}
&\int\sum_{s=t}^{T}\beta^{s-t}
\left(
\frac{\partial u_{is}}{\partial x_s}\cdot\dot x_s
+\frac{\partial u_{is}}{\partial\theta_s}\cdot\dot\theta_s
-\dot p_{is}
\right)\\
&\qquad\cdot d\mathbb P_{it}(\cdot\mid (h_t,(\tau_i^{t-1},s_i^{t-1},x_i^{t-1})),\tau_{it},s_{it}\,;\,\Gamma,(\widehat\tau_i,x_i(\cdot)),x_{-i}(\cdot),\nu_i)=0.
\end{aligned}
\]
Then
\begin{equation}\label{eq:pst-marginal-utility}
\begin{aligned}
\frac{dV_{it}^{**}(\tau_{it})}{d\tau_{it}}
&=\int\sum_{s=t}^{T}\beta^{s-t}
\frac{\partial u_{is}(x_s,\tau_{is},\theta_s)}{\partial\tau_{is}}I_{it}^{is}\\
&\quad\cdot d\mathbb P_{it}(\cdot\mid (h_t,(\tau_i^{t-1},s_i^{t-1},x_i^{t-1})),\tau_{it},s_{it}\,;\,\Gamma,(\widehat\tau_i,x_i(\cdot)),x_{-i}(\cdot),\nu_i).
\end{aligned}
\end{equation}

Unlike in the conventional dynamic screening model, private values alone do not eliminate non-type marginal effects arising from allocation, state, and transfer responses. Only their aggregate expected effect must vanish, not individual or pathwise effects. When this aggregate restriction holds, mimicking preserves economic decisions, and the controlled type process satisfies the PST-type restrictions, \eqref{eq:pst-marginal-utility} recovers the PST envelope. Proposition~\ref{prop:controlled-marginal-utility} allows private-action effects on signals, disclosure, states, continuation choices, and types.

\subsection{Characterization of JPBIC: First-Order Conditions and Integral Monotonicity}
\label{subsec:implementation}

For $i,t$ and $(h_t,(\tau_i^{t-1},s_i^{t-1},x_i^{t-1}),s_{it})$, consider the recommended continuation $\px_{i,t}^{T}(\cdot)$ and beliefs satisfying JPBE belief consistency. Let $V_{it}^{*}(\tau,\tau')$ be the action-optimized payoff of type $\tau$ reporting $\tau'$, and $V_{it}(\tau,(x_i,\px_{i,t+1}^{T}(\cdot)))$ with truthful future reporting suppressed. Differentiation in $\tau$ holds $\tau'$ and the other parameters unchanged.

In the Markov setting of \citet{PavanSegalToikka2014}, the current private state summarizes the payoff- and incentive-relevant effects of past private information. With endogenous information and hidden actions, current-type incentives need not control full-history reporting--action deviations. We thus impose the following conditional-law sufficiency condition.
\begin{assumption}[Conditional Reporting Sufficiency]\label{ass:reporting-sufficiency}
For every $i,t,(h_t,\vartheta_i^t)$, including off-path histories, and every continuation strategy pair $(\widehat\vartheta_{i,t}^{T}(\cdot),x_{i,t}^{T}(\cdot))$, there exists a \emph{single} continuation strategy pair $(\widehat\vartheta_{i,t}^{\prime T}(\cdot),x_{i,t}^{\prime T}(\cdot))$ satisfying:

\noindent\textbf{(i) Current-type reporting.} At every continuation history $(h_r,\vartheta_i^r)$ and every $r=t,\ldots,T$,
\begin{equation}\label{eq:conditional-reporting-current-type}
\widehat\vartheta_{ir}'(h_r,\vartheta_i^r)
=\bigl(\tau_i^{r-1},\widehat\tau_{ir}'(h_r,\vartheta_i^r),s_i^r,x_i^{r-1}\bigr).
\end{equation}

\noindent\textbf{(ii) Conditional-law equivalence.} For every $A\in\mathcal B(W)$,
\begin{equation}\label{eq:conditional-reporting-sufficiency}
\begin{aligned}
&\mathbb P_{it}\bigl(A\mid h_t,\vartheta_i^t;\Gamma,\widehat\vartheta_{i,t}^{T}(\cdot),x_{i,t}^{T}(\cdot),\nu_i\bigr)\\
&\quad=\mathbb P_{it}\bigl(A\mid h_t,\vartheta_i^t;\Gamma,\widehat\vartheta_{i,t}^{\prime T}(\cdot),x_{i,t}^{\prime T}(\cdot),\nu_i\bigr).
\end{aligned}
\end{equation}
\end{assumption}

\begin{theorem}[First-Order and Integral-Monotonicity Characterization of JPBIC]
\label{thm:integral-monotonicity}
Suppose Assumptions~\ref{ass:regularity}, \ref{ass:controlled-history}.2, \ref{ass:envelope-regularity}, and \ref{ass:reporting-sufficiency} hold. Then a direct mechanism $\Gamma$ with JPBE-consistent beliefs $\nu$ is JPBIC if and only if the following three conditions hold:
\begin{enumerate}
\item \textbf{Choice obedience (CO).} At each post-report history $(h_t,\vartheta_i^t,\widehat\vartheta_{it},\px_{t},p_t)$,
\begin{equation}\label{eq:production-ic-characterization}
\px_{it}\in\arg\max_{x_i\in\overline X_{it}}V_{it}(\tau_{it},(x_i,\px_{i,t+1}^{T}(\cdot))).
\end{equation}
Under Assumption~\ref{ass:controlled-differentiability}, CO is equivalent to the CO-FOC of Proposition~\ref{prop:choice-obedience} at $\px_{it}$ for all $x_i\in\overline X_{it}$. At interior points with linear directional derivatives, equality holds; Proposition~\ref{prop:choice-obedience} also gives the score representation under its additional conditions.
\item \textbf{Reporting-envelope condition (RE-FOC).} $V_{it}^{*}(\tau,\tau)$ is absolutely continuous, with
\begin{equation}\label{eq:pbic-diagonal-envelope}
\frac{d}{d\tau}V_{it}^{*}(\tau,\tau)=\partial_1V_{it}^{*}(\tau,\tau),\qquad\text{a.e. }\tau\in\T_{it}.
\end{equation}
\item \textbf{Global integral monotonicity (GIM).}
\begin{equation}\label{eq:global-integral-ic}
\int_{\tau'}^{\tau}\bigl[\partial_1V_{it}^{*}(z,z)-\partial_1V_{it}^{*}(z,\tau')\bigr]dz\geq0,\qquad \forall\tau,\tau'\in\T_{it}.
\end{equation}
\end{enumerate}
At $t=T$, CO reduces to terminal action optimality.
\end{theorem}

\begin{proof}
Fix $i,t$ and the non-type private-history coordinates, including off path, and write $U(\tau)=V_{it}^{*}(\tau,\tau)$. Under RE-FOC,
\begin{equation}\label{eq:gim-payoff-identity}
\begin{aligned}
U(\tau)-V_{it}^{*}(\tau,\tau')
&=U(\tau')-V_{it}^{*}(\tau',\tau')
  +\int_{\tau'}^{\tau}[U'(z)-\partial_1V_{it}^{*}(z,\tau')]\,dz\\
&=\int_{\tau'}^{\tau}[\partial_1V_{it}^{*}(z,z)-\partial_1V_{it}^{*}(z,\tau')]\,dz.
\end{aligned}
\end{equation}
Thus, under RE-FOC, GIM is equivalent to $V_{it}^{*}(\tau,\tau)\ge V_{it}^{*}(\tau,\tau'),  \forall \tau,\tau'\in\T_{it}$.
\emph{Necessity.} Theorem~\ref{thm:recursive-ic} gives CO and current-type reporting IC. By \eqref{eq:recursive-reporting}, $U(\tau)=V_{it}^{**}(\tau)$. Proposition~\ref{prop:envelope} gives RE-FOC by sandwiching the fixed-report value between the truthful fixed-strategy payoff and $V_{it}^{**}$, with equality at $\tau$. Then \eqref{eq:gim-payoff-identity} gives GIM.

\emph{Sufficiency.} RE-FOC and GIM imply current-type reporting IC. For finite $T$, backward induction establishes optimality against the continuation reporting--action deviations satisfying Assumption~\ref{ass:reporting-sufficiency}(i), including at off-path histories. At $t=T$, current-type reporting IC and CO give joint optimality. Suppose the conclusion holds at every date-$(t+1)$ private history, including histories reached after deviations. Fix a current-type report $\tau'$, a current private action $x_i$, and any subsequent restricted reporting--action plan. Assumption~\ref{ass:controlled-history}.2 and the induction hypothesis permit replacing that future plan with truthful history reporting and recommended actions without reducing continuation payoff. Conditional on the fixed current report and history, CO then gives
\[
\begin{aligned}
&V_{it}\bigl(\tau,(\widehat\tau_{i,t+1}^{T}(\cdot),(x_i,x_{i,t+1}^{T}(\cdot)))\bigr)\\
&\quad\le V_{it}\bigl(\tau,(\tau_{i,t+1}^{T}(\cdot),(x_i,\px_{i,t+1}^{T}(\cdot)))\bigr)
\le V_{it}\bigl(\tau,(\px_{it},\px_{i,t+1}^{T}(\cdot))\bigr).
\end{aligned}
\]
The first inequality uses continuation optimality after the chosen action; the second is CO at the current post-report history. Consequently, for every restricted continuation deviation with current report $\tau'$,
\[
\begin{aligned}
&V_{it}^{-}\bigl(\tau,((\tau',\widehat\tau_{i,t+1}^{T}(\cdot)),x_{i,t}^{T}(\cdot))\bigr)
\le V_{it}^{*}(\tau,\tau')\\
&\qquad\le V_{it}^{*}(\tau,\tau)
=V_{it}^{-}\bigl(\tau,(\vartheta_{i,t}^{T}(\cdot),\px_{i,t}^{T}(\cdot))\bigr).
\end{aligned}
\]
The first inequality follows from the continuation comparison and action optimization, the second from current-type IC, and the equality from continuation optimality and CO. This completes the induction. For $T=\infty$, truncate deviations by restoring truthful and obedient continuation after finitely many periods. Backward induction and Assumption~\ref{ass:regularity}(3) then yield the bound for arbitrary continuation deviations.

Now take any full-history reporting--action continuation deviation $(\widehat\vartheta_{i,t}^{T}(\cdot),x_{i,t}^{T}(\cdot))$. Assumption~\ref{ass:reporting-sufficiency}(i) yields one counterpart $(\widehat\vartheta_{i,t}^{\prime T}(\cdot),x_{i,t}^{\prime T}(\cdot))$ that satisfies the restricted reporting rule at every continuation history. Part~(ii) and the continuation-payoff definition imply, suppressing the fixed history arguments,
\[
\begin{aligned}
&V_{it}^{-}\bigl((\widehat\vartheta_{i,t}^{T}(\cdot),x_{i,t}^{T}(\cdot))\bigr)
=V_{it}^{-}\bigl((\widehat\vartheta_{i,t}^{\prime T}(\cdot),x_{i,t}^{\prime T}(\cdot))\bigr)\\
&\qquad\le V_{it}^{-}\bigl((\vartheta_{i,t}^{T}(\cdot),\px_{i,t}^{T}(\cdot))\bigr).
\end{aligned}
\]
Because the deviation was arbitrary, truthful reporting and recommended actions attain the full-history continuation optimum in \eqref{eq:recursive-reporting}. CO gives \eqref{eq:recursive-action}; Theorem~\ref{thm:recursive-ic} and belief consistency yield JPBIC.
\end{proof}

\begin{remark}[First-order incentives and dynamic implementability]
Unlike in the conventional dynamic screening model of
\citet{PavanSegalToikka2014}, our characterization combines reporting
incentives with choice obedience and accounts for private-action
reoptimization and its informational consequences. Conditional
reporting sufficiency (Assumption~\ref{ass:reporting-sufficiency})
requires a single restricted counterpart preserving each
full-history deviation's conditional economic-path law, rather
than merely a Markov restriction, and does not follow from fixing
the additional information and action margins.
Corollary~\ref{cor:dynamic-monotonicity} extends four
sufficient criteria, \emph{single crossing}, \emph{average monotonicity},
\emph{ex-post monotonicity}, and \emph{strong monotonicity}, to
deviation-adjusted values. These criteria ensure GIM, not CO.
Section~\ref{sec:optimal-screening} builds on this characterization
to study optimal contracting.
\begin{corollary}[Sufficient conditions for global integral monotonicity]
\label{cor:dynamic-monotonicity}
Suppose CO (equivalently CO-FOC under Assumption~\ref{ass:controlled-differentiability}) and RE-FOC in Theorem~\ref{thm:integral-monotonicity} hold. Under Assumptions~\ref{ass:regularity} and \ref{ass:envelope-regularity}, each of the following conditions is sufficient for GIM:
\begin{enumerate}
\item \textbf{Single crossing.}
\begin{equation}\label{eq:dynamic-single-crossing}
[\partial_1 V_{it}^{*}(z,z)-\partial_1 V_{it}^{*}(z,\tau')](z-\tau')\geq0,
\quad \text{a.e. }z\in\T_{it},\ \forall\tau'\in\T_{it}.
\end{equation}
\item \textbf{Average monotonicity.}
\begin{equation}\label{eq:dynamic-average-monotonicity}
\partial_1 V_{it}^{*}(z,\tau'_1)\leq\partial_1 V_{it}^{*}(z,\tau'_2),
\quad \text{a.e. }z\in\T_{it},\ \forall\tau'_1<\tau'_2\text{ in }\T_{it}.
\end{equation}
\item \textbf{Ex-post monotonicity.} For a.e. $z\in\T_{it}$ and every $\tau'\in\T_{it}$, suppose $V_{it}^{*}(z,\tau')$ is attained and the induced paths admit a common primitive-shock coupling as in Assumption~\ref{ass:controlled-impulse-regularity}, with conditional shock law independent of $\tau'$ and locally of $z$. Holding the selected action optimizer and primitive shocks fixed, define
\[
g_{is}(z,\tau'):=\left.\frac{\partial}{\partial a}[u_{is}(\tilde x_s(a),\tilde\tau_{is}(a),\tilde\theta_s(a))-\tilde p_{is}(a)]\right|_{a=z},
\]
suppressing the shock argument. Under pathwise differentiability, integrable domination for differentiation under expectation and discounted summation, and the fixed-report action envelope,
\begin{equation}\label{eq:dynamic-expost-representation}
\partial_1 V_{it}^{*}(z,\tau')=\mathbb E\left[\left.\sum_{s=t}^{T}\beta^{s-t}g_{is}(z,\tau')\right|(h_t,(\tau_i^{t-1},s_i^{t-1},x_i^{t-1})),s_{it},z\right].
\end{equation}
On a common full-measure set of types and shocks,
\[
\sum_{s=t}^{T}\beta^{s-t}g_{is}(z,\tau'_1)\leq\sum_{s=t}^{T}\beta^{s-t}g_{is}(z,\tau'_2),\quad \forall\tau'_1<\tau'_2\text{ in }\T_{it}.
\]
\item \textbf{Strong monotonicity.} Under the regularity and coupling conditions of (3), on a common full-measure set of types and shocks,
\[
g_{is}(z,\tau'_1)\leq g_{is}(z,\tau'_2),\quad \forall s\geq t,\quad \forall\tau'_1<\tau'_2\text{ in }\T_{it}.
\]
\end{enumerate}
These conditions are sufficient, not necessary, for GIM; unrestricted full-history JPBIC remains subject to Theorem~\ref{thm:recursive-ic}.
\end{corollary}

\begin{proof}
Fix $i,t,(h_t,(\tau_i^{t-1},s_i^{t-1},x_i^{t-1}),s_{it})$ and $\tau_{it},\tau'\in\T_{it}$. Under (1), the integrand in \eqref{eq:global-integral-ic} is nonnegative for $z>\tau'$ and nonpositive for $z<\tau'$, implying GIM with the appropriate integration limits. Condition (2) implies (1) by comparing $z$ and $\tau'$. Under (3), taking conditional expectations under the common shock law and using \eqref{eq:dynamic-expost-representation} yields (2). Under (4), $\beta^{s-t}>0$ and dominated summation yield (3). These conditions do not replace \eqref{eq:recursive-reporting}.
\end{proof} \begin{remark}[Monotonicity and endogenous information production]
These conditions link dynamic implementability (\citealp{PavanSegalToikka2014}; \citealp{Rochet1987}) to monotone comparative statics (\citealp{MilgromShannon1994}). With endogenous information, false reports affect actions, signals, disclosure, and continuation payoffs; monotonicity therefore concerns deviation-adjusted marginal utility.
\end{remark} For $\tau_{it}\ne\tau'$, define the
\emph{signed average marginal difference}
\[
\mathcal M_{it}(\tau_{it},\tau'\mid (h_t,(\tau_i^{t-1},s_i^{t-1},x_i^{t-1})),s_{it})
=
\frac{1}{\tau_{it}-\tau'}
\int_{\tau'}^{\tau_{it}}
\left[
\partial_1 V_{it}^{*}(\tau,\tau)
-
\partial_1 V_{it}^{*}(\tau,\tau')
\right]d\tau.
\]
GIM is equivalent to
\[
(\tau_{it}-\tau')
\mathcal M_{it}(\tau_{it},\tau'\mid (h_t,(\tau_i^{t-1},s_i^{t-1},x_i^{t-1})),s_{it})
\geq0.
\] Under CO, RE-FOC, and regularity, GIM characterizes global current-type IC within the specified deviation family; full-history IC additionally requires \eqref{eq:recursive-reporting}. Corollary~\ref{cor:dynamic-monotonicity} gives sufficient, not necessary, conditions: GIM may hold despite sign changes in the integrand. GIM concerns implementability; distributional regularity (e.g., monotone hazard rates) governs rent weights in optimal screening. Section~\ref{sec:optimal-screening} applies these incentive characterizations to optimal contracting.
\end{remark}

The following proposition gives sufficient conditions for Assumption~\ref{ass:reporting-sufficiency}.

\begin{proposition}[Sufficient Conditions for Conditional Reporting Sufficiency]\label{prop:reporting-sufficient}
For every $i,t,(h_t,\vartheta_i^t)$ and $(\widehat\vartheta_{i,t}^{T}(\cdot),x_{i,t}^{T}(\cdot))$, suppose a single causal counterpart $(\widehat\vartheta_{i,t}^{\prime T}(\cdot),x_{i,t}^{\prime T}(\cdot))$ satisfies R1--R3 at all paired histories, including off path and after deviations:
\begin{enumerate}[label=\textup{(\roman*)},leftmargin=*,itemsep=0.12em,topsep=0.12em]
\item \textbf{Current-type reporting (R1).} At each $(h_r,\vartheta_i^r)$,
\[
\widehat\vartheta_{ir}'(h_r,\vartheta_i^r)
=\bigl(\tau_i^{r-1},\widehat\tau'_{ir}(h_r,\vartheta_i^r),s_i^r,x_i^{r-1}\bigr),\quad
\widehat\tau'_{ir}(h_r,\vartheta_i^r)\in\T_{ir}.
\]

\item \textbf{Local rule equivalence (R2).} For paired joint report histories $\widehat\vartheta^r,\widehat\vartheta^{\prime r}$, with other agents' strategies fixed,
{\small
\[
\begin{aligned}
(\mathcal X_r,\mathcal P_r)(\cdot\mid h_r,\widehat\vartheta^r)
&=(\mathcal X_r,\mathcal P_r)(\cdot\mid h_r,\widehat\vartheta^{\prime r}),\\
\mathcal D_{r+1}(\cdot\mid \px^r,p^r,d^r,\widehat\vartheta^r,y_{r+1})
&=\mathcal D_{r+1}(\cdot\mid \px^r,p^r,d^r,\widehat\vartheta^{\prime r},y_{r+1}),\quad r<T.
\end{aligned}
\]
}

\item \textbf{Actual-action replication (R3).} At every paired post-report history,
{\small
\[
x_{ir}'(h_r,\vartheta_i^r,\widehat\vartheta_{ir}',\px_{r},p_r)
=x_{ir}(h_r,\vartheta_i^r,\widehat\vartheta_{ir},\px_{r},p_r).
\]
}
The pairing preserves other agents' private histories and fixed reporting--action strategies.
\end{enumerate}
Then Assumption~\ref{ass:reporting-sufficiency} holds.
\end{proposition}

\begin{proof}
Fix $i,t,(h_t,\vartheta_i^t)$ and the two continuations in the proposition. R1 gives \eqref{eq:conditional-reporting-current-type} for every $r\ge t$, proving Assumption~\ref{ass:reporting-sufficiency}(i).

For part~(ii), both continuations start from the same conditional law $\left.\nu_{it}(\cdot\mid h_t,\vartheta_i^t)\right|_{\mathscr B_t^{-}}.$ Fix $r\ge t$ and suppose their conditional economic-prefix laws agree on $\mathscr B_r^{-}$. Conditional on the realized coordinates generating $\mathscr B_r^{-}$, R2 equates the allocation--transfer kernels $(\mathcal X_r,\mathcal P_r)$. Given the same $(\px_{r},p_r)$, R3 equates agent $i$'s actual action, while the other agents retain identical reporting--action strategies at paired private histories. The conditional law of the realized allocation $x_r$ therefore agrees. By the chain rule, the conditional joint laws of $(\tilde{\px}_{r},\tilde p_r,\tilde x_r)$ agree on $\mathscr B_r^{-}$.

For $r<T$, Section~\ref{subsec:technology} gives the ordered kernel $\mathcal K_{r+1}=\mathcal Q_{r+1}\mathcal Y_{r+1}\mathcal D_{r+1}\mathcal F_{r+1}^{\tau}.$ The common $\mathcal Q_{r+1}$, $\mathcal Y_{r+1}$, and $\mathcal F_{r+1}^{\tau}$ condition on the same realized economic history, including $x^r$; R2 equates $\mathcal D_{r+1}$ at paired report histories. Thus the conditional law of $(\tilde\theta_{r+1},\tilde y_{r+1},\tilde s_{r+1},\tilde d_{r+1},\tilde\tau_{r+1})$ agrees, implying equality of the economic-prefix laws on $\mathscr B_{r+1}^{-}$. R1--R3 preserve the pairing at each successor history, including following deviations. This proves the induction; at $r=T$, R2--R3 also match terminal decisions and actions.

The resulting finite-dimensional economic-path distributions coincide. Their cylinders generate $\mathcal B(W)$, also for $T=\infty$, so \eqref{eq:conditional-reporting-sufficiency} holds for every $A\in\mathcal B(W)$. This is Assumption~\ref{ass:reporting-sufficiency}(ii).
\end{proof}

Pointwise invariance of allocation--transfer and disclosure rules to the replaced non-type report coordinates, with $\widehat\tau'_{ir}=\widehat\tau_{ir}$, is a sufficient case of R2; only local equivalence of the paired continuations is required.

\subsection{Correlated information and peer-report incentives}
\label{subsec:disclosure-peer-information}

Correlated signals permit peer reports to discipline current-type reporting. We characterize proper-score incentives, finite-type predictive separation, and their continuous-type limit, while retaining private-action obedience and committed continuation incentives. For fixed-history type comparisons, $V_{it}^{*}((h_t,(\tau_i^{t-1},s_i^{t-1},x_i^{t-1})),\tau_{it},\cdot)$ abbreviates the action-optimized value at the actual augmented history $\vartheta_{it}$, with all non-type coordinates fixed; full-history reports still range over $\mathcal V_{it}$.

Fix $i,t,h_t$ and $\tau\in\T_{it}$. Define the \emph{conditional peer-type prediction}
$b_{it}^{-i}(\cdot\mid h_t;\tau)=\mathbb P(\tilde\tau_{-i,t}\in\cdot\mid h_t,\tilde\tau_{it}=\tau)\in\Delta(\T_{-i,t})$.
The conditional peer-\emph{type} law is induced by the primitives and mechanism of Sections~\ref{sec:environment}--\ref{sec:mechanisms} under the candidate truthful assessment. 

\emph{Predictive sufficiency} requires, at each relevant $(h_t,\vartheta_i^t)$ and under the unilateral deviations considered, $\mathbb P\bigl(\hat\tau_{-i,t}\in\cdot\mid h_t,\vartheta_i^t\bigr)=b_{it}^{-i}(\cdot\mid h_t;\tau_{it})$.
The equality is required under each relevant deviation-induced law of peer reports; it is not implied by correlation in $\tilde\tau_t$.

Let $\operatorname{sc}_{it}(b,m_{-i})$ be a strictly proper scoring rule with finite expected scores. The expected advantage of announcing the true conditional prediction rather than the one associated with a false type is the \emph{peer-information divergence}
\begin{equation}\label{eq:peer-information-divergence}
\begin{aligned}
\Delta_{it}^{\mathrm{peer}}(\tau_{it},\hat\tau_{it}\mid h_t)
=\int\Big[&\operatorname{sc}_{it}\bigl(b_{it}^{-i}(\cdot\mid h_t;\tau_{it}),m_{-i}\bigr)\\
&-\operatorname{sc}_{it}\bigl(b_{it}^{-i}(\cdot\mid h_t;\hat\tau_{it}),m_{-i}\bigr)\Big]
 b_{it}^{-i}(dm_{-i}\mid h_t;\tau_{it}).
\end{aligned}
\end{equation}
Strict propriety gives $\Delta_{it}^{\mathrm{peer}}(\tau_{it},\hat\tau_{it}\mid h_t)\geq0$, with strict inequality if
$b_{it}^{-i}(\cdot\mid h_t;\tau_{it})\ne b_{it}^{-i}(\cdot\mid h_t;\hat\tau_{it})$.

The mechanism can reward the reported prediction through the transfer
\[
p_{it}'=p_{it}-w_{it}(h_t)\operatorname{sc}_{it}\bigl(b_{it}^{-i}(\cdot\mid h_t;\hat\tau_{it}),\hat\tau_{-i,t}\bigr),
\qquad w_{it}(h_t)\geq0.
\]
Although $\operatorname{sc}_{it}$ depends on $\hat\tau_t$, not directly on $x_{it}$, $p_{it}'$ enters subsequent public histories. Impose \emph{continuation neutrality}: for every $i,t,(h_t,\vartheta_i^t,\widehat\vartheta_{it},\px_{t},p_t)$ and every reporting--action continuation, the score-augmented and original mechanisms induce the same non-scoring conditional continuation payoff, jointly across scored agents, including after false reports and reoptimized actions. Score transfers satisfy Assumption~\ref{ass:regularity}.

Using the pre-score action-optimized value $V_{it}^{*}$ and truthful reporting rule $\vartheta_{i,t}^{T}(\cdot)$, define the \emph{minimal uniform current-report scoring weight}
\begin{equation}\label{eq:peer-score-critical}
w_{it}^*(h_t)=\sup_{\substack{(h_t,\vartheta_i^t,\widehat\vartheta_{it},\px_{t},p_t)\text{ at }h_t\\\hat\tau_{it}\ne\tau_{it}}}
\frac{\left[\begin{aligned}&V_{it}^{*}((h_t,(\tau_i^{t-1},s_i^{t-1},x_i^{t-1})),\tau_{it},\widehat\vartheta_{i,t}^{T}(\cdot))\\[-2pt]&\quad-V_{it}^{*}((h_t,(\tau_i^{t-1},s_i^{t-1},x_i^{t-1})),\tau_{it},\vartheta_{i,t}^{T}(\cdot))\end{aligned}\right]_+}
{\Delta_{it}^{\mathrm{peer}}(\tau_{it},\hat\tau_{it}\mid h_t)}.
\end{equation}
A positive numerator divided by zero is interpreted as $+\infty$; $0/0$ and the supremum of an empty set are zero.

\begin{proposition}[Peer-Scoring Incentives for Truthful Reporting]
\label{prop:peer-discipline}
Under the above peer-score specification, Assumption~\ref{ass:regularity}, predictive sufficiency, strict propriety, and continuation neutrality:
\begin{enumerate}
\item \textbf{Reporting incentives.} If $w_{it}^*(h_t)<\infty$, every weight $w_{it}(h_t)\geq w_{it}^*(h_t)$ satisfies, at each corresponding private history and for every $\hat\tau_{it}\in\T_{it}$,
\[
\begin{aligned}
&V_{it}^{*}((h_t,(\tau_i^{t-1},s_i^{t-1},x_i^{t-1})),\tau_{it},\widehat\vartheta_{i,t}^{T}(\cdot))\\[-2pt]
&\quad-V_{it}^{*}((h_t,(\tau_i^{t-1},s_i^{t-1},x_i^{t-1})),\tau_{it},\vartheta_{i,t}^{T}(\cdot))\\[-2pt]
&\leq w_{it}(h_t)\Delta_{it}^{\mathrm{peer}}(\tau_{it},\hat\tau_{it}\mid h_t).
\end{aligned}
\]
Thus every specified score-augmented current-report gain is nonpositive.
\item \textbf{Joint implementation conditional on residual incentives.} Part~(1) controls current-type deviations only. JPBIC also requires residual full-history reporting incentive compatibility \eqref{eq:recursive-reporting}, choice obedience \eqref{eq:recursive-action}, and consistent beliefs $\nu^*$; together they imply $(\hat\vartheta=\vartheta,x(\cdot),\nu^*)\in\operatorname{JPBE}(\Gamma).$
\end{enumerate}
\end{proposition}
\begin{proof}
Predictive sufficiency and continuation neutrality give the score-adjusted deviation gain
\[
\begin{aligned}
&V_{it}^{*}((h_t,(\tau_i^{t-1},s_i^{t-1},x_i^{t-1})),\tau_{it},\widehat\vartheta_{i,t}^{T}(\cdot))\\[-2pt]
&\quad-V_{it}^{*}((h_t,(\tau_i^{t-1},s_i^{t-1},x_i^{t-1})),\tau_{it},\vartheta_{i,t}^{T}(\cdot))\\[-2pt]
&\quad -w_{it}(h_t)\Delta_{it}^{\mathrm{peer}}(\tau_{it},\hat\tau_{it}\mid h_t)\leq0
\end{aligned}
\]
by \eqref{eq:peer-score-critical}. Finiteness of $w_{it}^*(h_t)$ also covers zero divergence, proving part~(1). Part~(2) follows from Theorem~\ref{thm:recursive-ic} only when its remaining history-reporting and contingent-action conditions are also imposed under belief consistency.
\end{proof}

For a finite peer-report space, the quadratic score
$\operatorname{sc}_{it}(b,m_{-i})=2b(m_{-i})-\sum_{m_{-i}'}b(m_{-i}')^2$
gives
\begin{equation}\label{eq:peer-quadratic-divergence}
\Delta_{it}^{\mathrm{peer}}(\tau_{it},\hat\tau_{it}\mid h_t)
=\left\|b_{it}^{-i}(\cdot\mid h_t;\tau_{it})-b_{it}^{-i}(\cdot\mid h_t;\hat\tau_{it})\right\|^2.
\end{equation}
Predictive separation requires, for every $\tau_{it}\ne\hat\tau_{it}$,
\[
 b_{it}^{-i}(\cdot\mid h_t;\tau_{it})\ne b_{it}^{-i}(\cdot\mid h_t;\hat\tau_{it})
 \quad\text{implies}\quad \Delta_{it}^{\mathrm{peer}}(\tau_{it},\hat\tau_{it}\mid h_t)>0.
\]
For finite $\T_{it}$ at fixed $h_t$, $\min_{\tau_{it}\ne\hat\tau_{it}}\Delta_{it}^{\mathrm{peer}}(\tau_{it},\hat\tau_{it}\mid h_t)>0$. With uniformly bounded positive non-scoring gains, \eqref{eq:peer-score-critical} gives $w_{it}^*(h_t)<\infty$.

For a sufficient condition, specialize to finite ordered $\T_{it}$, $|\T_{it}|<\infty$ for every $i$, and full support at fixed $h_t$: $\mathbb P(\tilde\tau_t=a\mid h_t)>0,\;\forall a\in\prod_{i\in N}\T_{it}.$ These are conditional probability masses, not the densities of Section~\ref{sec:environment}. For any two profiles $a,b$, define their \emph{componentwise join and meet} by $(a\vee b)_j=\max\{a_j,b_j\}$ and $(a\wedge b)_j=\min\{a_j,b_j\}$ for every agent $j$. The profiles are \emph{incomparable} if neither $a_j\le b_j$ for all $j$ nor $b_j\le a_j$ for all $j$. The types are \emph{strictly affiliated} conditional on $h_t$ if, for all incomparable $\tau_t,\tau_t'$, 
\begin{equation}\label{eq:strict-affiliation}
\begin{aligned}
&\mathbb P(\tilde\tau_t=\tau_t\vee\tau_t'\mid h_t)
 \mathbb P(\tilde\tau_t=\tau_t\wedge\tau_t'\mid h_t)\\
&\qquad>\mathbb P(\tilde\tau_t=\tau_t\mid h_t)
 \mathbb P(\tilde\tau_t=\tau_t'\mid h_t).
\end{aligned}
\end{equation}
The conditional predictions used by the scoring rule must agree with this joint distribution:
\begin{equation}\label{eq:affiliated-peer-prediction}
b_{it}^{-i}(\tau_{-i,t}\mid h_t;\tau_{it})
=\frac{\mathbb P(\tilde\tau_t=(\tau_{it},\tau_{-i,t})\mid h_t)}
{\mathbb P(\tilde\tau_{it}=\tau_{it}\mid h_t)}.
\end{equation}
Maintain predictive sufficiency at all relevant private histories, and impose
the condition that if $|\T_{it}|>1$, then $|\T_{jt}|>1$ for some $j\ne i$.
Positive non-scoring reporting gains in \eqref{eq:peer-score-critical} are uniformly bounded over private histories sharing $h_t$.

\begin{corollary}[Peer discipline under strictly affiliated types]
\label{cor:peer-affiliation}
Under Assumption~\ref{ass:regularity} and the finite-type specification above, strict affiliation implies:
\begin{enumerate}
\item \textbf{Predictive separation.} For every $i$ and $\tau_{it}\ne\hat\tau_{it}$,
\[
b_{it}^{-i}(\cdot\mid h_t;\tau_{it})\ne
b_{it}^{-i}(\cdot\mid h_t;\hat\tau_{it}),
\qquad \Delta_{it}^{\mathrm{peer}}(\tau_{it},\hat\tau_{it}\mid h_t)>0.
\]
\item \textbf{Finite scoring threshold.} If the peer score is continuation neutral and its transfers satisfy Assumption~\ref{ass:regularity}, then
$w_{it}^*(h_t)<\infty,\,w_{it}(h_t)\geq w_{it}^*(h_t)$
The reporting inequality in Proposition~\ref{prop:peer-discipline}(1) follows at every corresponding private history and false report. Full-history reporting IC \eqref{eq:recursive-reporting}, CO \eqref{eq:recursive-action}, and consistent $\nu^*$ remain required for JPBIC, as in Proposition~\ref{prop:peer-discipline}(2).
\end{enumerate}
\end{corollary}

\begin{proof}
Fix agent $i$ and two distinct types $\tau_{it}<\hat\tau_{it}$. By the nondegenerate-peer condition, choose $j\ne i$ and $a_j<a_j'$ in $\T_{jt}$; fix $a_{-ij}\in\prod_{k\ne i,j}\T_{kt}$. Put $a_{-i}=(a_j,a_{-ij})$ and $a_{-i}'=(a_j',a_{-ij})$. The two joint profiles $(\tau_{it},a_{-i}')$ and $(\hat\tau_{it},a_{-i})$ are incomparable, with componentwise meet $(\tau_{it},a_{-i})$ and join $(\hat\tau_{it},a_{-i}')$. Strict affiliation and full support therefore imply
\[
\frac{\mathbb P(\tilde\tau_t=(\hat\tau_{it},a_{-i}')\mid h_t)}
{\mathbb P(\tilde\tau_t=(\tau_{it},a_{-i}')\mid h_t)}
>
\frac{\mathbb P(\tilde\tau_t=(\hat\tau_{it},a_{-i})\mid h_t)}
{\mathbb P(\tilde\tau_t=(\tau_{it},a_{-i})\mid h_t)}.
\]
If $b_{it}^{-i}(\cdot\mid h_t;\tau_{it})=b_{it}^{-i}(\cdot\mid h_t;\hat\tau_{it})$, \eqref{eq:affiliated-peer-prediction} would make both ratios equal to $\mathbb P(\tilde\tau_{it}=\hat\tau_{it}\mid h_t)/\mathbb P(\tilde\tau_{it}=\tau_{it}\mid h_t)$, contradicting the strict inequality. Thus $b_{it}^{-i}(\cdot\mid h_t;\tau_{it})\ne b_{it}^{-i}(\cdot\mid h_t;\hat\tau_{it})$, and strict propriety gives $\Delta_{it}^{\mathrm{peer}}(\tau_{it},\hat\tau_{it}\mid h_t)>0$, proving (1). Finiteness and the uniform non-scoring gain bound yield $w_{it}^*(h_t)<\infty$ by \eqref{eq:peer-score-critical}; Proposition~\ref{prop:peer-discipline} gives (2).
\end{proof}

For continuous types, local predictive differences may be only second order in scoring loss. If the predictive vector from a finite peer statistic is differentiable at interior $\tau_{it}$, then, as $h\downarrow0$,
\begin{equation}\label{eq:peer-local-divergence}
\Delta_{it}^{\mathrm{peer}}(\tau_{it},\tau_{it}-h\mid h_t)
=h^2\left\|\frac{\partial b_{it}^{-i}(\cdot\mid h_t;\tau_{it})}{\partial\tau_{it}}\right\|^2+o(h^2).
\end{equation}

\begin{proposition}[Limitations of Quadratic Peer Scoring]
\label{prop:peer-continuous-limit}
Under Assumption~\ref{ass:regularity}, the differentiable quadratic-score specification above, and continuation neutrality, suppose that, at some private history with interior $\tau_{it}$, $\widehat\vartheta_{i,t}^{T}(\cdot)$ reports $\tau_{it}-h$ at date $t$ and follows the fixed reporting continuation thereafter, and
\[
\begin{aligned}
&V_{it}^{*}((h_t,(\tau_i^{t-1},s_i^{t-1},x_i^{t-1})),\tau_{it},\widehat\vartheta_{i,t}^{T}(\cdot))\\[-2pt]
&\quad-V_{it}^{*}((h_t,(\tau_i^{t-1},s_i^{t-1},x_i^{t-1})),\tau_{it},\vartheta_{i,t}^{T}(\cdot))\\[-2pt]
&=mh+o(h),
\qquad m>0.
\end{aligned}
\]
Then, for every finite weight $w_{it}(h_t)\geq0$, there exists $\varepsilon>0$ such that for every $0<h<\varepsilon$,
\[
\begin{aligned}
&V_{it}^{*}((h_t,(\tau_i^{t-1},s_i^{t-1},x_i^{t-1})),\tau_{it},\widehat\vartheta_{i,t}^{T}(\cdot))\\[-2pt]
&\quad-V_{it}^{*}((h_t,(\tau_i^{t-1},s_i^{t-1},x_i^{t-1})),\tau_{it},\vartheta_{i,t}^{T}(\cdot))\\[-2pt]
&\quad -w_{it}(h_t)\Delta_{it}^{\mathrm{peer}}(\tau_{it},\tau_{it}-h\mid h_t)>0.
\end{aligned}
\]
Thus small enough downward reports remain profitable under any finite quadratic-score weight.
\end{proposition}

\begin{proof}
By \eqref{eq:peer-local-divergence}, for every $w_{it}(h_t)<\infty$,
\[
mh+o(h)-w_{it}(h_t)\Delta_{it}^{\mathrm{peer}}(\tau_{it},\tau_{it}-h\mid h_t)
=mh+o(h)-w_{it}(h_t)O(h^2)>0
\]
for sufficiently small $h>0$, since $m>0$.
\end{proof}

\begin{remark}[Peer prediction and dynamic screening]
Correlated peer reports can discipline reporting incentives \citep{CremerMcLean1988}; proper scores elicit predictive distributions \citep{MillerResnickZeckhauser2005}. Predictive separation can discipline finite-type reporting but does not itself ensure continuation or action incentives. With continuous types, finite quadratic-score weights cannot eliminate first-order gains from misreporting under the stated local conditions (Section~\ref{sec:incentives}). The score loss is second order, whereas the gain from the deviation is first order.
\end{remark}

\section{Optimal Dynamic Contracting with Screening and Hidden Actions}
\label{sec:optimal-screening}

We examine how hidden actions and endogenous information alter optimal dynamic contracts, the associated informational rents, and the costs of implementing economic outcomes. Conditional contracting values provide the basis for the information-design analysis in Section~\ref{sec:benchmark}.

\subsection{Optimal contracting and regime comparisons}
\label{subsec:optimal-screening}

The principal chooses a direct mechanism $\Gamma$ subject to JPBIC and interim participation. Recommended actions $\px(\cdot)$ are obeyed in the implementing assessment, while deviations $x_{it}\ne \px_{it}$ may affect realized allocations and continuation information. Participation occurs after $(\tau_{i1},s_{i1})$, before the initial report, with zero outside payoff. The principal solves (with $\Pi(\Gamma)=\int U^P\,d\mathbb P_W(\cdot\,;\,\Gamma,\vartheta,\px(\cdot))$)
\begin{align}
\sup_{\Gamma}\quad&\Pi(\Gamma)
\label{eq:principal-objective}\\
\text{subject to}\quad&
V_{i1}^{**}(\tau_{i1},s_{i1})\geq0,
\qquad
\forall i\in N,\ 
(\tau_{i1},s_{i1})\in\T_{i1}\times\Sg_{i1},
\qquad\text{(IR)}
\label{eq:participation}\\
&\left\{
\begin{aligned}
V_{it}^{**}(\tau_{it})&=V_{it}^{*}(\tau_{it},\vartheta_{i,t}^{T}(\cdot)),\\
\px_{it}&\in\arg\max_{x_i\in \overline X_{it}}V_{it}(\tau_{it},(\vartheta_{i,t+1}^{T}(\cdot),(x_i,\px_{i,t+1}^{T}(\cdot)))),
\end{aligned}
\right.
\qquad\text{(JPBIC)}
\label{eq:design-ic}
\end{align}
In \eqref{eq:design-ic}, $\tau_{it}$ varies with other private-history coordinates fixed; $V_{it}^{*}$ and $V_{it}^{**}$ are the full-history continuation values of \eqref{eq:report-deviation-direct}. Reporting incentives and choice obedience hold at relevant histories, including off path, under JPBE-consistent beliefs (Theorem~\ref{thm:recursive-ic}).

With common primitives, rule domains, information technologies and disclosure opportunities, and participation requirements, compare four contracting regimes: \textbf{$1B$}, observable incentive-relevant information and contractible information-producing choices; \textbf{$2B^{\mathrm{HI}}$}, private information with contractible choices; \textbf{$2B^{\mathrm{HA}}$}, observable incentive-relevant information with privately implemented choices; and \textbf{$3B$}, private information and privately implemented choices (cost interpretation in Subsection~\ref{subsec:screening-distortions}). The two second-best branches $2B^{\mathrm{HI}}$ and $2B^{\mathrm{HA}}$ are unordered, each between $1B$ and $3B$. For $j\in\{1B,\mathrm{HI},\mathrm{HA},3B\}$, let $\Pi_j^*=\Pi_j^*(\mathcal Y,\{\mathcal D\})$ denote the supremum of $\Pi(\Gamma)$ under regime $j$. The nested values are $\Pi_j^{**}(\mathcal Y,D):=\sup_{\mathcal D\in D}\Pi_j^*(\mathcal Y,\{\mathcal D\})$ and $\Pi_j^{***}(Y,D):=\sup_{\mathcal Y\in Y}\Pi_j^{**}(\mathcal Y,D)$. In particular, $\Pi_{3B}^{***}(Y,D)$ is the value of \eqref{eq:principal-objective}--\eqref{eq:design-ic}.

\begin{proposition}[Partial ordering of contracting regimes]
\label{prop:screening-benchmark-order}
Under Assumption~\ref{ass:regularity}, $\Pi_{1B}^{*}\ge \Pi_{\mathrm{HI}}^{*}\ge \Pi_{3B}^{*}$ and $\Pi_{1B}^{*}\ge \Pi_{\mathrm{HA}}^{*}\ge \Pi_{3B}^{*}$. The same rankings hold for $\Pi_j^{**}$ at a common $(\mathcal Y,D)$ and for $\Pi_j^{***}$ at common $(Y,D)$. Neither second-best branch generally dominates the other.
\end{proposition}

\begin{proof}
Additional observability can be ignored, and contractibility can enforce equilibrium actions, holding the remaining rules fixed. The rankings follow by relaxing reporting IC or choice obedience and are preserved by further optimization over the common disclosure rules and technologies. In Example~1, \eqref{eq:benchmark-transaction-interaction} with $\delta_t=0$ for all $t<T$ and positive screening loss gives $\Pi_{\mathrm{HI}}^*<\Pi_{\mathrm{HA}}^*$; the positive-interaction example in Remark~\ref{rem:interaction-frictions} gives the reverse strict inequality. Singleton information-rule classes preserve these comparisons at the nested levels.
\end{proof}

Hidden-action incentives turn the conventional dynamic screening optimum into a potentially unattainable upper bound. The contracting hierarchy becomes branched because relaxing reporting and action constraints expands implementability along distinct dimensions, neither generally subsuming the other. These rankings concern optimized payoffs, not pointwise allocations, information-producing choices, or disclosure rules, whose comparisons require additional restrictions.

\subsection{Dynamic information rents and virtual surplus}
\label{subsec:dynamic-virtual-surplus}

The dynamic envelope of Section~\ref{subsec:dynamic-envelope} converts the JPBIC constraints in Subsection~\ref{subsec:optimal-screening} into an initial-rent identity for the principal's objective. For a JPBIC $\Gamma$ with obedient $x^*(\cdot)=\px(\cdot)$, let $\tau_{i1}\mid s_{i1}$ have density $f_{i1}(\tau\mid s_{i1})>0$ and distribution $F_{i1}(\tau\mid s_{i1})$ on $\T_{i1}=[\underline\tau_{i1},\overline\tau_{i1}]$. Set $\Lambda_{i1}(\tau\mid s)=[1-F_{i1}(\tau\mid s)]/f_{i1}(\tau\mid s)$ for $\tau<\overline\tau_{i1}$. The derivative $\partial_{\tau}V_{i1}^{**}(\tau,s)$ exists a.e. for each $s$. Initial types and signals may be correlated across agents.

\begin{theorem}[Dynamic information rents and virtual surplus]
\label{thm:dynamic-virtual-surplus}
% Initial-signal-conditioned rent identity, consistent with the date-1 participation convention.
Under Assumptions~\ref{ass:regularity} and \ref{ass:envelope-regularity}, for every JPBIC direct mechanism satisfying initial interim participation and the initial-type distribution conditions above, the principal's expected payoff is
\begin{equation}\label{eq:dynamic-virtual-surplus-general}
\Pi
=\underbrace{\E\left[\sum_{t=1}^T\beta^{t-1}\left(u_t^{P}+\sum_{i\in N}u_{it}\right)\right]}_{\text{Expected total surplus}}
-\underbrace{\sum_{i\in N}\E\left[\Lambda_{i1}(\tau_{i1}\mid s_{i1})\partial_{\tau_{i1}}V_{i1}^{**}(\tau_{i1},s_{i1})\right]}_{\text{Dynamic information rents}}
-\underbrace{\sum_{i\in N}\E_{s_{i1}}[V_{i1}^{**}(\underline\tau_{i1},s_{i1})]}_{\text{Lowest-type rents}}.
\end{equation}
\end{theorem}

\begin{proof}
For each $s$, by Proposition~\ref{prop:envelope},
\[
V_{i1}^{**}(\tau,s)=V_{i1}^{**}(\underline\tau_{i1},s)
+\int_{\underline\tau_{i1}}^{\tau}\partial_{z}V_{i1}^{**}(z,s)\,dz.
\]
Under the stated conditional-density and integrability conditions, conditional Fubini yields
\[
\begin{aligned}
\E[V_{i1}^{**}(\tilde\tau_{i1},s_{i1})\mid s_{i1}=s]
={}&V_{i1}^{**}(\underline\tau_{i1},s)
+\int_{\T_{i1}}\partial_{z}V_{i1}^{**}(z,s)[1-F_{i1}(z\mid s)]\,dz\\
={}&V_{i1}^{**}(\underline\tau_{i1},s)
+\E[\Lambda_{i1}(\tilde\tau_{i1}\mid s)\left.\partial_{z}V_{i1}^{**}(z,s)\right|_{z=\tilde\tau_{i1}}\mid s_{i1}=s].
\end{aligned}
\]
Integrating over $s_{i1}$, summing over agents, and canceling transfers gives
\[
\Pi=\E\left[\sum_{t=1}^T\beta^{t-1}\left(u_t^{P}+\sum_{i\in N}u_{it}\right)\right]
-\sum_{i\in N}\E[V_{i1}^{**}(\tilde\tau_{i1},s_{i1})].
\]
Substitution establishes \eqref{eq:dynamic-virtual-surplus-general}.
\end{proof}

Identity~\eqref{eq:dynamic-virtual-surplus-general} separates total surplus (including direct information-production costs), dynamic information rents, and lowest-type rents. It permits correlated initial types and signals without requiring nonnegative marginal-rent adjustments, and yields the following revenue-equivalence result.

\begin{corollary}[Dynamic Revenue Equivalence]
\label{cor:dynamic-revenue-equivalence}
Let $\Gamma,\Gamma'$ be JPBIC mechanisms satisfying the conditions of Theorem~\ref{thm:dynamic-virtual-surplus}, with a common initial type--signal distribution (as specified in Section~2) and $\partial_\tau V_{i1}^{\Gamma}(\tau,s)=\partial_\tau V_{i1}^{\Gamma'}(\tau,s)$ for each $i$ and a.e. $(\tau,s)$. Then, for a.e. $s$ and all $\tau\in\T_{i1}$, $V_{i1}^{\Gamma}(\tau,s)-V_{i1}^{\Gamma'}(\tau,s)=V_{i1}^{\Gamma}(\underline\tau_{i1},s)-V_{i1}^{\Gamma'}(\underline\tau_{i1},s)$. If expected total surplus also agrees,
\begin{equation}\label{eq:dynamic-revenue-equivalence}
\Pi(\Gamma)-\Pi(\Gamma')
=-\sum_{i\in N}\E_{s_{i1}}\left[V_{i1}^{\Gamma}(\underline\tau_{i1},s_{i1})-V_{i1}^{\Gamma'}(\underline\tau_{i1},s_{i1})\right].
\end{equation}
Thus principal payoffs coincide if conditional lowest-type rents also agree.
\end{corollary}

\begin{proof}
By Proposition~\ref{prop:envelope} and $\partial_\tau V_{i1}^{\Gamma}=\partial_\tau V_{i1}^{\Gamma'}$ a.e., $V_{i1}^{\Gamma}-V_{i1}^{\Gamma'}$ is constant in $\tau$ for a.e. $s$. Applying \eqref{eq:dynamic-virtual-surplus-general} to $\Gamma$ and $\Gamma'$ and canceling the equal total-surplus and marginal-rent terms gives \eqref{eq:dynamic-revenue-equivalence}.
\end{proof}

Dynamic revenue equivalence survives when the stated distribution, rent, and surplus conditions hold. However, identical recommended allocations need not preserve these conditions when realized actions and information are endogenous. Revenue comparisons must therefore account for induced economic and informational consequences, rather than recommended allocations alone.

Under the controlled type-process restrictions and Assumption~\ref{ass:controlled-impulse-regularity}, Proposition~\ref{prop:controlled-marginal-utility} yields the following dynamic virtual surplus:
\begin{equation}\label{eq:dynamic-virtual-surplus}
\Phi_t=u_t^{P}+\sum_{i\in N}u_{it}
-\sum_{i\in N}\Lambda_{i1}(\tau_{i1}\mid s_{i1})
\left[
\left(\frac{\partial u_{it}}{\partial\tau_{it}}\right)I_{i1}^{it}
+\frac{\partial u_{it}}{\partial x_t}\cdot\dot x_t
+\frac{\partial u_{it}}{\partial\theta_t}\cdot\dot\theta_t
-\dot p_{it}
\right].
\end{equation}
Here $\Phi_t$ incorporates type propagation $I_{i1}^{it}$, allocation responses $\dot x_t$, state responses $\dot\theta_t$, and transfer responses $\dot p_{it}$, including information-production effects through $\dot x_t$. Thus \eqref{eq:dynamic-virtual-surplus-general} gives
\begin{equation}\label{eq:virtual-surplus-payoff}
\Pi
=\int\sum_{t=1}^T\beta^{t-1}\Phi_t\,
d\mathbb P(\cdot\,;\,\Gamma,x^*)
-\sum_{i\in N}\E_{s_{i1}}[V_{i1}^{**}(\underline\tau_{i1},s_{i1})].
\end{equation}

When the reduced marginal-payoff representation \eqref{eq:pst-marginal-utility} holds, equation~\eqref{eq:virtual-surplus-payoff} becomes
\begin{equation}\label{eq:reduced-virtual-surplus-payoff}
\begin{aligned}
\Pi
={}&\int\sum_{t=1}^T\beta^{t-1}
\left[
\underbrace{u_t^{P}+\sum_{i\in N}u_{it}}_{\text{Total surplus}}
-\underbrace{\sum_{i\in N}\Lambda_{i1}(\tau_{i1}\mid s_{i1})
\left(\frac{\partial u_{it}}{\partial\tau_{it}}\right)I_{i1}^{it}}_{\text{Dynamic information-rent adjustment}}
\right]d\mathbb P(\cdot\,;\,\Gamma,x^*)\\
&-\underbrace{\sum_{i\in N}\E_{s_{i1}}[V_{i1}^{**}(\underline\tau_{i1},s_{i1})]}_{\text{Lowest-type rents}}.
\end{aligned}
\end{equation}
The omitted non-type response terms vanish in the stipulated aggregate expectation, not necessarily pointwise. For linear private values, $u_{it}(x_t,\tau_{it},\theta_t)=\tau_{it}x_{it}+b_{it}(x_t,\theta_t)$, and \eqref{eq:reduced-virtual-surplus-payoff} becomes
\begin{equation}\label{eq:dynamic-virtual-type}
\begin{aligned}
\Pi
={}&\int\sum_{t=1}^T\beta^{t-1}
\left\{
u_t^{P}+\sum_{i\in N}
\left[
\left(\tau_{it}-\Lambda_{i1}(\tau_{i1}\mid s_{i1})I_{i1}^{it}\right)x_{it}
+b_{it}
\right]
\right\}d\mathbb P(\cdot\,;\,\Gamma,x^*)\\
&-\sum_{i\in N}\E_{s_{i1}}[V_{i1}^{**}(\underline\tau_{i1},s_{i1})].
\end{aligned}
\end{equation}

The \emph{dynamic virtual type} $\tau_{it}-\Lambda_{i1}(\tau_{i1}\mid s_{i1})I_{i1}^{it}$ adjusts $\tau_{it}$ by the initial inverse hazard $\Lambda_{i1}$ and its controlled propagation $I_{i1}^{it}$.\footnote{The inverse-initial-hazard--times--impulse-response adjustment has the same Myerson--PST structure as in \citet[Eq.~(13)]{PavanSegalToikka2014}. Here the initial hazard is conditional on $s_{i1}$, while the general representation \eqref{eq:dynamic-virtual-surplus} also accounts for endogenous information and privately implemented actions; the reduction requires the stated additional restrictions.} Here $b_{it}$ is type independent; direct information-production costs remain in $u_{it}$.

For $I_{i1}^{it}\ge0$, the dynamic virtual type does not exceed $\tau_{it}$. A nondecreasing conditional hazard implies nonincreasing $\Lambda_{i1}$, but not necessarily a monotone virtual-type schedule when $I_{i1}^{it}$ depends on history or the mechanism. Endogenous information and hidden actions can change both the virtual-surplus objective and its implementability constraints (Sections~\ref{sec:mechanisms}--\ref{sec:incentives}). Allocation, state, and transfer responses can generate additional marginal-rent effects, while information production and disclosure can affect $I_{i1}^{it}$. Even when the PST representation is recovered, choice obedience can preclude implementation of its conventional maximizer. Recovering the conventional objective therefore does not suffice to recover the conventional optimal contracting problem.

\subsection{Economic implications: incentive frictions and transaction costs}
\label{subsec:screening-distortions}

The Coase--Williamson transaction-cost perspective interprets the incentive restrictions of Sections~3--4 in terms of institutional costs. \emph{Informational costs} arise through dynamic reporting rents, with marginal weight $\Lambda_{i1}(\tau_{i1}\mid s_{i1})I_{i1}^{it}$. \emph{Hidden-action implementation costs} arise from choice obedience (Theorem~\ref{thm:recursive-ic}): they restrict attainable JPBIC outcomes rather than adding a separable term to \eqref{eq:dynamic-virtual-surplus-general}. Both differ from \emph{direct information-production expenditures}, which enter $u_{it}$ and consume resources. Thus, the three institutional costs operate through distinct channels: rent adjustments, implementability restrictions, and resource expenditures. Hidden-action implementation costs cannot generally be represented as a separate virtual-type adjustment, while direct expenditures reduce total surplus. Private choices $x_i(\cdot)$ can affect future information and reporting rents, while disclosure $\mathcal D$ changes incentives to acquire information. Under Assumptions~\ref{ass:regularity} and \ref{ass:envelope-regularity}, the four regimes of Proposition~\ref{prop:screening-benchmark-order}, with common primitives, feasible rules, participation requirements, and unrestricted transfers, have the following interpretations: \textbf{$1B$}, total-surplus maximization without reporting IC or choice obedience; \textbf{$2B^{\mathrm{HI}}$}, rent-adjusted surplus in \eqref{eq:dynamic-virtual-surplus-general} subject to dynamic reporting IC; \textbf{$2B^{\mathrm{HA}}$}, total surplus subject to sequential choice obedience but without reporting IC or screening rents; and \textbf{$3B$}, rent-adjusted surplus subject to joint reporting IC and choice obedience, including hidden-action reoptimization after false reports. For observable types, initial transfers contingent only on participation-stage information permit rent extraction without affecting continuation incentives; for private types, Theorem~\ref{thm:dynamic-virtual-surplus} provides the rent adjustment; Theorem~\ref{thm:recursive-ic} gives the recursive characterization of incentive restrictions, and Theorem~\ref{thm:integral-monotonicity} gives their first-order and integral-monotonicity characterization. Under the controlled-type conditions of Subsection~\ref{subsec:dynamic-envelope}, the HI and 3B objectives admit \eqref{eq:virtual-surplus-payoff}; when \eqref{eq:pst-marginal-utility} holds, they admit \eqref{eq:reduced-virtual-surplus-payoff}.

For fixed $(\mathcal Y,\mathcal D)$, the loss in optimized principal payoff admits two decompositions:
\begin{equation}\label{eq:benchmark-loss-decomposition}
\Pi_{1B}^{*}-\Pi_{3B}^{*}
=\underbrace{(\Pi_{1B}^{*}-\Pi_{\mathrm{HI}}^{*})}_{\text{Informational transaction cost}}
+\underbrace{(\Pi_{\mathrm{HI}}^{*}-\Pi_{3B}^{*})}_{\text{Additional implementation cost}},
\end{equation}
and
\begin{equation}\label{eq:benchmark-loss-decomposition-hp}
\Pi_{1B}^{*}-\Pi_{3B}^{*}
=\underbrace{(\Pi_{1B}^{*}-\Pi_{\mathrm{HA}}^{*})}_{\text{Implementation cost}}
+\underbrace{(\Pi_{\mathrm{HA}}^{*}-\Pi_{3B}^{*})}_{\text{Additional informational cost}}.
\end{equation}
Each difference is nonnegative by Proposition~\ref{prop:screening-benchmark-order}. The two branches introduce the frictions in opposite orders. These differences concern $\Pi_j^{*}$, not necessarily total-surplus losses; the pointwise rent adjustment in \eqref{eq:dynamic-virtual-surplus-general} may have either sign.

The \emph{nonadditive interaction} between the two incentive frictions is
\[
(\Pi_{\mathrm{HI}}^{*}-\Pi_{3B}^{*})-(\Pi_{1B}^{*}-\Pi_{\mathrm{HA}}^{*})
=(\Pi_{\mathrm{HA}}^{*}-\Pi_{3B}^{*})-(\Pi_{1B}^{*}-\Pi_{\mathrm{HI}}^{*})
=\Pi_{\mathrm{HI}}^{*}+\Pi_{\mathrm{HA}}^{*}-\Pi_{1B}^{*}-\Pi_{3B}^{*}.
\]
This interaction measures whether the incremental optimized-payoff loss from one friction increases or decreases in the presence of the other; Proposition~\ref{prop:screening-benchmark-order} does not determine its sign.\footnote{This accounting identity does not imply a nonzero interaction; the separable quadratic specification below has zero interaction. The differences compare optimized principal payoffs, not pointwise allocations, information production, or disclosure.}

The same decompositions hold for $\Pi_j^{**}$ and $\Pi_j^{***}$ under the corresponding common information-design opportunities. Reoptimizing information technology or disclosure may change the magnitudes and sign of the interaction; zero interaction under fixed rules does not imply zero interaction after reoptimization.

Screening and hidden-action costs need not be additive. Hidden-action constraints can redirect economic activity toward choices subject to screening distortions, thereby reinforcing institutional losses. Conversely, they can eliminate activities that generate screening rents, thereby offsetting informational costs. Example~1 illustrates additivity in the quadratic specification, whereas the subsequent examples establish interactions of either sign.

\smallskip
\noindent\textbf{Example 1 (Quadratic Contracting with Fixed Disclosure).}

Consider a finite-horizon ($T\ge2$), single-agent quadratic environment. We compare the four contracting regimes under fixed disclosure $\mathcal D$, which is optimized in Example~2. Let $x_t=(q_t,e_t)\in X_t=\mathbb R_+\times[0,\bar e_t]$ for $t<T$, with $X_T=\mathbb R_+\times\{0\}$. Output $q_t$ is contractible; effort $e_t$ is contractible in $1B,2B^{\mathrm{HI}}$ and hidden in $2B^{\mathrm{HA}},3B$. Before setting $q_t$, the principal observes neither $\theta_t$ nor raw evidence $y_t$; transfers can condition on them only through public disclosure $d_t$. The rule $\mathcal D_{t+1}$ has accuracy $\delta_t\in[0,1]$.

Let $\tau_1\sim F_1$ on $\T_1=[\underline\tau_1,\overline\tau_1]$ with $f_1>0$ and degenerate initial private signal. Types follow $\tau_{t+1}=\varrho\tau_t+\varepsilon_{t+1}$, $0\le\varrho\le1$, with bounded-support shocks independent across dates, of $\tau_1$, and of the state process. Then $I_{11}^{1t}=\varrho^{t-1}$; set $\Lambda_1=(1-F_1)/f_1$ nonincreasing. The period payoffs before transfers are $u_t(x_t,\tau_t)=\tau_tq_t-\frac{q_t^2}{2}-\frac{e_t^2}{2\kappa_t}, \; u_t^{P}(x_t,\theta_t)=\theta_tq_t,$ where $\kappa_t>0$ for $t<T$ and the effort-cost term is zero at $T$. The binary state space is $\Theta_t^{B}=\{-\theta_t^{B},\theta_t^{B}\}$, with $\theta_t^{B}>0$ and $\theta_t\in\Theta_t^{B}$; the two states are equally likely. States are independent across dates and of the type process. Effort at date $t$ determines the precision of evidence about $\theta_{t+1}$, produced after date-$t$ choices and disclosed before date-$t+1$ reports and output decisions. Specifically,
\begin{equation}\label{eq:benchmark-evidence}
\mathbb P\left(y_{t+1}=\theta_{t+1}/\theta_{t+1}^{B}\mid\theta_{t+1},e_t\right)
=\frac{1+\sqrt{1/4+e_t}}{2},\qquad
\mathbb P(d_{t+1}=y_{t+1}\mid y_{t+1})=\frac{1+\delta_t}{2},
\end{equation}
where $y_{t+1},d_{t+1}\in\{-1,1\}$, $\delta_t$ is independent of reports, and $\bar e_t\le1/2$. The public signal has distribution and posterior mean
\begin{equation}\label{eq:benchmark-signal-value}
\mathbb P(d_{t+1}=1\mid e_t)=\frac12,\qquad
\E[\theta_{t+1}\mid d_{t+1},e_t]
=\theta_{t+1}^{B}\delta_t\sqrt{1/4+e_t}\,d_{t+1}.
\end{equation}
State independence extends these identities to preceding histories; under hidden effort, posterior beliefs use equilibrium $e_t$.

For explicit interior output and effort solutions, suppose $\bigl(\tau_t-\Lambda_1(\tau_1)\varrho^{t-1}\bigr)>\theta_t^{B}$ throughout the type-history support, including its boundary, and $\kappa_t\beta{\theta_{t+1}^{B}}^2\delta_t^2/2<\bar e_t$ when $\delta_t>0$. When $\delta_t=0$, optimal effort is zero.

Although $e_t$ changes the informativeness of $d_{t+1}$, it does not change its marginal distribution, $\mathbb P(d_{t+1}=1\mid e_t)=1/2$. Together with exogenous types and independent states, this leaves the continuation distribution observable to the mechanism invariant to unilateral hidden-effort deviations. Since $e_t>0$ incurs a strictly positive cost, the unique obedient hidden-effort choice is $e_t=0$ after every report. When effort is contractible, the incremental expected information benefit at $t+1$, net of current effort cost, is
\begin{equation}\label{eq:benchmark-project-surplus}
\max_{e_t\in[0,\bar e_t]}
\left\{\frac{\beta{\theta_{t+1}^{B}}^2\delta_t^2}{2}e_t
-\frac{e_t^2}{2\kappa_t}\right\}.
\end{equation}
Accordingly, optimal effort under fixed disclosure satisfies
\begin{equation}\label{eq:benchmark-action-rules}
e_t^{1B}=e_t^{2B^{\mathrm{HI}}}
=\frac{\kappa_t\beta{\theta_{t+1}^{B}}^2\delta_t^2}{2},\qquad
e_t^{2B^{\mathrm{HA}}}=e_t^{3B}=0,\qquad t<T,
\end{equation}
and $e_T^k=0$ in every environment $k$. The recommended effort rule is independent of reports and types.

\paragraph{Full-history payoff bound.}
Consider the report-independent binary disclosure technology specified above. The principal observes only the
committed disclosed information, reports, recommended outputs and transfers; there
is no contractible observation of realized hidden effort, the underlying state,
or raw evidence. The private signals are degenerate, and the type shocks
are independent of states, evidence and effort. For each JPBIC contract considered in the full-history comparison, impose the conditional initial-type envelope, integrability, and initial interim participation conditions of Theorem~\ref{thm:dynamic-virtual-surplus}. These conditions are maintained for the comparison, not derived from quadratic payoffs alone. Let $\mathcal I_t$ denote the information at the date-$t$ output decision, including the observed mechanism history and truthful type history; define $\bar\theta_t=\E[\theta_t\mid\mathcal I_t]$ under implemented effort $e_t$.
Then the principal's expected value is bounded above by
\begin{equation}\label{eq:benchmark-full-history-upper}
\begin{aligned}
\Pi\leq{}&\mathbb E\left[\sum_{t=1}^T\frac{\beta^{t-1}}2
(a_t+\bar\theta_t)^2-\sum_{t=1}^{T-1}\beta^{t-1}\frac{e_t^2}{2\kappa_t}\right],\\
a_t={}&\begin{cases}
\tau_t,&\text{when types are observable},\\
\tau_t-\Lambda_1(\tau_1)\varrho^{t-1},&\text{when types are private}.
\end{cases}
\end{aligned}
\end{equation}
In a hidden-effort regime, every JPBIC contract satisfying the preceding conditions implements $e_t=0$ at every reached post-report history.
Maximizing the bound in \eqref{eq:benchmark-full-history-upper} over contractible effort, with hidden effort equal to zero, yields the values in \eqref{eq:benchmark-value-formulas}.

\noindent\emph{Derivation of the bound.}
For private $\tau_1$, Theorem~\ref{thm:dynamic-virtual-surplus} and $V_1^{**}(\underline\tau_1)\ge0$ give a full-history upper bound after subtracting the marginal rent. The additive type process implies
\[
\tau_s+\varrho^{s-1}(r-\tau_1)
=r\varrho^{s-1}+\sum_{k=2}^{s}\varrho^{s-k}\varepsilon_k,
\qquad \partial_{\tau_1}\tau_s=\varrho^{s-1},\quad s=1,\ldots,T.
\]
The first expression is the type history reported under the same primitive shocks when mimicking initial type $r$. Under the maintained initial-type envelope, a deviation mimicking the complete reporting and action plan for initial type $r$ along this coupled path holds recommended outputs and transfers fixed. Since $\partial_{\tau_s}u_s=q_s$, the associated initial marginal rent is
\[
\frac{dV_1^{**}(\tau_1)}{d\tau_1}
=\E\left[\left.\sum_{s=1}^{T}\beta^{s-1}\varrho^{s-1}q_s\,\right|\tau_1\right].
\]
The identity uses the maintained fixed-plan envelope; $I_{11}^{1s}=\varrho^{s-1}$ alone would not eliminate non-type marginal effects for arbitrary uncoupled continuation plans. For observable types, unrestricted transfers and participation yield the total-surplus bound. In both cases, the resulting period objective is $(a_t+\theta_t)q_t-q_t^2/2-e_t^2/(2\kappa_t)$.

Independence of $\theta_t$ from the type process and past states, with the specified binary observation and publicly known implemented effort in the contractible regime, determines $\bar\theta_t$. Conditional on $\mathcal I_t$,
\[
\E\!\left[(a_t+\theta_t)q_t-\frac{q_t^2}{2}\,\middle|\,\mathcal I_t\right]
=(a_t+\bar\theta_t)q_t-\frac{q_t^2}{2}
\leq\frac{(a_t+\bar\theta_t)^2}{2},
\]
with equality at $q_t=a_t+\bar\theta_t>0$. Since no additional contractible history conveys information about $\theta_t$ in this specification, this gives \eqref{eq:benchmark-full-history-upper}.
For hidden $e_t$, fix any post-report history and keep the subsequent reporting and action strategy, including its recorded histories, unchanged as a function of the agent's realized observations. By \eqref{eq:benchmark-signal-value} and the specified independence, $\mathbb P(d_{t+1}=1\mid e_t,h_t,\tau^t)=\frac12 =\mathbb P(d_{t+1}=1\mid 0,h_t,\tau^t),\; \forall e_t\in[0,\bar e_t].$ The same holds for the joint law of the agent's future payoff-relevant observations, reports, recommended outputs, and transfers: the type process and later disclosed information is independent of this unilateral effort change. The joint state--signal law, however, remains effort-dependent. Since $u_s$ contains no state term, the mimicking payoff comparison is
\[
\E\!\left[\sum_{s=t}^{T}\beta^{s-t}(u_s-p_s)\,\middle|\,e_t\right]
-\E\!\left[\sum_{s=t}^{T}\beta^{s-t}(u_s-p_s)\,\middle|\,e_t=0\right]
=-\frac{e_t^2}{2\kappa_t}<0,\qquad e_t>0.
\]
Thus $e_t=0$ at each reached post-report history. For contractible effort, \eqref{eq:benchmark-signal-value} gives
\[
\E[\bar\theta_{t+1}^2\mid e_t]
={\theta_{t+1}^{B}}^2\delta_t^2(1/4+e_t),\qquad
\partial_{e_t}\left(\frac{\beta{\theta_{t+1}^{B}}^2\delta_t^2}{2}e_t-\frac{e_t^2}{2\kappa_t}\right)
=\frac{\beta{\theta_{t+1}^{B}}^2\delta_t^2}{2}-\frac{e_t}{\kappa_t}.
\]
The second derivative is $-1/\kappa_t<0$; the stated interior condition yields \eqref{eq:benchmark-action-rules}.

\paragraph{Optimal output rules.}
Maximizing $(a_t+\bar\theta_t)q_t-q_t^2/2$ over $q_t\ge0$ gives $q_t=a_t+\bar\theta_t$ under the stated interior condition. At date $1$,
\[
q_1^{1B}=q_1^{2B^{\mathrm{HA}}}=\tau_1,\qquad
q_1^{2B^{\mathrm{HI}}}=q_1^{3B}=\bigl(\tau_1-\Lambda_1(\tau_1)\bigr),
\]
and, for $t\ge2$,
\begin{equation}\label{eq:benchmark-fixed-contracts}
\begin{aligned}
q_t^{1B}&=\tau_t+\theta_t^{B}\delta_{t-1}\sqrt{1/4+e_{t-1}^{1B}}\,d_t,
&\quad q_t^{2B^{\mathrm{HI}}}&=\bigl(\tau_t-\Lambda_1(\tau_1)\varrho^{t-1}\bigr)+\theta_t^{B}\delta_{t-1}\sqrt{1/4+e_{t-1}^{2B^{\mathrm{HI}}}}\,d_t,\\
q_t^{2B^{\mathrm{HA}}}&=\tau_t+\frac{\theta_t^{B}\delta_{t-1}}2d_t,
&\quad q_t^{3B}&=\bigl(\tau_t-\Lambda_1(\tau_1)\varrho^{t-1}\bigr)+\frac{\theta_t^{B}\delta_{t-1}}2d_t.
\end{aligned}
\end{equation}
The transfers below attain \eqref{eq:benchmark-full-history-upper}.

\paragraph{Implementation and attainment of the bound.}
We implement the output rules by verifying later-report IC, initial-report GIM, and hidden-action obedience, thereby attaining \eqref{eq:benchmark-full-history-upper} under full-history JPBIC.

Fix an initial report $r$ and subsequent public history. In either hidden-information environment, the date-$s$ output rule for $s\ge2$, conditional on that initial report, has the form
\[
q_s(r,z,d_s)=z+c_s(r,d_s),\qquad
c_s(r,d_s)=-\Lambda_1(r)\varrho^{s-1}
+\theta_s^{B}\delta_{s-1}\sqrt{1/4+e_{s-1}}\,d_s,
\]
where $z$ is the current report and the recommended effort $e_{s-1}$ is fixed independently of reports. The same construction covers either effort regime at its recommended effort. Choose the date-$s$ transfer charged to the agent to be $p_s(r,z,d_s)=-c_s(r,d_s)z$ for $s\ge2$.
For each true current type $\tau_s$, and holding the initial report and public history fixed, the agent's current payoff (including the recommended effort cost, which is independent of $z$) satisfies the exact identity
\begin{equation}\label{eq:benchmark-later-report-ic}
\bigl[u_s(q_s(r,\tau_s,d_s),e_s,\tau_s)-p_s(r,\tau_s,d_s)\bigr]
-\bigl[u_s(q_s(r,z,d_s),e_s,\tau_s)-p_s(r,z,d_s)\bigr]
=\frac12(z-\tau_s)^2\ge0.
\end{equation}
Continuation rules depend on $r$, subsequent current reports, and public disclosures, but not on intervening reports or their recorded allocations and payments; recorded histories remain observable. Given $r$, type and disclosure laws are invariant to a date-$s$ report. Therefore \eqref{eq:benchmark-later-report-ic} applies recursively at every post-report history, ruling out arbitrary subsequent report deviations.

Let $G(\tau,r)$ denote the agent's expected discounted utility from dates $1$ through $T$, including effort costs and subsequent transfers but excluding the date-$1$ transfer, when the initial true type is $\tau$, the initial report is $r$, and the agent subsequently reports truthfully and implements the recommended choices. At date $1$, set $q_1(r)=r-\Lambda_1(r)$. Under the common exogenous-shock coupling of the type process specified in Example~1, the envelope identity for the optimized continuation following initial report $r$ gives
\begin{equation}\label{eq:benchmark-initial-envelope}
\partial_\tau G(\tau,r)
= q_1(r)+\sum_{s=2}^{T}\beta^{s-1}\varrho^{s-1}
\E\bigl[q_s(r,\tau_s,d_s)\mid\tau_1=\tau\bigr].
\end{equation}
By \eqref{eq:benchmark-later-report-ic}, later reports optimize period payoffs, so the envelope differentiates type-dependent utility rather than optimal reports. Signal laws and effort are independent of the initial type and report. For $r'\ge r$, subtracting the two derivatives in \eqref{eq:benchmark-initial-envelope} yields
\begin{equation}\label{eq:benchmark-initial-single-crossing}
\begin{aligned}
\partial_\tau G(\tau,r')-\partial_\tau G(\tau,r)
={}&(r'-r)-\bigl[\Lambda_1(r')-\Lambda_1(r)\bigr]\\
&-\bigl[\Lambda_1(r')-\Lambda_1(r)\bigr]
\sum_{s=2}^{T}\beta^{s-1}\varrho^{2(s-1)}\ge0.
\end{aligned}
\end{equation}
The inequality follows from $\varrho\ge0$ and the nonincreasing inverse hazard rate. Define the initial truthful utility and initial transfer by $V_1^{*}(\tau)=\int_{\underline\tau_1}^{\tau} \partial_v G(v,v)\,dv,\; p_1(r)=G(r,r)-V_1^{*}(r).$ For every $\tau,r$, absolute continuity and \eqref{eq:benchmark-initial-single-crossing} imply the global integral-monotonicity identity
\begin{equation}\label{eq:benchmark-initial-gim}
V_1^{*}(\tau)-\bigl[G(\tau,r)-p_1(r)\bigr]
=\int_r^\tau\bigl[\partial_vG(v,v)-\partial_vG(v,r)\bigr]dv\ge0.
\end{equation}
Increasing differences gives the sign for either integration direction. Thus initial reporting is truthful, $V_1^{*}(\underline\tau_1)=0$, and the same type response $\varrho^{s-1}$ yields the reduced virtual-type adjustment.

In hidden-effort environments, $\mathbb P(d_{s+1}=1\mid e_s)=1/2$ by \eqref{eq:benchmark-signal-value}. Because later types, allocations, and payments depend on effort only through these observations, deviations leave continuation distributions unchanged and raise current costs for $e_s>0$; hence $e_s=0$ is optimal at every history. Contractible effort is enforced otherwise. Together with \eqref{eq:benchmark-later-report-ic} and \eqref{eq:benchmark-initial-gim}, Theorem~\ref{thm:recursive-ic} establishes JPBIC after arbitrary histories, with transfers implementing the stated output rules and initial rents.

\begin{proposition}[Dynamic distortion hierarchy]
\label{prop:benchmark-fixed-disclosure}
In the quadratic environment of Example 1, under the initial-type envelope, integrability, participation and monotonicity conditions stated above, the globally optimal effort and output rules for all four contracting regimes are given by \eqref{eq:benchmark-action-rules} and \eqref{eq:benchmark-fixed-contracts}, together with the displayed date-1 rules. The implementing transfers constructed above attain the full-history payoff bound \eqref{eq:benchmark-full-history-upper}; the four maximized principal values are given in \eqref{eq:benchmark-value-formulas}.
\end{proposition}

\paragraph{Distortions and transaction costs.}
\emph{Output distortions.} By \eqref{eq:benchmark-fixed-contracts} and \eqref{eq:benchmark-action-rules}, at a common $(\tau^T,d^T)$,
\begin{equation}\label{eq:benchmark-allocation-distortion}
q_t^{1B}-q_t^{2B^{\mathrm{HI}}}
=q_t^{2B^{\mathrm{HA}}}-q_t^{3B}
=\Lambda_1(\tau_1)\varrho^{t-1},\qquad t\le T.
\end{equation}
For $t\ge2$,
\[
q_t^{1B}-q_t^{2B^{\mathrm{HA}}}
=q_t^{2B^{\mathrm{HI}}}-q_t^{3B}
=\theta_t^{B}\delta_{t-1}
\left(\sqrt{1/4+e_{t-1}^{1B}}-\frac12\right)d_t.
\]
Thus the informational distortion is $\Lambda_1(\tau_1)\varrho^{t-1}\ge0$, whereas the hidden-action distortion has the sign of $d_t$ whenever $\delta_{t-1}e_{t-1}^{1B}>0$; no pointwise output ranking follows from the latter.

\emph{Optimal principal values.}
By $\E[d_t\mid\tau^T]=0$, $d_t^2=1$, and \eqref{eq:benchmark-signal-value}, conditional quadratic maximization yields $\frac12(\tau_t-\Lambda_1(\tau_1)\varrho^{t-1})^2$ in the private-type regimes (and $\tau_t^2/2$ otherwise), plus the posterior-variance contribution. After substituting \eqref{eq:benchmark-action-rules} and effort costs, with integrals over $\mathbb P^\tau$,
\begin{equation}\label{eq:benchmark-value-formulas}
\begin{aligned}
\Pi_{1B}^{*}(\delta)&=\int\sum_{t=1}^T\frac{\beta^{t-1}\tau_t^2}{2}\,d\mathbb P^\tau
+\sum_{t=1}^{T-1}\beta^t\frac{{\theta_{t+1}^{B}}^2\delta_t^2}{2}(1/4+e_t^{1B})
-\sum_{t=1}^{T-1}\beta^{t-1}\frac{(e_t^{1B})^2}{2\kappa_t},\\
\Pi_{\mathrm{HI}}^{*}(\delta)&=\int\sum_{t=1}^T\frac{\beta^{t-1}(\tau_t-\Lambda_1(\tau_1)\varrho^{t-1})^2}{2}\,d\mathbb P^\tau
+\sum_{t=1}^{T-1}\beta^t\frac{{\theta_{t+1}^{B}}^2\delta_t^2}{2}(1/4+e_t^{2B^{\mathrm{HI}}})
-\sum_{t=1}^{T-1}\beta^{t-1}\frac{(e_t^{2B^{\mathrm{HI}}})^2}{2\kappa_t},\\
\Pi_{\mathrm{HA}}^{*}(\delta)&=\int\sum_{t=1}^T\frac{\beta^{t-1}\tau_t^2}{2}\,d\mathbb P^\tau
+\sum_{t=1}^{T-1}\beta^t\frac{{\theta_{t+1}^{B}}^2\delta_t^2}{8},\\
\Pi_{3B}^{*}(\delta)&=\int\sum_{t=1}^T\frac{\beta^{t-1}(\tau_t-\Lambda_1(\tau_1)\varrho^{t-1})^2}{2}\,d\mathbb P^\tau
+\sum_{t=1}^{T-1}\beta^t\frac{{\theta_{t+1}^{B}}^2\delta_t^2}{8}.
\end{aligned}
\end{equation}

\emph{Transaction costs.}
Subtracting \eqref{eq:benchmark-value-formulas} and using
\[
\tau_t^2-\left(\tau_t-\Lambda_1(\tau_1)\varrho^{t-1}\right)^2
=\Lambda_1(\tau_1)\varrho^{t-1}\left[2\tau_t-\Lambda_1(\tau_1)\varrho^{t-1}\right]\ge0
\]
under the stated positivity and monotonicity conditions gives
\begin{equation}\label{eq:benchmark-transaction-interaction}
\begin{aligned}
\Pi_{1B}^{*}(\delta)-\Pi_{\mathrm{HI}}^{*}(\delta)
&=\Pi_{\mathrm{HA}}^{*}(\delta)-\Pi_{3B}^{*}(\delta)
=\frac12\int\sum_{t=1}^T\beta^{t-1}
\left[\tau_t^2-(\tau_t-\Lambda_1(\tau_1)\varrho^{t-1})^2\right]d\mathbb P^\tau,\\
\Pi_{1B}^{*}(\delta)-\Pi_{\mathrm{HA}}^{*}(\delta)
&=\Pi_{\mathrm{HI}}^{*}(\delta)-\Pi_{3B}^{*}(\delta)
=\sum_{t=1}^{T-1}\beta^{t-1}
\frac{\kappa_t\beta^2{\theta_{t+1}^{B}}^4\delta_t^4}{8},\\
\Pi_{\mathrm{HI}}^{*}(\delta)+\Pi_{\mathrm{HA}}^{*}(\delta)
&-\Pi_{1B}^{*}(\delta)-\Pi_{3B}^{*}(\delta)=0.
\end{aligned}
\end{equation}
The first loss is strictly positive if $\Lambda_1(\tau_1)\varrho^{t-1}>0$ with positive probability for some $t$; the second is strictly positive if $\delta_t>0$ for some $t<T$. Hence $\Pi_{\mathrm{HI}}^{*}(\delta)-\Pi_{\mathrm{HA}}^{*}(\delta)$ has no fixed sign. Zero interaction here follows from the separability of $\Lambda_1(\tau_1)\varrho^{t-1}$ and $e_t$; it does not hold generally.

\begin{remark}[Interaction of incentive frictions]\label{rem:interaction-frictions}
The regime ordering of Proposition~\ref{prop:screening-benchmark-order} does not determine the sign of $\Pi_{\mathrm{HI}}^{*}+\Pi_{\mathrm{HA}}^{*}-\Pi_{1B}^{*}-\Pi_{3B}^{*}$. Consider two two-date, one-agent specifications with $\tau\sim U[0,1]$, degenerate states, uninformative kernels, and zero terminal payoffs; all decisions occur at date~1. Gross and net payoffs are $(u_1,u^P)$ and $(u_1-p,u^P+p)$. Actions are contractible in $1B,2B^{\mathrm{HI}}$ and hidden with no informative public consequences in $2B^{\mathrm{HA}},3B$; the agent's participation payoff is zero.

\emph{Positive interaction.} Let $q,e\in\{0,1\}$, $q+e\le1$, with contractible $q$ and $(u_1,u^P)=(\tau q-e,3e)$. The total surplus satisfies $\tau q+2e\le2$, attained by $(q,e,p)=(0,1,-1)$ in $1B$ and $2B^{\mathrm{HI}}$. Under hidden $e$, $e=0$ strictly dominates $e=1$, so $\Pi_{\mathrm{HA}}^{*}=\E\tau=1/2$. For $3B$, global reporting IC and participation imply $(V_1^{*})'(\tau)=q(\tau)$ a.e. and $V_1^{*}(0)\ge0$, giving
\[
\E p=\int_0^1(2\tau-1)q(\tau)\,d\tau-V_1^{*}(0)
\le\int_{1/2}^1(2\tau-1)\,d\tau=\frac14.
\]
The posted-price contract $q(\tau)=\mathbf1\{\tau\ge1/2\}$, $p(\tau)=\frac12\mathbf1\{\tau\ge1/2\}$ attains the bound. Therefore,
\[
(\Pi_{1B}^{*},\Pi_{\mathrm{HI}}^{*},\Pi_{\mathrm{HA}}^{*},\Pi_{3B}^{*})
=(2,2,\tfrac12,\tfrac14),\quad
\Pi_{\mathrm{HI}}^{*}+\Pi_{\mathrm{HA}}^{*}-\Pi_{1B}^{*}-\Pi_{3B}^{*}=\tfrac14>0.
\]

\emph{Negative interaction.} Let $e\in\{0,1\}$ and $(u_1,u^P)=((\tau-2)e,4e)$. In $1B$, $(e,p)=(1,\tau-2)$ gives $\Pi_{1B}^{*}=\E(2+\tau)=5/2$. In $2B^{\mathrm{HI}}$, global reporting IC and participation imply $(V_1^{*})'(\tau)=e(\tau)$ a.e. and $V_1^{*}(0)\ge0$, so
\[
\E(4e+p)=\int_0^1(1+2\tau)e(\tau)\,d\tau-V_1^{*}(0)
\le\int_0^1(1+2\tau)\,d\tau=2,
\]
attained by $(e,p)=(1,-2)$. Under hidden $e$, $(\tau-2)e<0$ for $e=1$ and all $\tau\in[0,1]$; hence $e=0$ after every report and $\Pi_{\mathrm{HA}}^{*}=\Pi_{3B}^{*}=0$. Consequently,
\[
(\Pi_{1B}^{*},\Pi_{\mathrm{HI}}^{*},\Pi_{\mathrm{HA}}^{*},\Pi_{3B}^{*})
=(\tfrac52,2,0,0),\quad
\Pi_{\mathrm{HI}}^{*}+\Pi_{\mathrm{HA}}^{*}-\Pi_{1B}^{*}-\Pi_{3B}^{*}=-\tfrac12<0.
\]
The displayed contracts satisfy reporting IC, participation, and hidden-action obedience after every report. Bounded finite-horizon payoffs ensure integrability and attainment.
\end{remark}

\section{Joint Optimal Information Technology and Disclosure Design}
\label{sec:optimal-disclosure}
\label{sec:benchmark}

Disclosure is not a direct argument of primitive payoff functions; through the induced information process, it affects continuation values, incentives, and implementable outcomes.\footnote{This separation between economic outcome design and information design is consistent with the integrated information-design perspective in \citet{BergemannMorris2019}; our setting additionally allows endogenous information production and privately implemented actions.}

Information production and disclosure shape continuation incentives and the evolution of economic opportunities. Building on the preceding implementation and optimal-contracting results, we study their joint design with contracting under dynamic incentive constraints.

\subsection{Information-Production Possibilities and Feasible Disclosure}
\label{subsec:disclosure-joint-design}

For $t<T$, the information technology $\mathcal Y_{t+1}$ and disclosure kernel $\mathcal D_{t+1}$ induce the joint distribution of disclosed information $d_{t+1}$ and private signals $s_{t+1}$:
\begin{equation}\label{eq:disclosure-public-experiment}
\begin{aligned}
&\mathbb P_{t+1}^{\mathcal Y,\mathcal D}(ds,dd\mid\theta^{t+1},\tau^t,x^t,h_t,\hat\vartheta^t,\px_{t},p_t)\\
&\quad=\int_{\overline{Y}_{t+1}}
\mathcal Y_{t+1}(dy,ds\mid\theta^{t+1},\tau^t,x^t,d^t)
\mathcal D_{t+1}(dd\mid h_t,\hat\vartheta^t,\px_{t},p_t,y).
\end{aligned}
\end{equation}
The identity holds for realized $x^t$, including deviations, while $\mathcal D_{t+1}$ conditions only on mechanism-observable information. Conditional probabilities follow $\mathbb P_W$ given $(h_t,\hat\vartheta^t,\px_{t},p_t)$.

For a fixed disclosure class $D$, define the set of \emph{feasible joint information kernels} by
\begin{equation}\label{eq:dynamic-information-feasibility}
FI(\mathcal Y):=\left\{\bigl(\mathbb P_{t+1}^{\mathcal Y,\mathcal D}\bigr)_{t<T}:\mathcal D\in D\right\}.
\end{equation}
\emph{Dynamic information feasibility} requires the same $\mathcal D\in D$ throughout the conditioning domains in \eqref{eq:disclosure-public-experiment}, including reporting and choice deviations. Whereas \citet{DovalEly2020} and \citet{MakrisRenou2023} characterize equilibrium outcomes across information structures, here the structures must additionally be generated by the specified information-production technology and committed disclosure rules. Feasibility of separately specified posteriors does not by itself establish feasibility of the committed dynamic design. When comparing different disclosure classes, write $FI(\mathcal Y;D)$ for this set.

\begin{proposition}[Information Feasibility and Bayes Plausibility]
\label{prop:disclosure-garbling}
Under the preceding information technology and disclosure specification, every disclosure rule $\mathcal D$ satisfies:
\begin{enumerate}
\item \textbf{Blackwell garbling.} For every $(\theta^{t+1},\tau^t,x^t)$,
\[
\begin{aligned}
&\mathbb P_{t+1}^{\mathcal Y,\mathcal D}(dd\mid\theta^{t+1},\tau^t,x^t,h_t,\hat\vartheta^t,\px_{t},p_t)\\
&\quad=\int_{\overline{Y}_{t+1}}\mathcal D_{t+1}(dd\mid h_t,\hat\vartheta^t,\px_{t},p_t,y)
\,\mathcal Y_{t+1}(dy,{S}_{t+1}\mid\theta^{t+1},\tau^t,x^t,d^t).
\end{aligned}
\]
Conversely, any such factorization defines a disclosure rule when its kernel belongs to the specified disclosure class.

\item \textbf{Bayes plausibility.} For every measurable $B\subseteq\Theta^{t+1}\times\mathcal T^t\times\overline X^t$,
\begin{equation}\label{eq:disclosure-bayes-plausibility}
{\small
\begin{aligned}
\int_{\overline{D}_{t+1}}\mathbb P(B\mid h_t,\hat\vartheta^t,\px_{t},p_t,d;\Gamma)\,\mathbb P_{t+1}^{\mathcal Y,\mathcal D}(dd\mid h_t,\hat\vartheta^t,\px_{t},p_t;\Gamma) =\mathbb P(B\mid h_t,\hat\vartheta^t,\px_{t},p_t;\Gamma).
\end{aligned}
}
\end{equation}
\item \textbf{Insufficiency of Bayes plausibility.} Bayes plausibility does not imply the Blackwell factorization in part~(1).
\end{enumerate}
\end{proposition}

\begin{proof}
For part~(1), marginalizing \eqref{eq:disclosure-public-experiment} over $s$ gives the stated factorization. Part~(2) follows from the law of iterated expectations. For part~(3), a conditionally uninformative technology cannot generate a nontrivial posterior split, even if it satisfies Bayes plausibility.
\end{proof}

\subsection{Existence and Continuity of Optimal Design}
\label{subsec:joint-production-design}

Choosing $\mathcal Y$ changes information-production possibilities; choosing $x_{it}$ is an action within that technology. The values in Section~\ref{sec:optimal-screening} retain IR, JPBIC and all payoff costs.

\begin{assumption}[Regularity for joint optimal design]\label{ass:joint-design-regularity}
The technology and disclosure classes $Y,D$ and the space of allocation--transfer rule pairs are nonempty compact metric spaces. Let $\mathscr C(\mathcal Y,\mathcal D)$ denote the allocation--transfer pairs satisfying IR and JPBIC under $(\mathcal Y,\mathcal D)$. The graph of $\mathscr C$ is nonempty and closed in the product of these spaces, and $\Pi(\Gamma)$ is continuous on this graph.
\end{assumption}

\begin{theorem}[Existence and Continuity of Optimal Values]
\label{thm:joint-design-existence}
Under Assumption~\ref{ass:joint-design-regularity}, let
$\operatorname{dom}f:=\{z:f(z)\in\R\}$, treating
$\Pi_{3B}^{*}$ and $\Pi_{3B}^{**}$ as functions of
$(\mathcal Y,\mathcal D)$ and $\mathcal Y$, respectively,
for fixed $Y,D$.
\begin{enumerate}
\item
$\Pi_{3B}^{***}(Y,D)=
\max_{\mathcal Y\in Y}\Pi_{3B}^{**}(\mathcal Y,D)\in\R$.

\item The following domains are compact:
\[
\operatorname{dom}\Pi_{3B}^{*}
=\{(\mathcal Y,\mathcal D)\in Y\times D:
\mathscr C(\mathcal Y,\mathcal D)\ne\varnothing\},
\;\;
\operatorname{dom}\Pi_{3B}^{**}
=\{\mathcal Y\in Y:\exists\,\mathcal D\in D,\
\mathscr C(\mathcal Y,\mathcal D)\ne\varnothing\}.
\]
On these domains, respectively,
\[
\Pi_{3B}^{*}(\mathcal Y,\{\mathcal D\})
=\max_{(\mathcal X,\mathcal P)\in
\mathscr C(\mathcal Y,\mathcal D)}\Pi(\Gamma),
\quad
\Pi_{3B}^{**}(\mathcal Y,D)
=\max_{\mathcal D\in D}
\Pi_{3B}^{*}(\mathcal Y,\{\mathcal D\}),
\]
with value $-\infty$ otherwise.

\item $\Pi_{3B}^{*}$ and $\Pi_{3B}^{**}$ are upper
semicontinuous on $Y\times D$ and $Y$, respectively.
On their finite-valued domains, continuity at
$(\mathcal Y,\mathcal D)$ and $\mathcal Y$, respectively,
is equivalent to
\[
\begin{gathered}
\forall(\mathcal Y^k,\mathcal D^k)\to
(\mathcal Y,\mathcal D),\
\exists(\mathcal X^k,\mathcal P^k)\in
\mathscr C(\mathcal Y^k,\mathcal D^k):
\ \liminf_k\Pi(\Gamma^k)\geq
\Pi_{3B}^{*}(\mathcal Y,\{\mathcal D\});\\
\forall\mathcal Y^k\to\mathcal Y,\
\exists\mathcal D^k\in D,\
(\mathcal X^k,\mathcal P^k)\in
\mathscr C(\mathcal Y^k,\mathcal D^k):
\ \liminf_k\Pi(\Gamma^k)\geq
\Pi_{3B}^{**}(\mathcal Y,D).
\end{gathered}
\]
Here the parameter sequences lie in the respective domains,
feasibility holds for all $k$, and
$\Gamma^k=(\mathcal V^T;\allowbreak\mathcal X^k,
\allowbreak\mathcal P^k,\allowbreak\mathcal Y^k,
\allowbreak\mathcal D^k)$.

\item The sets
$\arg\max_{\mathcal Y\in Y}\Pi_{3B}^{**}(\mathcal Y,D)$
and
$\arg\max_{\mathcal D\in D}
\Pi_{3B}^{*}(\mathcal Y,\{\mathcal D\})$,
$\mathcal Y\in\operatorname{dom}\Pi_{3B}^{**}$,
are nonempty and compact.

\item On the respective finite-valued domains,
\[
(\mathcal Y,\mathcal D)\mapsto
\arg\max_{(\mathcal X,\mathcal P)\in
\mathscr C(\mathcal Y,\mathcal D)}\Pi(\Gamma),
\quad
\mathcal Y\mapsto
\arg\max_{\substack{\mathcal D\in D,\
(\mathcal X,\mathcal P)\in
\mathscr C(\mathcal Y,\mathcal D)}}\Pi(\Gamma)
\]
are upper hemicontinuous wherever $\Pi_{3B}^{*}$ and
$\Pi_{3B}^{**}$ are continuous, respectively; so is
$\mathcal Y\mapsto\arg\max_{\mathcal D\in D}
\Pi_{3B}^{*}(\mathcal Y,\{\mathcal D\})$
wherever $\Pi_{3B}^{**}$ is continuous.
\end{enumerate}
\end{theorem}

\begin{proof}
By Assumption~\ref{ass:joint-design-regularity}, the graph, its sections at fixed parameters and its projections are compact. Continuity of $\Pi$ and the Weierstrass theorem give (1)--(2). For either conditional value, take a sequence of parameters converging to a limit and, along a subsequence attaining the limit superior of the values, select maximizing mechanisms whenever the value is finite. If the limit superior is $-\infty$, the upper semicontinuity inequality is immediate. Otherwise compactness yields a convergent subsequence of mechanisms, whose limit is in the graph of $\mathscr C$ at the limiting parameters by closedness. Continuity of $\Pi$ bounds the limit superior by the value there. The inequalities in (3) are therefore sufficient for continuity on the finite-valued domains, since each selected mechanism's payoff is bounded above by the corresponding optimized value. Necessity follows by selecting maximizing mechanisms. Upper semicontinuity and attainment give (4). For (5), any convergent subsequence of optimal rules at converging parameters has a limit in the graph of $\mathscr C$ by closedness; continuity of $\Pi$ and of the corresponding optimal value makes the limit optimal. Compactness of the common rule spaces gives upper hemicontinuity, which is preserved under projection onto the disclosure coordinate.
\end{proof}

The conditions in Theorem~\ref{thm:joint-design-existence}(3) are sequential forms of transfer lower semicontinuity along feasible paths (\citealp{TianZhou1992,TianZhou1995}). A sufficient condition for $\Pi_{3B}^{*}(\mathcal Y,\{\mathcal D\})$ is that, for every $\epsilon>0$, some $(\mathcal X,\mathcal P)\in\mathscr C(\mathcal Y,\mathcal D)$ yields $\Pi(\Gamma)\geq\Pi_{3B}^{*}(\mathcal Y,\{\mathcal D\})-\epsilon$ and belongs to $\mathscr C(\mathcal Y',\mathcal D')$ for all nearby $(\mathcal Y',\mathcal D')\in\operatorname{dom}\Pi_{3B}^{*}$. For $\Pi_{3B}^{**}(\mathcal Y,D)$, the analogous condition holds with some $\mathcal D\in D$ fixed as $\mathcal Y$ varies in $\operatorname{dom}\Pi_{3B}^{**}$ and $\Pi_{3B}^{**}(\mathcal Y,D)-\epsilon$ as the payoff bound. Payoff continuity and $\epsilon\downarrow0$ yield (3).

In finite-horizon environments with finitely many histories and outcomes, continuity of the induced one-step probabilities and stage-payoff entries in the rule coordinates suffices for payoff continuity: $\Pi$ is a finite sum of products of these probabilities and payoffs. These are additional conditions on the design problem, not consequences of continuation-payoff regularity. In dynamic environments, unbounded transfers and changes in kernel supports can affect compactness and closedness, respectively; in particular, Bayesian consistency alone does not ensure closedness of the off-path restrictions in Definition~\ref{def:multiagent-pbe}.

\subsection{Institutional Value of Information Technologies}\label{subsec:institutional-dominance}

For a common disclosure class $D$, define \emph{institutional dominance} by
\begin{equation}\label{eq:institutional-dominance}
\mathcal Y'\succeq\mathcal Y\quad\text{if and only if}\quad FI(\mathcal Y)\subseteq FI(\mathcal Y').
\end{equation}
Institutional dominance compares attainable joint information laws, not selected disclosure rules.

For comparisons of two designs $(\mathcal Y,\mathcal D)$ and $(\mathcal Y',\mathcal D')$, keep the primitive transition kernels, payoff functions, decision histories, and contractual restrictions common. For each allocation--transfer pair satisfying the common contractual restrictions, require that every belief system consistent with Definition~\ref{def:multiagent-pbe}'s belief restrictions under truthful reporting and recommended choices in one mechanism admit a belief system satisfying those restrictions in the other, and conversely, with the same marginal distributions of past and current economic uncertainty excluding produced information $y_2^t$ at every pre- and post-report history. Specifically, these marginals are over $(\theta^t,\tau^t,s^t,\px^{t-1},x^{t-1},p^{t-1},d^t)$ before reports and additionally $(\px_t,p_t)$ after reports. This condition concerns entire belief systems, including histories at which Bayes' rule does not determine beliefs.

\begin{theorem}[Institutional Value and Dominance of Information Technologies]\label{prop:disclosure-capacity}
Under the common primitives and contractual restrictions described above:
\begin{enumerate}
\item Institutional dominance is a preorder, with $\mathcal Y\succeq\mathcal Y'\text{ and }\mathcal Y'\succeq\mathcal Y\iff FI(\mathcal Y)=FI(\mathcal Y')$.
\item If $\mathbb P_{t+1}^{\mathcal Y,\mathcal D}=\mathbb P_{t+1}^{\mathcal Y',\mathcal D'}$ for all $t<T$ at every history in the common conditioning domain, and the two designs satisfy the stated belief-system correspondence, then $\mathscr C(\mathcal Y,\mathcal D)=\mathscr C(\mathcal Y',\mathcal D')$. Principal payoffs coincide for each pair in the common allocation--transfer set; consequently, $\Pi_{3B}^{*}(\mathcal Y,\{\mathcal D\})=\Pi_{3B}^{*}(\mathcal Y',\{\mathcal D'\}).$ \item For fixed $\mathcal Y$, $D\subseteq D'$ implies $\Pi_{3B}^{**}(\mathcal Y,D)\leq\Pi_{3B}^{**}(\mathcal Y,D')$. If $FI(\mathcal Y;D)\subseteq FI(\mathcal Y';D')$ and every $\mathcal D\in D$ has a replicating $\mathcal D'\in D'$ satisfying the belief-system correspondence in (2), then $\Pi_{3B}^{**}(\mathcal Y,D)\leq\Pi_{3B}^{**}(\mathcal Y',D').$ Replication in both directions with the stated correspondence implies $\Pi_{3B}^{**}(\mathcal Y,D)=\Pi_{3B}^{**}(\mathcal Y',D').$
\item If $\mathcal Y^*\in Y$ institutionally dominates every $\mathcal Y\in Y$ and the corresponding replications satisfy the condition in (3) for the common class $D$, then $\Pi_{3B}^{**}(\mathcal Y^*,D)=\Pi_{3B}^{***}(Y,D)$. Under Assumption~\ref{ass:joint-design-regularity}, a joint optimum with technology $\mathcal Y^*$ is attained.
\end{enumerate}
\end{theorem}

\begin{proof}
Set inclusion gives (1). For (2), integrating $y_{t+1}$ out of \eqref{eq:induced-kernel-composition} leaves the same transition law for the remaining coordinates under both designs. With common initial distributions and allocation--transfer rules, iteration of these kernels, together with the same reporting and choice strategies, gives identical laws of economic paths excluding produced information and of the associated report histories. The assertion holds under arbitrary strategy profiles and, if $T=\infty$, follows for the path laws from equality of all finite-dimensional distributions. Starting at any decision history, the specified belief-system correspondence gives identical marginal distributions of the past and current coordinates entering the subsequent kernels and strategies. Iteration then gives the same conditional distributions of payoff-relevant coordinates under every continuation deviation. Thus the inequalities in Definition~\ref{def:multiagent-pbe} and the participation constraints coincide for corresponding beliefs. This proves both directions of implementability; equality of the principal's payoff follows from the common economic-path law and payoff functions. Part~(3) follows from $D\subseteq D'$ for fixed technology and, for different technologies, from part~(2) applied to the replicating disclosure rules and taking suprema. Part (4) follows by taking the supremum over $Y$; under Assumption~\ref{ass:joint-design-regularity}, Theorem~\ref{thm:joint-design-existence} ensures attainment at the dominant technology.
\end{proof}

For $\mathcal Y'=(\mathcal Y'_{t+1})_{t<T}$, suppose a disclosure-stage kernel $\mathcal B_{t+1}$ satisfies, at every primitive and realized-allocation history,
\[
\begin{aligned}
\mathcal Y_{t+1}(dy,ds\mid\theta^{t+1},\tau^t,x^t,d^t)
=\int_{\overline{Y}'_{t+1}}&\mathcal B_{t+1}(dy\mid \px^t,p^t,d^t,\hat\vartheta^t,y')\\
&\cdot \mathcal Y'_{t+1}(dy',ds\mid\theta^{t+1},\tau^t,x^t,d^t).
\end{aligned}
\]
The equality holds jointly on $\overline{Y}_{t+1}\times\Sg_{t+1}$ at every primitive and realized-allocation history. Since $\mathcal B_{t+1}$ uses only disclosure-stage information and $y'$, the composed joint information kernel preserves private-signal correlations along truthful and deviating paths; matching raw-information marginals alone would not suffice.
Let $ D'$ be the class of disclosure rules considered under $\mathcal Y'$. Suppose $ D'$ contains, for every $\mathcal D\in D$, the compositions
\[
\mathcal D'_{t+1}(dd\mid \px^t,p^t,d^t,\hat\vartheta^t,y')=\int_{\overline{Y}_{t+1}}\mathcal D_{t+1}(dd\mid \px^t,p^t,d^t,\hat\vartheta^t,y)\mathcal B_{t+1}(dy\mid \px^t,p^t,d^t,\hat\vartheta^t,y').
\]
The primitive transition kernels and contractual restrictions remain common.
Substitution of the two identities into \eqref{eq:disclosure-public-experiment} and integration gives equality of the joint private-signal and disclosure kernels at every conditioning history. Thus these identities imply $FI(\mathcal Y;D)\subseteq FI(\mathcal Y';D')$. The dynamic value comparison additionally requires the belief-system correspondence in Theorem~\ref{prop:disclosure-capacity}, including off-path beliefs; public-information marginals alone do not preserve continuation incentives. Such marginals need not preserve private-signal correlations, and even joint-kernel replication does not by itself establish the required off-path belief correspondence. Neither institutional dominance nor the enlargement of $D$ orders the value of mandatory full disclosure relative to a particular garbling.\footnote{If a change in technology entails additional costs not represented by the common payoff functions, the inclusion of information-kernel sets alone does not establish a net-payoff comparison.}

For fixed $\mathcal D$, Theorem~\ref{thm:recursive-ic} characterizes JPBIC and participation through \eqref{eq:recursive-action}, \eqref{eq:recursive-reporting}, and \eqref{eq:participation}. Commitment to the entire contingent mechanism retains these constraints at every decision history. The comparison concerns attainable joint information kernels after deviations, linking Blackwell garbling and Bayes plausibility \citep{KamenicaGentzkow2011,BergemannMorris2019} to dynamic implementation \citep{PavanSegalToikka2014,Sannikov2008}.

\subsection{Characterization of Optimal Information Design}
\label{subsec:disclosure-characterization}

Consider the finite-horizon version of the model with finite state, type, signal, allocation, payment, raw-information and disclosed-information spaces. Fix allocation--transfer rules satisfying the specified contractual restrictions, truthful reporting, recommended choice strategies and a belief system $\nu$ on the common decision-history spaces. For each information design, let $\Delta^{\mathrm{IC}}(\Gamma;\nu)$ consist of
\[
\begin{aligned}
&V_{it}^{-}((\vartheta_{i,t}^{T}(\cdot),\px_{i,t}^{T}(\cdot)))-V_{it}^{-}((\widehat\vartheta_{i,t}^{T}(\cdot),\widehat x_{i,t}^{T}(\cdot))),\\
&V_{it}((\vartheta_{i,t+1}^{T}(\cdot),\px_{i,t}^{T}(\cdot)))-V_{it}((\widehat\vartheta_{i,t+1}^{T}(\cdot),\widehat x_{i,t}^{T}(\cdot))),
\end{aligned}
\]
where the respective pre- and post-report history arguments are suppressed, and the components range over every $i,t$, history and unilateral continuation deviation in Definition~\ref{def:multiagent-pbe}, evaluated in the direct mechanism. Let $\Delta^{\mathrm{IR}}(\Gamma;\nu)$ consist of $V_{i1}^{-}(\tau_{i1},s_{i1},(\vartheta_{i,1}^{T}(\cdot),\px_{i,1}^{T}(\cdot)))$ for all $i\in N$, $(\tau_{i1},s_{i1})\in\T_{i1}\times\Sg_{i1}$, using the initial-history convention of \eqref{eq:participation}. Use $\mathbb P(\cdot;\Gamma)$ for the joint path and report-history law induced by these strategies. Let $\Delta^{\mathrm B}(\Gamma;\nu)$ consist of
\begin{equation}\label{eq:design-bayes-residual}
\mathbb P(\{w\},h;\Gamma)-\nu_{it}(\{w\}\mid h)\mathbb P(h;\Gamma),
\end{equation}
for every $w\in W$ and direct pre- or post-report history $h$ of every $i,t$ in the finite model. These equalities express Bayes consistency on the path of play; off-path support restrictions are imposed separately.

Restrict the candidate information designs to a set of kernel sequences on which the fixed $\nu$ respects observations, primitive restrictions and off-path kernel supports. Then implementation with this assessment and IR is equivalent to
\begin{equation}\label{eq:design-linear-constraints}
\Delta^{\mathrm{IC}}(\Gamma;\nu)\geq0,\quad
\Delta^{\mathrm{IR}}(\Gamma;\nu)\geq0,\quad
\Delta^{\mathrm B}(\Gamma;\nu)=0.
\end{equation}
Indeed, sequential rationality identifies the implemented interim payoffs with the optimized values in \eqref{eq:participation}. With multipliers $\lambda^{\mathrm{IC}},\lambda^{\mathrm{IR}}\geq0$ and unrestricted $\lambda^{\mathrm B}$ of the corresponding dimensions, define
\begin{equation}\label{eq:information-design-lagrangian}
\begin{aligned}
\mathcal L(\Gamma;\nu,\lambda)={}&\Pi(\Gamma)+\lambda^{\mathrm{IC}}\!\cdot\Delta^{\mathrm{IC}}(\Gamma;\nu)\\
&+\lambda^{\mathrm{IR}}\!\cdot\Delta^{\mathrm{IR}}(\Gamma;\nu)+\lambda^{\mathrm B}\!\cdot\Delta^{\mathrm B}(\Gamma;\nu),
\end{aligned}
\end{equation}
where $\lambda=(\lambda^{\mathrm{IC}},\lambda^{\mathrm{IR}},\lambda^{\mathrm B})$. Candidate mechanisms retain fixed allocation--transfer rules and strategies.

\begin{theorem}[Characterization of Optimal Information Design]\label{thm:information-design-characterization}
Under the preceding finite-model and assessment restrictions, let $\mathscr K$ denote the candidate set of mechanisms and let $\mathscr K_0:=\{\Gamma\in\mathscr K:\Gamma\text{ satisfies }\eqref{eq:design-linear-constraints}\}$.
\begin{enumerate}
\item Fix $t<T$ and all information kernels except $\mathcal Y_{t+1}$ or $\mathcal D_{t+1}$. If $\mathscr K$ is a nonempty polytope in the varying kernel entries and $\mathscr K_0\ne\varnothing$, $\max_{\Gamma\in\mathscr K_0}\Pi(\Gamma)$ is a linear program. For $\Gamma^*\in\mathscr K_0$, with $\lambda$ subject to the sign restrictions in \eqref{eq:information-design-lagrangian},
$\Gamma^*\in\arg\max_{\Gamma\in\mathscr K_0}\Pi(\Gamma)$ if and only if $\exists\,\lambda$ such that $\Gamma^*\in\arg\max_{\Gamma\in\mathscr K}\mathcal L(\Gamma;\nu,\lambda)$ and
\begin{equation}\label{eq:information-design-complementarity}
\lambda^{\mathrm{IC}}\cdot\Delta^{\mathrm{IC}}(\Gamma^*;\nu)=
\lambda^{\mathrm{IR}}\cdot\Delta^{\mathrm{IR}}(\Gamma^*;\nu)=0.
\end{equation}
A joint optimum satisfies this condition on every such $\mathscr K$ containing it, for any implementing belief system $\nu$.
\item If the candidate set in (1) for $\mathcal D_{t+1}$ is a product of probability simplices over allowed outcomes at each conditioning history in \eqref{eq:disclosure-kernel}, then $\Gamma^*\in\arg\max_{\Gamma\in\mathscr K}\mathcal L(\Gamma;\nu,\lambda)$ if and only if, at every such history and for all allowed $d,d'$,
$\mathcal D_{t+1}^*(d\mid\px^t,p^t,d^t,\hat\vartheta^t,y_{t+1})>0$ implies
\begin{equation}\label{eq:optimal-disclosure-support}
\frac{\partial\mathcal L}{\partial\mathcal D_{t+1}(d\mid\px^t,p^t,d^t,\hat\vartheta^t,y_{t+1})}
\geq
\frac{\partial\mathcal L}{\partial\mathcal D_{t+1}(d'\mid\px^t,p^t,d^t,\hat\vartheta^t,y_{t+1})},
\end{equation}
where the derivatives denote the coefficients of the conditional linear Lagrangian. Together with $\Gamma^*\in\mathscr K_0$ and \eqref{eq:information-design-complementarity}, this characterizes conditional optimal disclosure.
\item Allow all information kernels to vary in $\mathscr K$. For $\Gamma^*\in\mathscr K_0$ and multipliers as in (1) satisfying \eqref{eq:information-design-complementarity},
\[
\Gamma^*\in\arg\max_{\Gamma\in\mathscr K}\mathcal L(\Gamma;\nu,\lambda)
\implies \Gamma^*\in\arg\max_{\Gamma\in\mathscr K_0}\Pi(\Gamma).
\]
Separate kernel optimality generally does not suffice.
\end{enumerate}
\end{theorem}

\begin{proof}
For fixed strategies, each finite path probability is a product of primitive, allocation--transfer and information-kernel entries, with at most one entry of each period's technology and disclosure kernel in any product. Each conditional continuation payoff is a finite sum of such products for the subsequent path, weighted by the fixed marginal beliefs at its initial decision history. Since the same representation holds for every unilateral continuation deviation, $\Pi(\Gamma)$, $\Delta^{\mathrm{IC}}(\Gamma;\nu)$, $\Delta^{\mathrm{IR}}(\Gamma;\nu)$ and $\Delta^{\mathrm B}(\Gamma;\nu)$ are linear in each kernel separately (using the probability normalization). The feasible set in (1) is therefore a polytope. Linear-programming strong duality and complementary slackness give the necessary and sufficient conditions, with the candidate-set inequalities retained in the maximization of the Lagrangian. A joint optimum cannot be improved by a change of one kernel that preserves its contract and implementing assessment. This proves (1). In (2), the linear Lagrangian separates over the simplices; a probability vector maximizes its row's linear function precisely when its support lies in the coefficient-maximizing outcomes. For (3), for every candidate $\Gamma$ satisfying \eqref{eq:design-linear-constraints},
\[
\Pi(\Gamma)\leq\mathcal L(\Gamma;\nu,\lambda)\leq\mathcal L(\Gamma^*;\nu,\lambda)=\Pi(\Gamma^*),
\]
where the last equality follows from \eqref{eq:design-linear-constraints} and complementary slackness. Jointly varying kernels yields products of their entries, so the resulting expressions are linear in each kernel separately but generally nonlinear jointly. For example, $(0,0)$ is coordinatewise optimal but not globally optimal for $\max_{(a,b)\in[0,1]^2}ab$. Hence separate optimality alone does not yield global optimality.
\end{proof}

In the static model of \citet{TianOID2026}, conditional technology design is multilinear in independent marginal kernels, whereas conditional disclosure design is linear. Here, even disclosure kernels interact across periods: current disclosure changes future decision histories and the returns to subsequent information policies. Under the stated finite-model, fixed-contract and fixed-assessment restrictions, the objective and constraint residuals are linear in each kernel separately but generally nonlinear jointly; periodwise optimality need not imply joint optimality. Varying beliefs adds products in \eqref{eq:design-bayes-residual}, and changes in kernel supports may violate off-path restrictions. Neither conditional duality nor the fixed-assessment global sufficient condition characterizes unrestricted joint design. A deterministic Lagrangian maximizer in (2) need not satisfy \eqref{eq:design-linear-constraints}; optimal disclosure may require randomization.

If $Y$ is convex in a topological vector space and $\Pi_{3B}^{**}(\cdot,D)$ is real-valued and directionally differentiable along directions $\mathcal Y-\mathcal Y^*$, $\mathcal Y\in Y$, every optimal technology $\mathcal Y^*$ satisfies
\[
D_{\mathcal Y-\mathcal Y^*}\Pi_{3B}^{**}(\mathcal Y^*,D)\leq0,\qquad \forall\mathcal Y\in Y.
\]
Concavity makes this condition sufficient, and strict concavity ensures at most one optimizer. Likewise, for fixed $\mathcal Y$, strict concavity of the real-valued $\Pi_{3B}^{*}(\mathcal Y,\{\cdot\})$ on a convex $D$ ensures at most one optimal disclosure rule. Existence follows under Theorem~\ref{thm:joint-design-existence}'s conditions.

By definition,
\[
\Pi_{3B}^{***}(Y,D)\geq\Pi_{3B}^{**}(\mathcal Y,D)\geq\Pi_{3B}^{*}(\mathcal Y,\{\mathcal D\}),\qquad \mathcal Y\in Y,\ \mathcal D\in D.
\]
A no-information technology with finite value thus provides a lower bound when it belongs to $Y$.

For a committed scalar family, Theorem~\ref{thm:dynamic-virtual-surplus} identifies the marginal trade-off among reoptimized total surplus, screening rents, and participation rents under JPBIC.

Let $\mathcal D^\delta\in D$, $\delta\in[0,1]$, be a committed family. Assume each fixed-disclosure problem attains an optimal JPBIC, individually rational contract inducing $\mathbb P^\delta$, with value $\Pi_{3B}^{*}(\mathcal Y,\{\mathcal D^\delta\})$. Under the initial-type conditions of Theorem~\ref{thm:dynamic-virtual-surplus}, set
\[
\begin{aligned}
\mathcal S(\delta)&=\int_{W}\sum_{t=1}^{T}\beta^{t-1}\left(u_t^{P}+\sum_{i\in N}u_{it}\right)d\mathbb P^\delta,\\
\mathcal R_{\mathrm{IC}}(\delta)&=\sum_{i\in N}\int_{W}\Lambda_{i1}(\tau_{i1}\mid s_{i1})\partial_{\tau_{i1}}V_{i1}^{\delta}(\tau_{i1},s_{i1})\,d\mathbb P^\delta,\\
\mathcal R_{0}(\delta)&=\sum_{i\in N}\E_{s_{i1}}\left[V_{i1}^\delta(\underline\tau_{i1},s_{i1})\right].
\end{aligned}
\]
Superscript $\delta$ denotes the selected optimal contract.\footnote{Smooth dependence of a disclosure kernel on $\delta$ alone does not ensure differentiability of reoptimized value components; the latter is maintained here.} Assume $\mathcal S,\mathcal R_{\mathrm{IC}},\mathcal R_0$ are differentiable on $(0,1)$ with finite endpoint one-sided derivatives; concavity is not assumed.

Under Assumptions~\ref{ass:regularity} and \ref{ass:envelope-regularity} and the preceding scalar-family conditions,
\[
\Pi_{3B}^{*}(\mathcal Y,\{\mathcal D^\delta\})=\mathcal S(\delta)-\mathcal R_{\mathrm{IC}}(\delta)-\mathcal R_{0}(\delta),
\]
and hence, for $\delta\in(0,1)$,
\begin{equation}\label{eq:disclosure-rent-derivative}
\frac{d}{d\delta}\Pi_{3B}^{*}(\mathcal Y,\{\mathcal D^\delta\})
=\mathcal S'(\delta)-\mathcal R_{\mathrm{IC}}'(\delta)-\mathcal R_{0}'(\delta).
\end{equation}
The identity also holds for one-sided endpoint derivatives. Neither rent derivative has a predetermined sign.

Equation~\eqref{eq:disclosure-rent-derivative} measures the net marginal value of disclosure \emph{after contract reoptimization}, not the statistical value of information alone. Disclosure can raise total surplus without raising the principal's optimized payoff if additional screening and participation rents offset the gains; conversely, rent reductions can benefit the principal without increasing surplus. Information provision can itself serve as an intertemporal incentive instrument \citep{Ball2023}; with dynamic screening and endogenous information production, its value also depends on reporting rents and private-action implementation. Correlated information and cross-agent effects enter $\partial_{\tau_{i1}} V_{i1}^{\delta}$ through information production, type transitions, and reporting incentives.\footnote{Lowest-type continuation rents need not vanish without binding participation.}

For the following scalar characterization, assume $\delta\mapsto \Pi_{3B}^{*}(\mathcal Y,\{\mathcal D^\delta\})$ is continuous and concave on $[0,1]$. The endpoints represent nondisclosure and full disclosure only where applicable. The optimized scalar-family value equals $\Pi_{3B}^{**}(\mathcal Y, D)$ when $\{\mathcal D^\delta:\delta\in[0,1]\}= D$; for a proper subfamily it need not attain that value.

\paragraph{Optimal disclosure intensity.}

Under Assumption~\ref{ass:regularity} and the preceding scalar-family and concavity conditions, $\delta^*\in[0,1]$ solves
\begin{equation}\label{eq:optimal-disclosure-argmax}
\delta^*\in\arg\max_{\delta\in[0,1]}\Pi_{3B}^{*}(\mathcal Y,\{\mathcal D^\delta\})
\end{equation}
if and only if one of the following holds:
\begin{enumerate}
\item \textbf{Lower endpoint:} $\delta^*=0$ and $\mathcal S'_+(0)\leq{\mathcal R_{\mathrm{IC}}}'_+(0)+{\mathcal R_{0}}'_+(0)$.
\item \textbf{Interior:} $0<\delta^*<1$ and $\mathcal S'(\delta^*)=\mathcal R_{\mathrm{IC}}'(\delta^*)+\mathcal R_{0}'(\delta^*)$.
\item \textbf{Upper endpoint:} $\delta^*=1$ and $\mathcal S'_-(1)\geq{\mathcal R_{\mathrm{IC}}}'_-(1)+{\mathcal R_{0}}'_-(1)$.
\end{enumerate}
An interior optimum is selected when $\mathcal S'_+(0)>{\mathcal R_{\mathrm{IC}}}'_+(0)+{\mathcal R_0}'_+(0)$ and $\mathcal S'_-(1)<{\mathcal R_{\mathrm{IC}}}'_-(1)+{\mathcal R_0}'_-(1)$; strict concavity of the scalar optimized value makes it unique.

This follows from \eqref{eq:disclosure-rent-derivative} and concavity.

Parameter $\delta$ may index an entire ex ante committed, history-dependent rule. The scalar characterization compares fully reoptimized mechanisms subject to participation and JPBIC; historywise perturbations are covered only when they preserve all preceding constraints and the required concavity.

For a jointly implementable committed scalar family, suppose
\[
\Pi_{3B}^{*}(\mathcal Y,\{\mathcal D^\delta\})=\Pi_{3B}^{*}(\mathcal Y,\{\mathcal D^0\})+\delta\Delta,
\qquad\delta\in[0,1].
\]
The optimal disclosure set is
\[
\arg\max_{\delta\in[0,1]}\Pi_{3B}^{*}(\mathcal Y,\{\mathcal D^\delta\})=
\begin{cases}
\{0\},&\Delta<0,\\
[0,1],&\Delta=0,\\
\{1\},&\Delta>0.
\end{cases}
\]

Along induced histories, a sign change in marginal value can switch opacity to transparency only within a jointly implementable committed family.

The bang--bang conclusion requires a constant marginal optimized payoff, not merely linear variation of the disclosure experiment. Earlier JPBIC can preclude profitable historywise changes. Under \eqref{eq:dynamic-virtual-type}, disclosure may also change controlled impulse responses, virtual types, and allocations subject to global reporting and choice obedience.

Dynamic screening under given information \citep{PavanSegalToikka2014} and information design \citep{KamenicaGentzkow2011,BergemannMorris2019} meet here through the dependence of continuation rents and private-action incentives on $\mathcal D$. The rent identity requires neither concavity nor a constant marginal optimized payoff; the scalar characterization and bang--bang result use their respective additional restrictions.

\subsection{Institutional Substitution and Incentive Separation}
\label{subsec:disclosure-nonseparation-analysis}

Disclosure affects future economic opportunities through $\mathcal Q_{t+1}$, $\mathcal Y_{t+1}$ and $\mathcal F_{t+1}^{\tau}$, and decisions through information histories. Replicating current expected rewards by transfers, as in the static payoff-replication analysis of \citet{TianOID2026}, need not preserve these opportunities or continuation incentives.

For fixed $\mathcal Y$ and common allocation--transfer rules, consider two disclosure rules. Suppose, for every fixed allocation--transfer pair, their decision histories and continuation deviations coincide and their belief systems satisfying the belief restrictions of Definition~\ref{def:multiagent-pbe} correspond in both directions. If all pre- and post-report recommended-strategy minus deviation payoffs coincide under corresponding beliefs, JPBIC is invariant by that definition. If in addition the corresponding interim participation inequalities have the same truth values, $\mathscr C(\mathcal Y,\mathcal D)=\mathscr C(\mathcal Y,\mathcal D')$. \citet{ElyGeorgiadisRayo2025} show that withholding performance information can strengthen dynamic effort incentives. The preceding conditions identify when joint reporting and action incentives instead remain invariant. Neither fixed-deviation invariance nor JPBIC invariance alone ensures equality of implementable contract sets, and even common implementable sets need not preserve optimal contracts if the principal's payoff ordering changes. Thus preserving deviation incentives is distinct from preserving implementability, which in turn is distinct from preserving optimal contracting. Contracting and disclosure can be optimized separately if $\mathscr C(\mathcal Y,\mathcal D)=\mathscr C(\mathcal Y,\mathcal D')$ for all $\mathcal D,\mathcal D'\in D$ and the principal payoff is additively separable in $(\mathcal X,\mathcal P)$ and $\mathcal D$.

The interaction of $\mathcal Y_{t+1}$, $\mathcal D_{t+1}$, and privately implemented $x_{it}$ can change both choice responses and full-history reporting incentives. We examine local derivatives, cross-agent responses, and global JPBIC in turn.

As in Section~\ref{subsec:production-value}, fix all non-type private-history coordinates; $V_{it}((h_t,\vartheta_i^t,\widehat\vartheta_{it},\px_{t},p_t),\tau_{it},\cdot)$ abbreviates $V_{it}(h_t,\vartheta_i^t,\widehat\vartheta_{it},\px_{t},p_t,\cdot)$ at that history. The report space remains $\mathcal V_{it}$ and is not reduced to current types.

Fix $t<T$, $(h_t,\vartheta_i^t,\widehat\vartheta_{it},\px_{t},p_t)$, other agents' strategies, and all rules except $\mathcal D_{t+1}^{\delta}$, $\delta\in[0,1]$. Use $\big|_{\mathcal D_{t+1}^{\delta}}$ for evaluation of the post-report value $V_{it}((h_t,\vartheta_i^t,\widehat\vartheta_{it},\px_{t},p_t),\allowbreak \tau_{it},\allowbreak (\widehat\vartheta_{i,t+1}^{T}(\cdot),\allowbreak (x_i,\allowbreak x_{i,t+1}^{T}(\cdot))))$ of \eqref{eq:action-deviation-value} under disclosure $\mathcal D_{t+1}^{\delta}$.

Consider a scalar privately implemented choice $x_i$. With fixed history and continuation-strategy arguments suppressed, derivatives of $V_{it}$ are evaluated at $(x_{it}^*(\delta),\mathcal D_{t+1}^{\delta})$. Suppose $V_{it}$ is twice continuously differentiable in $(x_i,\delta)$ near an interior best response $x_{it}^*(\delta)$, with $\partial_{x_i}V_{it}=0$ and $\partial_{x_ix_i}^2V_{it}<0$.

\paragraph{Local choice responses.}
Under the above specification and Assumptions~\ref{ass:regularity} and \ref{ass:controlled-differentiability}, at each interior $\delta\in(0,1)$ along the local differentiable interior best-response branch,
\begin{equation}\label{eq:disclosure-action-response}
\frac{dx_{it}^*(\delta)}{d\delta}=-\frac{\partial_{x_i\delta}^2V_{it}}{\partial_{x_ix_i}^2V_{it}}.
\end{equation}
Thus $dx_{it}^*(\delta)/d\delta\ne0$ if and only if $\partial_{x_i\delta}^2V_{it}\ne0$.

The result follows by implicit differentiation of $\partial_{x_i}V_{it}=0$.

Equation~\eqref{eq:disclosure-action-response} is a local agent response, not principal reoptimization. For vector choices, $dx_{it}^*/d\delta=-[\nabla_{x_ix_i}^2V_{it}]^{-1}\nabla_{x_i\delta}^2V_{it}$ when the choice Hessian is nonsingular, under the analogous interior differentiability conditions.

Cross-agent information effects couple choice incentives. Restrict date-$t$ strategies to a differentiable finite-dimensional family $x_t(\cdot)=(x_{it}(\cdot))_{i\in N}$, fixing later strategies and the relevant private histories. At an interior $x_t^*(\delta)$ satisfying $\nabla_{x_{it}(\cdot)}V_{it}=0$ for all $i\in N$ and nonsingular $D_{x_t(\cdot)}(\nabla_{x_{it}(\cdot)}V_{it})_{i\in N}$, implicit differentiation yields
\begin{equation}\label{eq:coupled-action-response}
\frac{dx_t^*(\delta)}{d\delta}
=-\left[D_{x_t(\cdot)}\bigl(\nabla_{x_{it}(\cdot)}V_{it}\bigr)_{i\in N}\right]^{-1}
\frac{\partial}{\partial\delta}\bigl(\nabla_{x_{it}(\cdot)}V_{it}\bigr)_{i\in N},
\end{equation}
where all derivatives are evaluated at the specified private histories, $x_t^*(\delta)$, and $\mathcal D_{t+1}^{\delta}$. Off-diagonal Jacobian blocks measure cross-agent marginal-payoff effects. Equation~\eqref{eq:coupled-action-response} is local and finite-dimensional; full-history incentives remain governed by Theorem~\ref{thm:recursive-ic}.

The same interaction can generate \emph{continuation externalities} even when each agent's privately implemented choice is individually optimal. Fix $\mathbb P_{it}(\cdot\mid h_t,\vartheta_i^t,\widehat\vartheta_{it},\px_{t},p_t\,;\,\Gamma,x(\cdot),\nu_i^*,x_i)$ of Section~\ref{subsec:mechanisms-direct}, with $\mathcal D_{t+1}=\mathcal D_{t+1}^{\delta}$ and all other agents' choices held fixed. Whenever differentiation with respect to $x_i$ is justified, the marginal effect on continuation total surplus, less agent $i$'s own continuation payoff, satisfies
\begin{equation}\label{eq:cross-agent-production-externality}
\begin{aligned}
&\frac{\partial}{\partial x_i}\int\sum_{r=t+1}^{T}\beta^{r-t-1}\left(v_r+\sum_{j\in N}u_{jr}\right)d\mathbb P_{it}
-\frac{\partial}{\partial x_i}\int\sum_{r=t+1}^{T}\beta^{r-t-1}(u_{ir}-p_{ir})d\mathbb P_{it}\\
&\qquad=\frac{\partial}{\partial x_i}\int\sum_{r=t+1}^{T}\beta^{r-t-1}\left(v_r+\sum_{j\ne i}u_{jr}+p_{ir}\right)d\mathbb P_{it}.
\end{aligned}
\end{equation}
The sign of this difference is unrestricted.

Fix past and current marginal beliefs, allocation--transfer rules, subsequent strategies, and a one-period report--choice deviation at $(h_t,\vartheta_i^t)$. For $\mathcal E\in\{\mathcal D,\mathcal D'\}$, let $\mathbb P_{it}^{\mathcal E}$ and $\widehat{\mathbb P}_{it}^{\mathcal E}$ denote the conditional laws of Section~\ref{subsec:mechanisms-direct} at this history under disclosure $\mathcal E$, respectively for the recommended strategies and the fixed deviation. By \eqref{eq:continuation-plan}, for $r\in\{t,t+1\}$ define
\[
\Delta_{ir}^{\mathcal E}:=\int\sum_{s=r}^{T}\beta^{s-r}(u_{is}-p_{is})\,d\mathbb P_{it}^{\mathcal E}
-\int\sum_{s=r}^{T}\beta^{s-r}(u_{is}-p_{is})\,d\widehat{\mathbb P}_{it}^{\mathcal E}.
\]
Thus $\Delta_{it}^{\mathcal E}$ is the difference of the two pre-report continuation payoffs $V_{it}^{-}$, and $\Delta_{i,t+1}^{\mathcal E}$ is the corresponding expected continuation gain, conditional at date $t$. Compare disclosure rules differing only at or after the disclosure stage following date-$t$ choices, so current-payoff differences coincide.

\paragraph{Invariance of incentives.}

Under the above fixed-deviation specification and Assumption~\ref{ass:regularity}, $\Delta_{it}^{\mathcal D}-\Delta_{it}^{\mathcal D'}=\beta(\Delta_{i,t+1}^{\mathcal D}-\Delta_{i,t+1}^{\mathcal D'})$. Thus $\Delta_{it}^{\mathcal D}=\Delta_{it}^{\mathcal D'}$ if and only if $\Delta_{i,t+1}^{\mathcal D}=\Delta_{i,t+1}^{\mathcal D'}$.

The identity follows from equality of current-payoff differences; the equivalence follows from $\beta>0$.

Fixed-deviation invariance does not imply optimal-contract separability (Section~\ref{subsec:disclosure-nonseparation-analysis}).

\begin{remark}[Disclosure and incentives for information-producing choices]
Hidden-action contracting analyzes the implementation of privately chosen economic inputs, while dynamic feedback design examines how information release affects continuation incentives \citep{LaffontTirole1988,ElyGeorgiadisRayo2025}. The present setting links these incentives to the production of raw evidence, correlated private signals, and evolving screening types. Complementing the joint design of monitoring and compensation in \citet{GeorgiadisSzentes2020}, the local response, strategic interaction, and continuation-externality formulas identify how disclosure affects information-producing incentives as well as information use. Individually optimal choices need not internalize effects on others' future payoffs and the principal's continuation surplus. These channels must be evaluated through the JPBIC and optimal-contracting conditions; local responses alone do not establish nonseparability of the optimized problem.
\end{remark}

\smallskip
\noindent\textbf{Example 2 (Continuation of Example 1: Optimal Disclosure in the Quadratic Environment).}

Disclosure is optimized using Example 1's fixed-disclosure solutions, with all other primitives unchanged.

The principal commits at date $0$ to report-independent $\delta_t\in[0,1]$, $t<T$, at zero cost. Maintain Example 1's envelope, integrability, attainment, positive-output and monotonicity conditions, with $\kappa_t\beta{\theta_{t+1}^{B}}^2/2<\bar e_t\le1/2$.
\paragraph{Optimizing disclosure.}
By Proposition~\ref{prop:benchmark-fixed-disclosure}, conditional maximization of \eqref{eq:benchmark-project-surplus} gives
\[
\max_{e_t\in[0,\bar e_t]}
\left\{\frac{\beta(\theta_{t+1}^{B})^2\delta_t^2}{2}\left(\frac14+e_t\right)
-\frac{e_t^2}{2\kappa_t}\right\}.
\]
The conditions $\beta(\theta_{t+1}^{B})^2\delta_t^2/2-e_t/\kappa_t=0$ and $-1/\kappa_t<0$ yield $e_t^{1B}=e_t^{2B^{\mathrm{HI}}}=\kappa_t\beta(\theta_{t+1}^{B})^2\delta_t^2/2$ under the maintained interior bounds. Under hidden effort, the fixed-disclosure hidden-action deviation argument gives $e_t^{2B^{\mathrm{HA}}}=e_t^{3B}=0$ for all $\delta_t$.

The corresponding optimized date-$t$ informational contributions are
\[
\frac{\beta{\theta_{t+1}^{B}}^2\delta_t^2}{8}
+\frac{\kappa_t\beta^2{\theta_{t+1}^{B}}^4\delta_t^4}{8},
\qquad
\frac{\beta{\theta_{t+1}^{B}}^2\delta_t^2}{8}.
\]
Both expressions are strictly increasing in $\delta_t\in[0,1]$ since $\beta,\kappa_t,\theta_{t+1}^B>0$. By additive separability of \eqref{eq:benchmark-value-formulas} and JPBIC implementation for all $\delta\in[0,1]^{T-1}$:

\begin{enumerate}
\item \textbf{Optimal disclosure.} Full disclosure of the available evidence is uniquely optimal in every contracting environment:
\begin{equation}\label{eq:benchmark-disclosure-regimes}
\delta_t^{k,*}=1,\qquad
k\in\{1B,2B^{\mathrm{HI}},2B^{\mathrm{HA}},3B\},
\qquad t<T.
\end{equation}

\item \textbf{Optimal effort.} At the optimal disclosure rule,
\begin{equation}\label{eq:disclosure-benchmark-action}
e_t^{1B,*}=e_t^{2B^{\mathrm{HI}},*}
=\frac{\kappa_t\beta{\theta_{t+1}^{B}}^2}{2},
\qquad
e_t^{2B^{\mathrm{HA}},*}=e_t^{3B,*}=0,
\qquad t<T,
\end{equation}
with $e_T^{k,*}=0$ in every environment.

\item \textbf{Optimal output.} At date $1$,
\begin{equation}\label{eq:joint-disclosure-contracts}
q_1^{1B}=q_1^{2B^{\mathrm{HA}}}=\tau_1,
\qquad
q_1^{2B^{\mathrm{HI}}}=q_1^{3B}=\bigl(\tau_1-\Lambda_1(\tau_1)\bigr).
\end{equation}
For $t\ge2$,
\begin{equation}\label{eq:disclosure-benchmark-allocations}
\begin{aligned}
q_t^{1B}&=\tau_t+\theta_t^{B}\sqrt{1/4+\kappa_{t-1}\beta{\theta_t^{B}}^2/2}\,d_t,\\
q_t^{2B^{\mathrm{HI}}}&=\bigl(\tau_t-\Lambda_1(\tau_1)\varrho^{t-1}\bigr)+\theta_t^{B}\sqrt{1/4+\kappa_{t-1}\beta{\theta_t^{B}}^2/2}\,d_t,\\
q_t^{2B^{\mathrm{HA}}}&=\tau_t+\frac{\theta_t^{B}}{2}d_t,\\
q_t^{3B}&=\bigl(\tau_t-\Lambda_1(\tau_1)\varrho^{t-1}\bigr)+\frac{\theta_t^{B}}{2}d_t.
\end{aligned}
\end{equation}
These allocations satisfy the reporting and choice-obedience implementation conditions of Proposition~\ref{prop:benchmark-fixed-disclosure}.
\end{enumerate}

\noindent\textbf{Maximized principal values under optimal disclosure.}
By \eqref{eq:benchmark-value-formulas}, $\max_{\delta\in[0,1]^{T-1}}\Pi_k^{*}(\delta)=\Pi_k^{*}(1)$, where $k\in\{1B,2B^{\mathrm{HI}},2B^{\mathrm{HA}},3B\}$ and $1$ denotes $\delta_t=1$ for all $t<T$. The information-rent loss is the same with or without hidden effort, while the hidden-effort loss is the same with or without private information.

\paragraph{Economic implications.}
For any fixed $\delta=(\delta_t)_{t<T}$, the gain $\Pi_k^{*}(1)-\Pi_k^{*}(\delta)$ is
\[
\begin{aligned}
&\sum_{t=1}^{T-1}\frac{\beta^t{\theta_{t+1}^{B}}^2}{8}\bigl[(1-\delta_t^2)+\kappa_t\beta{\theta_{t+1}^{B}}^2(1-\delta_t^4)\bigr] &&(k=1B,2B^{\mathrm{HI}}),\\
&\sum_{t=1}^{T-1}\frac{\beta^t{\theta_{t+1}^{B}}^2}{8}(1-\delta_t^2) &&(k=2B^{\mathrm{HA}},3B).
\end{aligned}
\]
Under positive information value, full disclosure strictly improves the principal's value relative to no disclosure or any genuinely partial policy. All four regimes select full disclosure, but its productive consequences differ: higher $\delta_t$ raises contractible effort, whereas hidden effort remains $e_t=0$. Identical optimal information policies need not imply identical economic outcomes. Hidden-action constraints eliminate the effort-mediated component of the disclosure gain without eliminating the direct value of revealing evidence. Screening distortions persist. The quadratic informational benefit is convex in disclosure accuracy, so this example does not rely on the scalar characterization's concavity restriction. The output distortions and transaction-cost comparisons in \eqref{eq:benchmark-allocation-distortion}--\eqref{eq:benchmark-transaction-interaction} continue to apply.

\smallskip
\noindent\textbf{Example 3 (Gaussian Joint Information Design and Contracting).}

The principal can improve information by investing in precision and inducing effort; their joint cost determines whether information production is worthwhile. Consider a two-period ($T=2$), single-agent Gaussian environment under the four contracting regimes of Proposition~\ref{prop:screening-benchmark-order}. The design classes are $Y=\{\mathcal Y^\chi:\chi\ge0\}$ and $D=\{\mathcal D^\delta:\delta\in[0,1]\}$. At date 1, the principal commits to $\chi$ (cost $c\chi^2/2$), $\delta$, and a contract. The agent observes $\tau\sim U[1,2]$, reports $\widehat\tau$, and chooses $e\ge0$ (cost $e^2/(2\kappa)$). The independent state is $\theta\sim N(0,P_0)$; raw evidence $y_2=\sqrt{\chi e}\,\theta+\xi$ generates the date-2 disclosed information $d_2=\sqrt\delta\,y_2+\sqrt{1-\delta}\,\zeta$, with independent standard-normal noises. At date 2, signed output $q\in\mathbb R$ and unrestricted transfer $p$ yield agent payoff $\tau q-q^2/2-p-e^2/(2\kappa)$ and principal payoff $\theta q+p-c\chi^2/2$, where $c,\kappa,P_0>0$ and discounting is one.

\noindent\textbf{Reduced objective and information choice.}
For any fixed $\delta,\chi,e$, Gaussian updating gives
\[
\E[\theta\mid d_2]=\frac{P_0\sqrt{\delta\chi e}}{1+P_0\delta\chi e}\,d_2,
\qquad
\E\bigl[\E[\theta\mid d_2]^2\bigr]
=P_0-(P_0^{-1}+\delta\chi e)^{-1}.
\]
Conditional maximization of total surplus (observable type) or virtual surplus (private type, by the reporting envelope and integration by parts) yields
\[
q=\begin{cases}
\tau+\E[\theta\mid d_2],& k=1B,2B^{\mathrm{HA}},\\
2\tau-2+\E[\theta\mid d_2],& k=2B^{\mathrm{HI}},3B.
\end{cases}
\]
The respective type components of the bounds are $\E[\tau^2]/2=7/6$ and $\E[(2\tau-2)^2]/2=2/3$, each supplemented by half the posterior-mean second moment. These are upper bounds for every incentive-compatible and individually rational contract; in the hidden-effort environments the argument does not presume that effort is contractible or independent of type. The information-dependent part of either bound, after subtracting effort and technology costs, is
\[
\frac12\left[P_0-(P_0^{-1}+\delta\chi e)^{-1}\right]
-\frac{e^2}{2\kappa}-\frac{c\chi^2}{2}.
\] For fixed $\chi,\delta$, the expected information-dependent objective under any type-dependent effort rule is bounded by its maximum over $e\ge0$, attained by a constant effort rule.
For $\delta\chi e>0$,
\[
\frac{e^2}{2\kappa}+\frac{c\chi^2}{2}
\ge\sqrt{\frac c\kappa}\,\chi e
=\sqrt{\frac c\kappa}\,\frac{\delta\chi e}\delta,
\]
with equality if and only if $e=\sqrt{\kappa c}\,\chi$. Thus full disclosure minimizes the cost of any strictly positive effective precision. The remaining objective is strictly concave in the effective precision $\chi e$:
\[
\max_{\chi,e\ge0}\left\{\frac12\bigl[P_0-(P_0^{-1}+\chi e)^{-1}\bigr]-\sqrt{\frac c\kappa}\,\chi e\right\}.
\]
Its derivative with respect to $\chi e$ is $\tfrac12(P_0^{-1}+\chi e)^{-2}-\sqrt{c/\kappa}$, with second derivative $-(P_0^{-1}+\chi e)^{-3}<0$. The optimizing information choices are
\begin{equation}\label{eq:gaussian-joint-solution}
\begin{aligned}
\delta^*&=1,\\
\chi^*&=(\kappa c)^{-1/4}\sqrt{\left[\sqrt{\frac12\sqrt{\frac\kappa c}}-P_0^{-1}\right]_+},\\
 e^*&=(\kappa c)^{1/4}\sqrt{\left[\sqrt{\frac12\sqrt{\frac\kappa c}}-P_0^{-1}\right]_+},
\end{aligned}
\end{equation}
where $[u]_+=\max\{u,0\}$. These choices maximize the reduced objective in all four environments, subject to implementability of the resulting allocation and effort. If $e^*=\chi^*=0$, disclosure has no payoff effect; $\delta^*=1$ is then an optimal selection, not the unique policy.

\noindent\textbf{Implementing contracts.}
In $1B$ and $2B^{\mathrm{HI}}$, set $e=e^*$. In $1B$, $q=\tau+\E[\theta\mid d_2]$ and $p=\tau q-q^2/2-(e^*)^2/(2\kappa)$ yield zero net utility. In $2B^{\mathrm{HI}}$, $q(\widehat\tau,d_2)=2(\widehat\tau-1)+\E[\theta\mid d_2]$ and
$p(\widehat\tau,d_2)=-\tfrac12q(\widehat\tau,d_2)^2-1+\widehat\tau^2-\frac{(e^*)^2}{2\kappa}.$
At $e=e^*$ the expected net utility from reporting $\widehat\tau$ is $(\tau-1)^2-(\widehat\tau-\tau)^2$, so reporting is truthful, and the lowest type has zero rent. The same virtual-surplus allocation is obtained as in $3B$.

For hidden effort and positive $\chi^*,e^*$, the transfer includes the score term $-\frac{e^*}{\kappa\chi^*P_0}(d_2^2-1)$; at zero information production this term is omitted. Indeed, at $\delta^*=1$, $\E[d_2^2\mid e]=1+\chi^*eP_0, \; \frac{e^*}{\kappa\chi^*P_0}\chi^*P_0=\frac{e^*}{\kappa}\;(\chi^*>0).$ In $2B^{\mathrm{HA}}$, with observable $\tau$, set $q=\tau+\E[\theta\mid d_2]$ and choose
$p=\tau q-\tfrac12q^2-\frac{e^*}{\kappa\chi^*P_0}(d_2^2-1)+\frac{(e^*)^2}{2\kappa}.$
The expected net utility from any $e\ge0$ is $-(e-e^*)^2/(2\kappa)$, including at $\chi^*=e^*=0$ with the score omitted. In $3B$, use
\begin{equation}\label{eq:gaussian-joint-contract}
\begin{aligned}
q(\widehat\tau,d_2)&=2(\widehat\tau-1)+\E[\theta\mid d_2],\\
p(\widehat\tau,d_2)&=-\tfrac12q(\widehat\tau,d_2)^2-1+\widehat\tau^2-\frac{e^*}{\kappa\chi^*P_0}(d_2^2-1)+(e^*)^2/(2\kappa).
\end{aligned}
\end{equation}
Under every joint deviation $(\widehat\tau,e)$ the agent's expected net payoff is exactly
$(\tau-1)^2-(\widehat\tau-\tau)^2-\frac{(e-e^*)^2}{2\kappa}.$
The squared losses establish joint IC, choice obedience, and lowest-type participation. All four contracts attain their respective upper bounds, establishing optimality of \eqref{eq:gaussian-joint-solution}.

\noindent\textbf{Maximized principal payoffs and economic interpretation.}
The maximized information-design gain, common to all four regimes, is
\[
\max_{\chi,e\ge0}\left\{\frac12\bigl[P_0-(P_0^{-1}+\chi e)^{-1}\bigr]-\sqrt{c/\kappa}\,\chi e\right\}
=\frac{\bigl(P_0-\sqrt{2\sqrt{c/\kappa}}\bigr)_+^2}{2P_0}.
\]
Thus
\[
\Pi_{1B}^{***}=\Pi_{\mathrm{HA}}^{***}=\Pi_{\mathrm{HI}}^{***}+\tfrac12=\Pi_{3B}^{***}+\tfrac12=\frac76+\frac{\bigl(P_0-\sqrt{2\sqrt{c/\kappa}}\bigr)_+^2}{2P_0}.
\]
Unlike Example~2, hidden effort causes no additional loss here: evidence-based transfers using $d_2^2$ implement efficient effort under the Gaussian observation structure and unrestricted transfers. Screening still entails an information-rent loss of $1/2$. Thus the economic cost of hidden actions depends not only on their observability but also on the available information technology and contractual instruments, not on a general equivalence of observable and hidden actions. Information-technology investment has strictly positive value if $P_0^2>2\sqrt{c/\kappa}$. Full disclosure reveals the available evidence, not the underlying state.

\section{Extensions and Applications}\label{sec:applications}
The applications illustrate how endogenous information changes established economic results. In interdependent-value auctions, acquisition can reverse revenue rankings and undermine allocation efficiency; in financial contracting, disclosure shapes production incentives, screening rents, and optimal disclosure within the specified contract family. Both applications rely on the joint-incentive characterization of Section~3 and illustrate complementary aspects of Sections~5--6.

\subsection{Interdependent-Value Auctions with Endogenous Information Acquisition and Disclosure}
\label{subsec:auction-application}

Consider a two-date auction for a single indivisible object, with $n\geq2$ risk-neutral bidders and primitive payoffs $u_i(x_i,\tau)=x_i v_i(\tau)$, with $x_i\in\{0,1\}$ and $\sum_{i=1}^n x_i=1$, where $v_i(\tau)$ is bidder $i$'s valuation function, allowing interdependent values. At date~1, bidder $i$ observes $\tau_i$ and privately chooses information acquisition $x_{i1}$ at cost $c_i(x_{i1})$. At date~2, $\mathcal Y_2$ generates the resulting signals, $\mathcal D_2$ governs public disclosure, and bidders submit bids $m_i$ for the single final allocation. Acquisition changes information, not primitive valuations. The auction recommends $\px_{i}(m)$, with $x_i=\px_{i}(m)$ under implementation. Suppressing terminal-period indices, the FP and SP payments are $p_i^{\mathrm{FP}}=\px_{i}(m)m_i$ and $p_i^{\mathrm{SP}}=\px_{i}(m)\max_{j\ne i}m_j$.

Without acquisition, with fixed disclosure and no other intertemporal payoff effects, the bidding problem reduces to the classical static setting.

\noindent\textbf{Classical interdependent-value benchmark.} For fixed information without acquisition or disclosure, let $v(t,r)=\mathbb E[v_i(\tau)\mid\tau_i=t,\max_{j\ne i}\tau_j=r]$ be the conditional valuation, $\Psi(r\mid t)=\mathbb P(\max_{j\ne i}\tau_j\le r\mid\tau_i=t)$ the conditional CDF, and $\psi(r\mid t)=\partial_r\Psi(r\mid t)$ its density. Under the classical monotonicity and regularity conditions, the symmetric bidding functions satisfy
\[
 m_i^{\mathrm{SP},*}(t)=v(t,t),\qquad
 \bigl(m_i^{\mathrm{FP},*}\bigr)'(t)=
 \bigl[v(t,t)-m_i^{\mathrm{FP},*}(t)\bigr]
 \frac{\psi(t\mid t)}{\Psi(t\mid t)},
\]
with the FP solution
\[
 m_i^{\mathrm{FP},*}(t)=\int_0^t v(r,r)\,d\!\left[\exp\!\left\{-\int_r^t\frac{\psi(z\mid z)}{\Psi(z\mid z)}\,dz\right\}\right].
\]
The exponential term is an auxiliary CDF, not a type distribution. The Milgrom--Weber (1982) revenue comparison holds under its fixed-information, affiliation, monotonicity, and equilibrium hypotheses. With endogenous acquisition, auction formats can change not only bidding incentives but also the information on which bids are based. Revenue comparisons therefore depend on both the use and production of information, even with unchanged primitive valuations.

A deviation to $m_i^{\mathrm{FP},*}(r)$ by a bidder with initial signal $t$ yields
\[
 \int_0^r\!\bigl[v(t,z)-m_i^{\mathrm{FP},*}(r)\bigr]\psi(z\mid t)\,dz.
\]
Its derivative at $r=t$ is zero precisely when the FP differential equation above holds; integration with the standard lower-boundary condition gives the displayed exponential representation. By contrast, under acquired private information the bidding objective conditions on the full private history, as in \eqref{eq:bidvalue} below.

The auction-specific JPBIC characterization specializes the general theory of Section~3.

Let $m_i^*(h_{i2};\Gamma^f)$ be equilibrium bidding. Given rivals' strategies, terminal bidding value is
\begin{equation}\label{eq:bidvalue}
\begin{aligned}
V_{i2}^{**}(h_{i2};\Gamma^f)
&=
\sup_{m_i\geq0}
\int
\left[
\px_{i}(m_i,m_{-i}^*)v_i(\tau)
-p_i(m_i,m_{-i}^*;f)
\right]\\
&\quad d\mathbb P_{i2}
(\cdot\mid h_{i2}\,;\,\Gamma^f,m_{-i}^*(\cdot),\nu_i^*).
\end{aligned}
\end{equation}
For fixed acquisition $x_{i1}$, gross continuation bidding value is
\[
\int V_{i2}^{**}(\tilde h_{i2};\Gamma^f)\,
 d\mathbb P_{i1}(\tilde h_{i2}\mid h_{i1};\Gamma^f,m_{-i}^*,\nu_i^*,x_{i1}).
\]

Terminal bidding and date-1 acquisition satisfy the following backward-induction conditions. If type reports are elicited, the full-history condition specializes Theorem~\ref{thm:recursive-ic} under Assumption~\ref{ass:controlled-history}.2.

\begin{proposition}[Characterization of JPBIC in Interdependent-Value Auctions]
\label{prop:recursive-auction}
Under Assumption~\ref{ass:regularity} and JPBE-consistent beliefs, terminal bidding and acquisition are sequentially rational if and only if conditions (1)--(2) hold. If the mechanism additionally elicits type histories and Assumption~\ref{ass:controlled-history}.2 holds, full-history truthful reporting additionally requires condition (3):
\begin{enumerate}
\item At each relevant terminal private history, $m_i^*(h_{i2};\Gamma^f)$ attains the supremum in \eqref{eq:bidvalue}.
\item For every $i$ and relevant $h_{i1}$,
\begin{equation}\label{eq:obed}
\px_{i1}\in
\arg\max_{x_{i1}\in X_{i1}}
\left\{-c_i(x_{i1})+\beta \mathbb E[V_{i2}^{**}(\tilde h_{i2};\Gamma^f)\mid h_{i1},x_{i1}]\right\}.
\end{equation}
\item If the mechanism additionally elicits private types, then for every $i$, relevant pre-report private history $h_{i1}^{-}$, true type $\tau_i\in\T_{i1}$, the global reporting condition is
\begin{equation}\label{eq:globalic}
V_{i1}^{**}(h_{i1}^{-},\tau_i;\Gamma^f)
=
V_{i1}^{*}(h_{i1}^{-},\tau_i,\vartheta_{i,1}^{T}(\cdot);\Gamma^f)
.
\end{equation}
where $V_{i1}^{*}$ optimizes information production and bidding after each reporting plan.
\end{enumerate}
\end{proposition}

When the information-preservation conditions of Section~\ref{subsec:direct-revelation} hold, equilibrium bidding induces the composite direct mechanism
\[
\px_{i}(\hat\tau;\Gamma^f)
=
\px_{i}\bigl(m^*(\hat\tau;\Gamma^f)\bigr),
\qquad
p_i(\hat\tau;\Gamma^f)
=
p_i\bigl(m^*(\hat\tau;\Gamma^f);f\bigr).
\]
This direct implementation preserves equilibrium bids and payments; without information preservation, use the indirect PBE characterization.

For a dominated acquisition law differentiable in $x_{i1}$ with score $\ell_{x_{i1}}$, the terminal envelope gives
\begin{equation}\label{eq:score}
\begin{aligned}
&\frac{\partial}{\partial x_{i1}}
\mathbb E[V_{i2}^{**}(\tilde h_{i2};\Gamma^f)\mid h_{i1},x_{i1}]\\
&\qquad=\int\left[\px_{i}(m_2^*)v_i(\tau)-p_i(m_2^*;f)\right]\ell_{x_{i1}}
\,d\mathbb P_{i1}
(\cdot\mid h_{i1}\,;\,\Gamma^f,m_{-i}^*(\cdot),\nu_i^*,x_{i1}).
\end{aligned}
\end{equation}
The score refers to the joint terminal law under actual acquisition.

An interior optimal choice consequently satisfies
\begin{equation}\label{eq:FOC}
c_i'(\px_{i1})
=
\beta
\left.\frac{\partial}{\partial x_{i1}}\mathbb E[V_{i2}^{**}(\tilde h_{i2};\Gamma^f)\mid h_{i1},x_{i1}]\right|_{x_{i1}=\px_{i1}}.
\end{equation}

We compare marginal acquisition values across formats, holding intrinsic valuations fixed.

\begin{proposition}[Information Acquisition in First- and Second-Price Auctions]
\label{prop:score}
Under the maintained conditions:
\begin{enumerate}
\item \textbf{Private values.} If $v_i(\tau)=v_i(\tau_i)$ is known to bidder $i$, and the underlying joint law of valuations and competing bids does not depend on his information choice, then
\[
\begin{aligned}
m_i^{\mathrm{SP},*}&=v_i(\tau_i),\\
\mathbb E[V_{i2}^{**}(\tilde h_{i2};\Gamma^{\mathrm{SP}})\mid h_{i1},x_{i1}]
&=\mathbb E[V_{i2}^{**}(\tilde h_{i2};\Gamma^{\mathrm{SP}})\mid h_{i1},x_{i1}'],
\end{aligned}
\quad\forall x_{i1},x_{i1}'\in X_{i1}.
\]
Consequently, every SP-optimal information input satisfies $x_{i1}^{\mathrm{SP},*}\in\arg\min_{x_{i1}\in X_{i1}}c_i(x_{i1})$; if the cost minimizer is unique, so is the SP-optimal input.
\item \textbf{First-price information value.} Fix the opponents' strategies and the underlying joint distribution of terminal payoff-relevant uncertainty. Suppose the private-signal experiment induced by $\mathcal Y'$ Blackwell-dominates that under $\mathcal Y$, and write $\mathbb E_{\mathcal Y}[V_{i2}^{**}(\tilde h_{i2};\Gamma^{\mathrm{FP}})\mid h_{i1}]$ for optimized gross bidding value under the private-signal experiment induced by $\mathcal Y$. Then
\[
\mathbb E_{\mathcal Y'}[V_{i2}^{**}(\tilde h_{i2};\Gamma^{\mathrm{FP}})\mid h_{i1}]
\geq \mathbb E_{\mathcal Y}[V_{i2}^{**}(\tilde h_{i2};\Gamma^{\mathrm{FP}})\mid h_{i1}].
\]
If the optimal gross payoff under $\mathcal Y$ is attained, the inequality is strict whenever no bidding rule measurable with respect to $\mathcal Y$ attains the optimized payoff under $\mathcal Y'$.
\item \textbf{Interdependent values.} If $v_i(\tau)$ depends on $\tau_{-i}$, the SP information value need not vanish. Suppose the two acquisition objectives are strictly concave on a common action interval with a common acquisition cost, and, fixing the same rivals' strategies and relevant history, satisfy
\[
\frac{\partial \mathbb E[V_{i2}^{**}(\tilde h_{i2};\Gamma^{\mathrm{FP}})\mid h_{i1},x_{i1}]}{\partial x_{i1}}
\geq
\frac{\partial \mathbb E[V_{i2}^{**}(\tilde h_{i2};\Gamma^{\mathrm{SP}})\mid h_{i1},x_{i1}]}{\partial x_{i1}}
\qquad\forall x_{i1}\in X_{i1}.
\]
Then their unique optimal inputs obey $x_{i1}^{\mathrm{FP},*}\geq x_{i1}^{\mathrm{SP},*}$. Reversing the marginal inequality reverses the input ordering.
\end{enumerate}
\end{proposition}

\begin{proof}
With known private value, SP truthful bidding is weakly dominant; under a fixed joint law of valuations and rival bids, extra information has zero gross value. Thus a cost-minimizing acquisition input is optimal.

By Blackwell's theorem, the optimized gross bidding values satisfy $\mathbb E_{\mathcal Y'}[V_{i2}^{**}(\tilde h_{i2};\Gamma^{\mathrm{FP}})\mid h_{i1}] \geq \mathbb E_{\mathcal Y}[V_{i2}^{**}(\tilde h_{i2};\Gamma^{\mathrm{FP}})\mid h_{i1}].$ Strictness follows from the stated criterion. For part~(3), the weakly ordered marginal gross values, common $c_i$, and strict concavity imply the ordering of unique maximizers by the one-dimensional optimality inequalities; SP dominance is not invoked.
\end{proof}

Parts (1)--(2) fix rivals' strategies and the underlying payoff-relevant uncertainty; equilibrium acquisition rankings additionally require equilibrium best-response comparisons. The acquisition ranking depends on marginal continuation values, not the payment rule alone. \citet{Persico2000} establishes stronger FP acquisition incentives in his informational environment; our comparison states a marginal-value condition for the ranking. Endogenous acquisition can also change the conditional laws governing bidder rents. With known private values and a fixed law of valuations and rival bids, truthful SP bidding eliminates the gross value of additional information; with interdependent values, this argument no longer suffices.

Under affiliated types and signals and symmetric valuations, the classical fixed-information efficiency argument relies on single crossing and \emph{bid-rank preservation}. Single crossing requires
\begin{equation}\label{eq:auction-single-crossing}
\frac{\partial v_i(\tau)}{\partial\tau_i}
\geq
\frac{\partial v_j(\tau)}{\partial\tau_i},
\qquad i\ne j.
\end{equation}
Under the classical regularity conditions, the highest type then has the highest realized valuation.

Acquired signals also affect bids, so efficiency additionally requires valuation-rank preservation, for example
\begin{equation}\label{eq:auction-rank-preservation}
v_i(\tau)>v_j(\tau)
\quad\text{implies}\quad
m_i^*>m_j^*
\qquad\text{almost surely}.
\end{equation}
Conditions~\eqref{eq:auction-single-crossing}--\eqref{eq:auction-rank-preservation} imply efficient allocation. Affiliation and single crossing order valuations but do not by themselves guarantee the bid-rank preservation condition after bidders choose $x_{i1}$ and receive private signals. 

For interdependent values, the private-value envelope of Section~5 must account for type-dependent conditional laws.

Assume the initial private signal is degenerate and the primitive acquisition cost $c_i$ does not depend directly on $\tau_i$. For each fixed contingent strategy, suppose the conditional law of terminal payoff-relevant uncertainty admits a positive differentiable density in $\tau_i$, whose logarithmic type derivative is evaluated holding the contingent strategy fixed while allowing type-dependent posteriors. Assume an integrable dominating bound on the derivative of the payoff integrand permitting differentiation under the integral.

\begin{proposition}[Interdependent-value envelope, seller surplus, and revenue ranking]
\label{prop:virtual}\label{cor:auction-revenue-comparison}
Under the maintained conditions:
\begin{enumerate}
\item \textbf{Envelope.} For a.e. $\tau_i$,
\begin{equation}\label{eq:envelope}
\begin{aligned}
\partial_{\tau_i}V_{i1}^{**}(\tau_i;\Gamma^f)
={}&
\beta\int
\px_{i}\frac{\partial v_i(\tau)}{\partial\tau_i}
\,d\mathbb P_{i1}\\
&+
\beta\int
\left[\px_{i}v_i(\tau)-p_i\right]
\frac{\partial}{\partial\tau_i}\log f_i(\omega\mid\tau_i;\Gamma^f)
\,d\mathbb P_{i1}.
\end{aligned}
\end{equation}
Both integrals are conditional on the true type and evaluated under the optimizing contingent strategy. The first term is the direct valuation effect; the second captures type-dependent posterior and information effects.

\item \textbf{Information rents.} Let $F_i$ and $f_i>0$ denote the initial marginal type distribution and density, and set $\Lambda_i=(1-F_i)/f_i$. Then
\begin{equation}\label{eq:rent}
\int V_{i1}^{**}(\tau_i;\Gamma^f)\,dF_i(\tau_i)
=
V_{i1}^{**}(\underline\tau_i;\Gamma^f)
+
\int\Lambda_i(\tau_i)
\partial_{\tau_i}V_{i1}^{**}(\tau_i;\Gamma^f)\,dF_i(\tau_i).
\end{equation}

\item \textbf{Seller surplus.}
\begin{equation}\label{eq:virtualsurplus}
\begin{aligned}
\Pi(\Gamma^f)
={}&
\int\left[
\beta\sum_i \px_{i}v_i(\tau)
-\sum_i c_i(x_{i1})
\right]d\mathbb P_{W}(\cdot\,;\Gamma^f)\\
&-
\sum_i\int\Lambda_i(\tau_i)
\partial_{\tau_i}V_{i1}^{**}(\tau_i;\Gamma^f)\,dF_i(\tau_i)
-
\sum_i V_{i1}^{**}(\underline\tau_i;\Gamma^f).
\end{aligned}
\end{equation}
\item \textbf{Revenue ranking.} Under efficient allocation in both formats and the preceding payoff regularity conditions, $\Pi(\Gamma^{\mathrm{FP}})\geq \Pi(\Gamma^{\mathrm{SP}})$ if and only if
\[
\begin{aligned}
&\sum_i\int\!\left[V_{i1}^{**}(\tau_i;\Gamma^{\mathrm{SP}})
-V_{i1}^{**}(\tau_i;\Gamma^{\mathrm{FP}})\right]dF_i(\tau_i)\\
&\geq \int\sum_i c_i(x_{i1})\,d\mathbb P_W(\cdot;\Gamma^{\mathrm{FP}})
-\int\sum_i c_i(x_{i1})\,d\mathbb P_W(\cdot;\Gamma^{\mathrm{SP}}).
\end{aligned}
\]
Reversing either displayed inequality reverses the other.
\end{enumerate}
\end{proposition}

\begin{proof}
Fix the optimizing contingent strategy at $\tau_i$. For its terminal payoff integrand, the product rule and the fixed-strategy conditional density give
\[
\begin{aligned}
&\partial_{\tau_i}\int [\px_{i}v_i(\tau)-p_i]\,f_i(\omega\mid\tau_i;\Gamma^f)\,d\omega\\
&=\int \px_{i}\partial_{\tau_i}v_i(\tau)\,d\mathbb P_{i1}
+\int[\px_{i}v_i(\tau)-p_i]\partial_{\tau_i}\log f_i(\omega\mid\tau_i;\Gamma^f)\,d\mathbb P_{i1}.
\end{aligned}
\]
Here $\px_{i}$ and $p_i$ denote outcomes of the \emph{fixed} contingent strategy; their type-dependent equilibrium changes are excluded from the derivative. The initial $c_i(x_{i1})$ has zero partial type derivative. The Section~4 envelope and the discount factor $\beta$ give \eqref{eq:envelope}. Absolute continuity and integration by parts imply $\int V_{i1}^{**}\,dF_i =V_{i1}^{**}(\underline\tau_i)+\int(1-F_i)\partial_{\tau_i}V_{i1}^{**}\,d\tau_i,$ which is \eqref{eq:rent}. Summing $\beta\sum_i \px_{i}v_i-\sum_i c_i(x_{i1})$ and subtracting the bidder utilities cancels transfers and yields \eqref{eq:virtualsurplus}.
\end{proof}

The likelihood-score term in \eqref{eq:envelope} captures type-dependent beliefs about rivals and acquired signals, beyond the direct valuation effect; the private-value reduction of Section~5 need not apply. Even with equal allocated value, acquisition costs and bidder rents can reverse the classical revenue ranking.

The seller's payoff identity gives
\begin{equation}\label{eq:auction-revenue-comparison}
\begin{aligned}
&\Pi(\Gamma^{\mathrm{FP}})-\Pi(\Gamma^{\mathrm{SP}})\\
={}&\int\beta\sum_i \px_{i}v_i(\tau)\,d\mathbb P_W(\cdot;\Gamma^{\mathrm{FP}})
-\int\beta\sum_i \px_{i}v_i(\tau)\,d\mathbb P_W(\cdot;\Gamma^{\mathrm{SP}})\\
&-\int\sum_i c_i(x_{i1})\,d\mathbb P_W(\cdot;\Gamma^{\mathrm{FP}})
+\int\sum_i c_i(x_{i1})\,d\mathbb P_W(\cdot;\Gamma^{\mathrm{SP}})\\
&-\sum_i\int\!\left[V_{i1}^{**}(\tau_i;\Gamma^{\mathrm{FP}})
-V_{i1}^{**}(\tau_i;\Gamma^{\mathrm{SP}})\right]dF_i(\tau_i).
\end{aligned}
\end{equation}

Under efficient allocation, the expected allocated values coincide, proving the revenue-ranking part of Proposition~\ref{prop:virtual}.

For a committed disclosure family $\mathcal D_2^\delta$, let $\Gamma^{f,\delta}$ denote the corresponding format-specific mechanism. Applying Section~6 to the selected auction equilibrium yields
\[
\begin{aligned}
\frac{d\Pi(\Gamma^{f,\delta})}{d\delta}
={}&\frac{d}{d\delta}\int\beta\sum_i \px_{i}v_i(\tau)\,d\mathbb P_W(\cdot;\Gamma^{f,\delta})
-\frac{d}{d\delta}\int\sum_i c_i(x_{i1})\,d\mathbb P_W(\cdot;\Gamma^{f,\delta})\\
&-\frac{d}{d\delta}\sum_i\int V_{i1}^{**}(\tau_i;\Gamma^{f,\delta})\,dF_i(\tau_i),
\end{aligned}
\]
whenever the selected equilibrium payoff is differentiable in $\delta$. If efficient allocation is maintained throughout this disclosure family, the allocated-value derivative vanishes. An interior optimum then satisfies
\[
\frac{d}{d\delta}\left[\int\sum_i c_i(x_{i1})\,d\mathbb P_W(\cdot;\Gamma^{f,\delta})
+\sum_i\int V_{i1}^{**}(\tau_i;\Gamma^{f,\delta})\,dF_i(\tau_i)\right]=0,
\]
provided the reoptimized seller value is differentiable and the original joint incentive constraints remain satisfied. The corresponding boundary conditions apply under concavity.

Equation~\eqref{eq:auction-revenue-comparison} shows that efficient allocation does not restore the classical revenue ranking: acquisition expenditures and bidder rents can differ even when allocated value is identical. The interior disclosure condition is family-specific; the common-value construction below isolates the rent channel with zero acquisition costs and establishes a global revenue bound.

Consider two bidders with independent persistent types and a common realized value,
\begin{equation}\label{eq:auction-counterexample-values}
\tau_1,\tau_2\overset{\mathrm{iid}}{\sim}U[0,1],
\qquad v_1(\tau)=v_2(\tau)=\tau_1+\tau_2.
\end{equation}
At date~1, each bidder chooses $x_{i1}\in\{0,1\}$. Choosing $x_{i1}=1$ reveals $\tau_j$ privately to bidder $i$ before bidding; choosing $x_{i1}=0$ yields no additional signal. Let $c_i(x_{i1})=0$, and take $\mathcal D_2$ to be uninformative. Bidders may condition their date-2 bids on all privately observed information. Ties are split uniformly, and the seller does not set a reserve price. For this comparison normalize the date-2 discount factor to $\beta=1$.

Under SP, consider the strategy $x_{i1}^{\mathrm{SP}}=0$ and $m_i^{\mathrm{SP}}=2\tau_i$. If the opponent bids $2\tau_j$, a bidder of type $t$ who submits $2r$, $r\in[0,1]$, obtains
\begin{equation}\label{eq:auction-sp-counterexample-payoff}
\int_0^r(t+u-2u)\,du=tr-\frac{r^2}{2},
\end{equation}
with $\arg\max_{r\in[0,1]}\{tr-r^2/2\}=t$. Even after privately learning $\tau_j=u$, winning at the opponent's bid generates surplus $t-u$; consequently, the optimal decision remains to win precisely when $t>u$. Additional information therefore cannot improve the optimized SP payoff, so zero acquisition is optimal. The expected seller revenue is
\begin{equation}\label{eq:auction-sp-counterexample-revenue}
\Pi(\Gamma^{\mathrm{SP}})=2\mathbb E[\min\{\tau_1,\tau_2\}]=\frac23.
\end{equation}

Under FP, consider $x_{i1}^{\mathrm{FP}}=1$ and $m_i^{\mathrm{FP}}=\tau_1+\tau_2$ for both bidders. Both then know the realized common value. Against an opponent bidding this value, lowering one's bid loses and raising it produces a negative payoff conditional on winning; bidding the common value yields zero. No deviation in the date-1 information choice can yield a positive continuation payoff either: the opponent's bid remains equal to the actual common value, and any winning bid must pay at least that value. Thus the proposed acquisition and bidding strategies are jointly optimal, and
\begin{equation}\label{eq:auction-fp-counterexample-revenue}
\Pi(\Gamma^{\mathrm{FP}})
=\mathbb E[\tau_1+\tau_2]=1.
\end{equation}

The constructed equilibria therefore give the revenue reversal
\begin{equation}\label{eq:auction-revenue-reversal}
\Pi(\Gamma^{\mathrm{FP}})-\Pi(\Gamma^{\mathrm{SP}})=\frac13>0.
\end{equation}

\begin{proof}[Verification of the revenue counterexample]
Equations~\eqref{eq:auction-sp-counterexample-payoff}--\eqref{eq:auction-fp-counterexample-revenue} verify acquisition and bidding optimality and the revenue difference. There is no type-report stage. Beliefs follow the independent prior and observed signals, with Bayesian updating on path and consistent prior--signal updating after unreached acquisition histories; bidding is terminal.
\end{proof}

Retain \eqref{eq:auction-counterexample-values} and the absence of outside seller subsidies. Let $\mathcal D^0$ be the uninformative disclosure rule and $m^0(\cdot)$ the corresponding FP equilibrium acquisition and bidding profile. The supremum below ranges over disclosure rules $\mathcal D$ and ex ante IR PBE acquisition and bidding profiles $m(\cdot)$ of $\Gamma^{\mathrm{FP},\mathcal D}$.

\noindent\textbf{Optimal disclosure.} Under the above common-value specification, the uninformative disclosure rule and constructed FP equilibrium attain the global seller-revenue bound:
\begin{equation}\label{eq:auction-optimal-uninformative}
\Pi(\Gamma^{\mathrm{FP},\mathcal D^0};m^0(\cdot))=\sup_{\mathcal D,m(\cdot)}\Pi(\Gamma^{\mathrm{FP},\mathcal D};m(\cdot))=\mathbb E[\tau_1+\tau_2]=1.
\end{equation}
Private acquisition already enables full surplus extraction in the constructed equilibrium, so informative public disclosure is unnecessary for attaining the revenue bound. This illustrates a limited substitution between private acquisition and public disclosure in seller value, not general incentive equivalence. This concerns the specified policy--equilibrium pair, without uniqueness or equilibrium-selection invariance.

\begin{proof}
For each policy and individually rational equilibrium, adding the two bidders' ex ante payoffs to the seller's expected receipts cancels auction transfers. Since precisely one bidder obtains the common-value object and all real acquisition costs are nonnegative,
$\Pi(\Gamma^{\mathrm{FP},\mathcal D};m(\cdot)) \leq\mathbb E[v_1(\tau)]=\mathbb E[\tau_1+\tau_2]=1.$
The previously constructed first-price equilibrium under uninformative disclosure has zero acquisition costs and zero bidder payoffs, and obtains receipts of $1$ by \eqref{eq:auction-fp-counterexample-revenue}. Hence the upper bound is attained.
\end{proof}

The two constructed equilibria share primitives and expected allocated value $\mathbb E[\tau_1+\tau_2]=1$, yet differ in acquisition and bidder rents. FP extracts the common value; SP leaves positive rents. This is an actual equilibrium-specific revenue reversal, not a claim about every equilibrium selection. Strict single crossing is not required in this common-value construction; it is imposed in the following efficiency example.

Revenue reversal need not imply inefficiency. To study allocation separately, let
\begin{equation}\label{eq:auction-gaussian-values}
\tau_1,\tau_2\overset{\mathrm{iid}}{\sim}U[0,1],\qquad
v_i(\tau)=\tau_i+b\tau_j,\quad 0<b<1,\quad j\neq i.
\end{equation}
The independent types are affiliated, and strict single crossing holds because $\partial v_i/\partial\tau_i=1>b=\partial v_j/\partial\tau_i$. Hence the higher-type bidder is the efficient recipient. Bidder $i$ may choose $x_{i1}\in[0,1]$ at cost $kx_{i1}^2/2$, where $k>0$. With probability $x_{i1}$ he privately observes $s_i=\tau_j+\varepsilon_i$, where the errors are independent $N(0,\sigma^2)$ variables with $\sigma>0$; otherwise he receives no signal. Public disclosure is uninformative.

The SP strategies
\begin{equation}\label{eq:auction-sp-efficient-strategies}
x_{i1}^{\mathrm{SP}}=0,\qquad m_i^{\mathrm{SP}}=(1+b)\tau_i
\end{equation}
form an equilibrium. Indeed, when the rival bids $(1+b)\tau_j$, bidder $i$ gains $\tau_i-\tau_j$ from winning at the rival's bid. Thus his optimal winning decision is independent of any signal about $\tau_j$ and is implemented by \eqref{eq:auction-sp-efficient-strategies}. Information acquisition has zero gross value and positive cost. Accordingly,
\begin{equation}\label{eq:auction-sp-efficient-outcome}
\int\beta\sum_i \px_{i}v_i(\tau)\,d\mathbb P_W(\cdot;\Gamma^{\mathrm{SP}})=\beta\frac{2+b}{3},\qquad
\Pi(\Gamma^{\mathrm{SP}})=\beta\frac{1+b}{3}.
\end{equation}

More generally, an FP equilibrium allocating the object efficiently would have to rank bids according to types almost surely. With independently generated Gaussian signals of full support, bid distributions associated with distinct types must then have ordered, nonoverlapping essential ranges. The ranges can be nondegenerate for at most countably many types, so bids are signal-independent for almost every type. Positive information costs then require zero acquisition, contradicting the strictly positive marginal acquisition value at the zero-acquisition FP candidate. This yields a conditional efficiency conclusion: an efficient symmetric pure-strategy FP equilibrium cannot exist in the stated class. Establishing that an inefficient FP equilibrium actually exists would require a separate existence argument; the proposition does not supply one.

\begin{proposition}[Failure of the Linkage Principle and First-Price Efficiency]\label{prop:auction-revenue-reversal}
In economic environments with endogenous information acquisition, the classical linkage principle and first-price auction-efficiency conclusions do not generally hold. Under the preceding Gaussian information specification, the second failure is quantified as follows.

Under \eqref{eq:auction-gaussian-values} and the specified information technology, every symmetric pure-strategy FP equilibrium with measurable bidding profile $m^*(\cdot)$, independent private signal shocks, and almost-surely decisive bidding satisfies
\begin{equation}\label{eq:auction-fp-inefficiency}
\int\beta\sum_i \px_{i}v_i(\tau)\,d\mathbb P_W(\cdot;\Gamma^{\mathrm{FP}},m^*(\cdot))<\beta\frac{2+b}{3}.
\end{equation}
Affiliation and strict single crossing still order valuations, but do not ensure the corresponding bid ranking. In the specified Gaussian environment, efficient FP allocation requires signal-independent bidding almost everywhere; costly signals would then be unused, yet at zero acquisition the marginal gross value of information is positive and its marginal cost is zero. Efficiency and acquisition incentives are therefore incompatible. The efficiency conclusion is conditional on equilibrium existence in the specified class; no equilibrium-existence claim is made.
\end{proposition}

\begin{proof}
Suppose, to the contrary, that an equilibrium in the stated class allocates to the higher type almost surely. Let $t,u\in[0,1]$ be the two types. Let $\mathbb P(m_i\in\cdot\mid\tau_i=t)$ denote the bid law integrating over the opponent type and private shocks, and $\mathbb P(m_i\in\cdot\mid\tau_i=t,\tau_j=u)$ its conditional law. Since the Gaussian signal likelihood is strictly positive on $\mathbb R$ for every $u\in[0,1]$, and $x_{i1}(t)$ does not vary with $u$, the conditional signal/no-signal laws are mutually absolutely continuous for fixed $t$. Their pushforwards under the measurable bid strategy satisfy
\[
\mathbb P(m_i\in\cdot\mid\tau_i=t,\tau_j=u)
\sim\mathbb P(m_i\in\cdot\mid\tau_i=t),\qquad u\in[0,1].
\]
The essential bid bounds therefore depend only on the bidder's own type. Conditional independence of the two private signals, efficient allocation, and almost-surely decisive bidding imply, for almost every $t<u$,
\begin{equation}\label{eq:auction-fp-essential-order}
\operatorname*{ess\,sup}_{\mathbb P(\cdot\mid\tau_i=t)}m_i\leq\operatorname*{ess\,inf}_{\mathbb P(\cdot\mid\tau_i=u)}m_i.
\end{equation}

For any $q\in\mathbb Q$, a positive-measure set of types satisfying
\[
\operatorname*{ess\,inf}_{\mathbb P(\cdot\mid\tau_i=t)}m_i<q
<\operatorname*{ess\,sup}_{\mathbb P(\cdot\mid\tau_i=t)}m_i
\]
would contain positive-measure pairs $t<u$ contradicting \eqref{eq:auction-fp-essential-order}. Countability of $\mathbb Q$ gives, for almost every $t$,
\[
\operatorname*{ess\,inf}_{\mathbb P(\cdot\mid\tau_i=t)}m_i
=\operatorname*{ess\,sup}_{\mathbb P(\cdot\mid\tau_i=t)}m_i.
\]
Hence the equilibrium bid law is degenerate for almost every type. When $x_{i1}(t)>0$, mutual absolute continuity of the signal laws implies that observing the signal does not change that bid almost surely. The bidder can therefore choose zero acquisition while making the same bid, preserving the allocation and transfer distribution against the opponent's equilibrium strategy but strictly reducing $kx_{i1}(t)^2/2$. Sequential rationality entails $x_{i1}(t)=0$ almost everywhere.

Let $m_i^*(t)$ denote the resulting deterministic symmetric bid. Efficiency makes the higher type win almost surely. Bidding as type $r$ yields gross payoff $\int_0^r[t+bu-m_i^*(r)]\,du$. Truthful bidding and the envelope identity give $\frac{d}{dt}V_{i2}^{**}(t)=t$ and $V_{i2}^{**}(0)=0$, hence $V_{i2}^{**}(t)=t^2/2$. Equating this value to $\int_0^t[t+bu-m_i^*(t)]\,du$ implies
\begin{equation}\label{eq:auction-fp-noacq-bid}
m_i^*(t)=\frac{1+b}{2}t\quad\text{for almost every }t.
\end{equation}

For $t\in(0,1)$, suppose the bidder acquires a Gaussian observation $s=u+\varepsilon$ of the opponent's type. Let $f(u\mid s)$ and $F(u\mid s)$ be its strictly positive posterior density and distribution function on $[0,1]$. Against \eqref{eq:auction-fp-noacq-bid}, a bid indexed by $r$ yields $\int_0^r[t+bu-(1+b)r/2]f(u\mid s)\,du$. Differentiating at $r=t$ gives
\begin{equation}\label{eq:auction-fp-signal-derivative}
\left.\frac{\partial}{\partial r}\int_0^r[t+bu-(1+b)r/2]f(u\mid s)\,du\right|_{r=t}
=\frac{1+b}{2}\bigl[t f(t\mid s)-F(t\mid s)\bigr].
\end{equation}
For each interior $t$, the right side is negative for sufficiently low $s$ and positive for sufficiently high $s$: as $s\to-\infty$, $F(t\mid s)\to1$ and $f(t\mid s)\to0$, while as $s\to+\infty$, $f(t\mid s)/F(t\mid s)\to\infty$. Thus, for every $t\in(0,1)$, the derivative in \eqref{eq:auction-fp-signal-derivative} has opposite signs on two signal events of strictly positive probability. The signal-conditional optimizer therefore strictly improves the expectation of gross FP payoff relative to the fixed bid $m_i^*(t)$. Acquiring the observation with sufficiently small positive acquisition probability, bidding optimally when informed and using $m_i^*(t)$ otherwise, yields a strictly positive first-order gross payoff gain and only a second-order cost $kx_{i1}^2/2$. This contradicts $x_{i1}(t)=0$ for almost every interior type. Therefore an almost-surely efficient equilibrium in the stated class cannot exist. Because the allocative loss from awarding the object to the lower type is $(1-b)|t-u|>0$ whenever $t\ne u$, any equilibrium in the stated class has expected discounted allocated value strictly below $\beta(2+b)/3$.
\end{proof}

\noindent\textbf{Related work.} \citet{FangMorris2006} establish related revenue and efficiency failures under multidimensional private values. Our Gaussian construction shows that affiliation and strict single crossing do not suffice for efficient FP equilibrium with endogenous acquisition in the stated class. The obstruction is the incompatibility between valuation-rank-preserving bidding and privately optimal acquisition.

\subsection{Financial Contracting: Joint Screening, Evidence Production, and Disclosure}\label{subsec:app-finance}

The auction application shows how endogenous acquisition changes equilibrium comparisons. Financial contracting illustrates a complementary role of Section~3: evidence-contingent transfers implement hidden production while preserving reporting incentives. As in \citet{LaffontTirole1986}, screening and hidden investment must be jointly addressed. Here disclosure additionally determines the conditional probability of revealing contractible evidence, shaping production incentives and screening rents. The evidence available for contracting is therefore itself an institutional choice. The analysis also connects verification \citep{Townsend1979,Innes1990}, dynamic agency \citep{DeMarzoSannikov2006}, and information design \citep{KamenicaGentzkow2011}. Transfers are unrestricted and fees collectible; limited liability and costly verification are excluded.

Consider the two-period entrepreneur--lender specialization of Section~2:
\[
\begin{gathered}
T=2,\quad \mathcal T_{11}=[\tau_L,\tau_H],\quad 0<\tau_L<\tau_H,\quad
\mathbb P_1(\tau_{11}=\tau_j)>0\ (j\in\{L,H\}),\\
x_{11}\in[0,1],\quad u_{11}=-\frac{x_{11}^2}{2\tau_{11}},\quad
\theta_2\in\{0,1\},\quad \mathcal Q_2(\theta_2=1\mid x_{11})=x_{11},\\
y_2=\theta_2\text{ under }\mathcal Y_2,\qquad u_2^P=v^F\theta_2,\quad v^F>0.
\end{gathered}
\]
The initial law satisfies $\mathbb P_1(\tau_{11}\in\{\tau_L,\tau_H\})=1$, and the financial-contracting report set is $\widehat\tau_{11}\in\{\tau_L,\tau_H\}$. The entrepreneur observes $\tau_{11}$, reports $\widehat\tau_{11}$, and privately chooses $x_{11}$. The initial state and signal are degenerate, with no terminal production choice. At date~0 the lender commits to report-independent $\mathcal D_2^\delta$, $\delta\in[0,1]$, incurring the date-1 cost $\nu_1\delta+K_1\delta^2/2$, where $\nu_1\geq0$ and $K_1>0$. The disclosure kernel satisfies
\begin{equation}\label{eq:app-fin-evidence}
 \mathcal D_2^\delta(d_2=1\mid y_2=1)=\delta,\qquad \mathcal D_2^\delta(d_2=\varnothing\mid y_2=1)=1-\delta,\qquad \mathcal D_2^\delta(d_2=\varnothing\mid y_2=0)=1.
\end{equation}
Thus $\mathbb P(d_2=1\mid x_{11};\mathcal D_2^\delta)=\delta x_{11}$: the hidden choice $x_{11}$ determines the state through $\mathcal Q_2$, evidence $y_2$ is generated by $\mathcal Y_2$, and $\mathcal D_2^\delta$ determines the probability of a contractible bonus event.

Fix $0<b_L<b_H$ with $\beta\tau_Hb_H\leq1$ and set $b(\tau_j)=b_j$ for $j\in\{L,H\}$. Each report $\widehat\tau_{11}\in\{\tau_L,\tau_H\}$ selects $b(\widehat\tau_{11})$, the fee $p_{11}(\widehat\tau_{11};\delta)$, and $p_{12}=-b(\widehat\tau_{11})\mathbf1\{d_2=1\}$. The recommended action is $\px_{11}(\tau_j)=\beta\tau_jb_j\delta$. By \eqref{eq:continuation-plan}, the action-optimized post-report payoff is
\begin{equation}\label{eq:app-fin-utility}
 V_{11}^{*}(\tau_{11};\widehat\tau_{11},\delta)=\max_{x_{11}\in[0,1]}\left\{\beta\delta b(\widehat\tau_{11})x_{11}-\frac{x_{11}^2}{2\tau_{11}}\right\}-p_{11}(\widehat\tau_{11};\delta)=\frac{\beta^2\tau_{11}b(\widehat\tau_{11})^2\delta^2}{2}-p_{11}(\widehat\tau_{11};\delta),
\end{equation}
with the global post-report optimum $x_{11}^*(\tau_{11};\widehat\tau_{11},\delta)=\beta\tau_{11}b(\widehat\tau_{11})\delta$. Set the fees to
\begin{equation}\label{eq:app-fin-fees}
 \begin{aligned}
 p_{11}(\tau_L;\delta)&=\frac{\beta^2\tau_Lb_L^2\delta^2}{2},\\
 p_{11}(\tau_H;\delta)&=\beta^2\delta^2\left[\frac{\tau_Lb_H^2}{2}+\frac{(\tau_H-\tau_L)(b_H^2-b_L^2)}4\right].
 \end{aligned}
\end{equation}

\begin{proposition}[Joint Incentive Compatibility in Financial Contracting]\label{prop:app-fin-ic}
Under the maintained conditions, the fee menu \eqref{eq:app-fin-fees} satisfies, for every $\delta\in[0,1]$:
\begin{enumerate}
\item \textbf{Choice obedience.} For every $(\tau_{11},\widehat\tau_{11})\in\{\tau_L,\tau_H\}^2$,
\[
 x_{11}^*(\tau_{11};\widehat\tau_{11},\delta)=\beta\tau_{11}b(\widehat\tau_{11})\delta;
 \qquad x_{11}^*(\tau_{11};\tau_{11},\delta)=\px_{11}(\tau_{11}).
\]
\item \textbf{Reporting IC.} For $\delta>0$ and $\widehat\tau_{11}\ne\tau_{11}$,
\[
 V_{11}^{*}(\tau_{11};\tau_{11},\delta)-V_{11}^{*}(\tau_{11};\widehat\tau_{11},\delta)=\frac{\beta^2(\tau_H-\tau_L)(b_H^2-b_L^2)}4\delta^2>0.
\]
At $\delta=0$, both reports yield zero utility.
\item \textbf{Participation and rents.} Under truthful reporting,
\[
 V_{11}^{*}(\tau_L;\tau_L,\delta)=0,\qquad V_{11}^{*}(\tau_H;\tau_H,\delta)=\frac{\beta^2(\tau_H-\tau_L)(b_H^2+b_L^2)}4\delta^2.
\]
\end{enumerate}
After a false report, $x_{11}^*(\tau_{11};\widehat\tau_{11},\delta)$ need not equal $\px_{11}(\widehat\tau_{11})$. Thus Section~3 reporting IC is verified after reoptimization of $x_{11}$; the discrete-type specification does not invoke the differentiable-type envelope of Section~4.
\end{proposition}
\begin{proof}
By \eqref{eq:app-fin-evidence}, $\mathbb P(d_2=1\mid x_{11};\mathcal D_2^\delta)=\delta x_{11}$. For $(\tau_{11},\widehat\tau_{11})\in\{\tau_L,\tau_H\}^2$, \eqref{eq:app-fin-utility} gives $\partial_{x_{11}}V_{11}=\beta\delta b(\widehat\tau_{11})-\frac{x_{11}}{\tau_{11}},\; \partial_{x_{11}x_{11}}^2V_{11}=-\frac1{\tau_{11}}<0.$ Since $0\le\beta\tau_{11}b(\widehat\tau_{11})\delta\le\beta\tau_Hb_H\le1$, $x_{11}^*(\tau_{11};\widehat\tau_{11},\delta) =\beta\tau_{11}b(\widehat\tau_{11})\delta, \; x_{11}^*(\tau_{11};\tau_{11},\delta)=\px_{11}(\tau_{11}).$ Substituting \eqref{eq:app-fin-fees} into \eqref{eq:app-fin-utility} gives parts~(2)--(3), including at $\delta=0$. The report-conditioned choice is reoptimized after either report; terminal messages are degenerate.
\end{proof}

Disclosure strengthens production incentives but raises the high type's information rent. Recommended effort increases linearly in $\delta$, whereas the high type's rent increases quadratically.

Using \eqref{eq:lifetime-payoffs}, the lender's payoff under the implemented menu is
\begin{equation}\label{eq:app-fin-reduced}
\begin{aligned}
\Pi(\Gamma^\delta)
&=\sum_{j\in\{L,H\}}\mathbb P_1(\tau_{11}=\tau_j)
\left[
\beta v^F x_{11}^*(\tau_j;\tau_j,\delta)
-\beta\delta b_j x_{11}^*(\tau_j;\tau_j,\delta)
+p_{11}(\tau_j;\delta)
\right]
-\nu_1\delta-\frac{K_1\delta^2}{2}\\
&=M_F\delta-\frac{N_F}{2}\delta^2,\\
M_F
&=\beta^2v^F\sum_{j\in\{L,H\}}\mathbb P_1(\tau_{11}=\tau_j)\tau_j b_j-\nu_1,\\
N_F
&=K_1+2\beta^2\sum_{j\in\{L,H\}}\mathbb P_1(\tau_{11}=\tau_j)\tau_j b_j^2
-2\sum_{j\in\{L,H\}}\mathbb P_1(\tau_{11}=\tau_j) p_{11}(\tau_j;1)>0.
\end{aligned}
\end{equation}
$N_F>0$ follows from \eqref{eq:app-fin-fees}. The objective incorporates $v^F$, $b_j$, $p_{11}(\tau_j;\delta)$, and disclosure costs $(\nu_1,K_1)$ after reporting and action obedience are implemented.

\begin{proposition}[Optimal disclosure in the standardized financial contract]\label{prop:app-fin-disclosure}
Under the maintained conditions and the implemented menu \eqref{eq:app-fin-fees}:
\begin{enumerate}
\item \textbf{Optimal disclosure.} The unique maximizer of \eqref{eq:app-fin-reduced} on $[0,1]$ is
\begin{equation}\label{eq:app-fin-policy}
\delta^*=
\begin{cases}
0,&M_F\le0,\\
M_F/N_F,&0<M_F<N_F,\\
1,&M_F\ge N_F.
\end{cases}
\end{equation}

\item \textbf{Interior comparative statics.} If $0<M_F<N_F$, then
\[
\frac{\partial\delta^*}{\partial v^F}
=\frac{\beta^2\sum_{j\in\{L,H\}}\mathbb P_1(\tau_{11}=\tau_j)\tau_j b_j}{N_F}>0,
\qquad
\frac{\partial\delta^*}{\partial\nu_1}
=-\frac1{N_F}<0.
\]
\end{enumerate}
\end{proposition}
\begin{proof}
Substituting \eqref{eq:app-fin-fees} into \eqref{eq:app-fin-reduced} gives
\[
\begin{aligned}
N_F=K_1+\beta^2\Bigl[&\mathbb P_1(\tau_{11}=\tau_L)\tau_Lb_L^2\\
&+\tfrac12\mathbb P_1(\tau_{11}=\tau_H)\bigl((3\tau_H-\tau_L)b_H^2+(\tau_H-\tau_L)b_L^2\bigr)\Bigr]>0.
\end{aligned}
\]
Consequently, $\partial_\delta \Pi(\Gamma^\delta)=M_F-N_F\delta,\; \partial_{\delta\delta}^2\Pi(\Gamma^\delta)=-N_F<0.$ Strict concavity gives the unique maximizer in \eqref{eq:app-fin-policy}. Differentiating its interior expression gives part~(2) of Proposition~\ref{prop:app-fin-disclosure}.
\end{proof}

Unlike Examples~2--3, disclosure can be absent, partial, or full because it strengthens productive effort while raising screening rents, effort costs, and disclosure expenditures. Higher $v^F$ raises the return to disclosure, whereas higher $\nu_1$ lowers it. The marginal lender payoff $M_F-N_F\delta$ captures this tradeoff; optimal disclosure depends on economic consequences rather than informativeness alone. These comparisons concern the specified bonus--fee family $\bigl(b(\cdot),p_{11}(\cdot;\delta),p_{12}\bigr)$, not unrestricted optimization over all financial mechanisms.

\section{Conclusion}\label{sec:conclusion}

When institutional rules alter individuals' incentives and information, the resulting behavioral responses may change the equilibrium relationships on which institutional performance depends. This fundamental insight, rooted in Hurwicz's structural-invariance problem and reinforced by the Lucas critique and Sargent's emphasis on equilibrium consistency, motivates an approach to institutional design that treats economic and informational responses as jointly endogenous. While studies of individual incentive and informational margins remain indispensable for identifying their separate effects, interactions among these margins call for a comprehensive theory of joint institutional design.

Our analysis establishes how jointly determined contracting, information technology, and disclosure reshape dynamic implementability, optimal institutional design, and the validity of established economic results. JPBE, JPBIC, and the Dynamic Revelation Principle connect institutional rules to their implementation through private reporting and hidden actions. Recursive and global incentive characterizations yield dynamic information rents and optimal contracting conditions. The resulting contracting hierarchy is branched rather than linear: neither second-best regime generally dominates the other, and the interaction of incentive frictions can have either sign. Intertemporal feedback changes optimal information design: periodwise optimality need not imply joint optimality, and current-payoff replication need not preserve continuation incentives. We identify conditions for institutional dominance and incentive separation. The auction and financial-contracting applications demonstrate how these interactions can reverse familiar revenue comparisons and generate distinct optimal disclosure regimes.

The present theory studies joint mechanism design under specified economic allocation constraints. At the system level, \citet{Tian2026} conceptualizes economic systems as dynamic joint adjustment processes of resource and information allocation linked by recursive feedback, highlighting the endogenous informational evolution of economic systems. Integrating this system-level perspective with the full range of informational and incentive interactions studied here remains a direction for further research. More broadly, the theory reveals how and why established economic results change when previously restricted institutional margins are jointly considered, and under what conditions they remain valid.

\end{document}